\documentclass[11pt]{iopart}
\pdfoutput=1
\expandafter\let\csname equation*\endcsname\relax
\expandafter\let\csname endequation*\endcsname\relax 
\usepackage[english]{babel}
\usepackage{amsmath}
\usepackage{amssymb}
\usepackage{graphicx}
\usepackage{color}
\usepackage[utf8]{inputenc}
\usepackage{amsfonts}
\usepackage{graphicx} 
\usepackage{amsthm}
\usepackage{placeins}
\usepackage{verbatim}
\usepackage{booktabs}
\usepackage{color}
\usepackage{bbold}
\usepackage{comment}
\usepackage{url}
\usepackage[all]{xy}
\usepackage[width=18.5cm, vscale=0.8]{geometry}
\newcommand{\be}{\begin{equation}}
\newcommand{\ee}{\end{equation}}
\newcommand{\bea}{\begin{eqnarray}}
\newcommand{\eea}{\end{eqnarray}}

\newcommand{\bA}{{\bf A}}
\newcommand{\bC}{{\bf C}}

\newcommand{\bD}{{\bf D}}
\newcommand{\bP}{{\bf P}}

\newcommand{\order}{{\mathcal{O}}}
\newcommand{\half}{{\frac{1}{2}}}
\usepackage{epsfig}
\newcommand{\bxi}{{\mbox{\boldmath $\xi$}}}
\newcommand{\bsigma}{{\mbox{\boldmath $\sigma$}}}
\newcommand{\bpsi}{{\mbox{\boldmath $\psi$}}}
\newcommand{\bphi}{{\mbox{\boldmath $\phi$}}}

\newcommand{\bm}{{\bf m}}
\newcommand{\bv}{{\bf v}}
\newcommand{\bb}{{\bf b}}
\newcommand{\bM}{{\bf M}}

\newcommand{\by}{{\bf y}}
\newcommand{\bz}{{\bf z}}
\newcommand{\bX}{{\bf X}}
\newcommand{\bn}{{\bf n}}
\newcommand{\hm}{\hat{\mu}}
\newcommand{\bra}{{\langle}}
\newcommand{\ket}{{\rangle}}

\newcommand{\atanh}{{\rm atanh}}
\newcommand{\insp}{{\frac{1}{\sqrt{2(1+k)}}}}
\newcommand{\insm}{{\frac{1}{\sqrt{2(1-k)}}}}
\newcommand{\detx}{{\sqrt{1-k^2}}}

\begin{document} 
\title[The role of idiotypic interactions 
in the adaptive immune system]{The role of idiotypic interactions 
in the adaptive immune system: a belief-propagation approach }
\author{Silvia Bartolucci$^{1}$, Alexander Mozeika$^{2}$ and 
Alessia Annibale $^{1,2}$}
\address{$^1$ Department of Mathematics, King's College London, The Strand,
London WC2R 2LS, UK}
\address{$^2$ Institute for Mathematical and Molecular Biomedicine, King's 
College London, Hodgkin Building, London SE1 1UL, UK}

\begin{abstract}
In this work we use belief-propagation techniques to study the equilibrium behaviour of a minimal model for the immune system comprising interacting T and B clones. 
We investigate the effect of the so-called {\em idiotypic interactions} among 
complementary B clones on the system's activation.
Our result shows that B-B interactions increase the system's resilience to 
noise, making clonal activation more stable, while increasing 
the cross-talk between different clones. 
We derive analytically the noise level at which a B clone gets activated, in 
the absence of cross-talk, 
and find that this increases with the strength of idiotypic interactions and 
with the number of T cells signalling the B clone. 
We also derive, analytically and numerically, via population dynamics, 
the critical line where clonal cross-talk arises. 
Our approach allows us to derive the B clone size distribution, 
which can be experimentally measured and gives important information about 
the adaptive immune system response to antigens and vaccination.
\end{abstract}

\section{Introduction}
The immune system is a complex collection of biological structures (clones, organs, molecules, etc.) that protects the organism from a variety of agents causing diseases \cite{Abbas}.
The first line of defence, called {\em innate immune system}, is composed by anatomical barriers, which include physical (e.g. epithelium), chemical (e.g. gut flora) and biological (e.g. saliva) barriers; it produces a generic response to a large class of  different pathogens.
The {\em adaptive immune system} produces a more targeted response to specific external invaders, the antigens. 
The cells composing this part of the immune system, namely B and T lymphocytes, 
present receptors on their surfaces that are able to bind only to specific antigens \cite{Burnet}. Hence, the resulting immune response is highly specialised and more effective.
Moreover, the adaptive immunity keeps an immunological memory of the specific antigen: 
in subsequent encounters with the same antigen, the system produces stronger and faster immune responses.
When a pathogen enters the organism, T and B clones independently recognise a specific antigen via their receptors; T clones regulate the immune response by sending excitatory or inhibitory signals in the form of signalling proteins, the {\em cytokines}, to the B clones: these confirmatory messages are essential to initiate the antibodies production \cite{Abbas}. B clones are indeed the only clones that can produce antibodies, proteins able to identify and neutralise the pathogens.
Once the immune response has been activated, B clones start to proliferate and clonally expand, producing clones of identical cells sharing the same antigen receptors.
 
Recently, following the development of experimental techniques such as high-throughput sequencing data \cite{hts1} new 
interesting observables have become available, among which the clonal size distribution. All clones in the immune systems undergo a 
birth-death process, which is regulated and modified to achieve the desired immune response. Monitoring the clone size distribution is thus
important to understand the immune system response to antigens and vaccinations \cite{Bdist1}.
Experiments and analytical work are now available for different cell types, both for humans and other species \cite{Bdist1,hts2,Bdist2}.

\begin{figure}[htb!]
\centering
\includegraphics[width=0.6\textwidth]{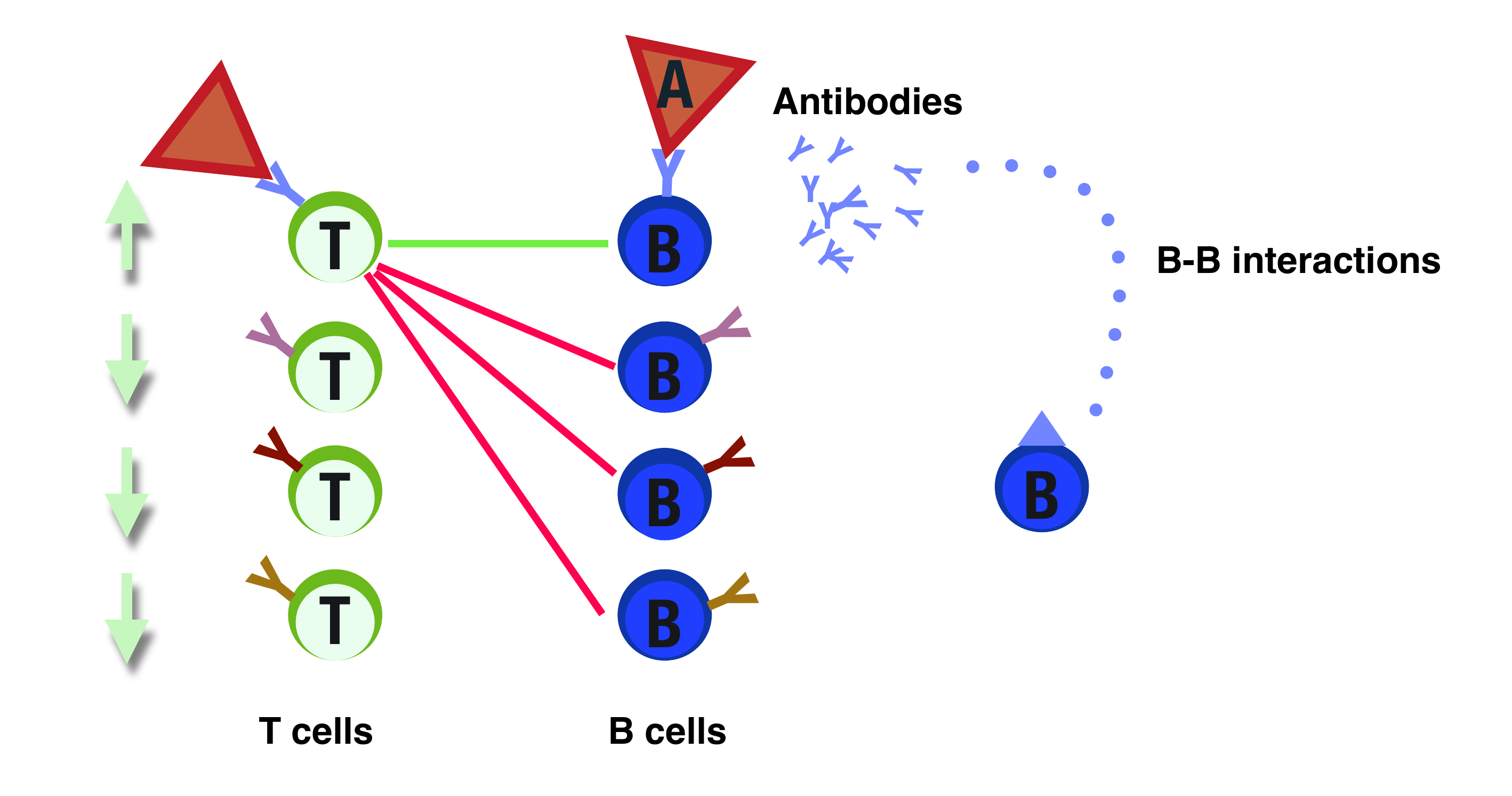}
\caption{Scheme of B-T and B-B interactions: B and T clones are activated by an antigen (A) and T clones start sending cytokines to promote (green link) or suppress (red link) the activation of the relevant B clones able to react to the  antigen. Once activated, the B cell starts clonally expanding, i.e. identically copying itself, producing antibodies. In this process, also B clones with complementary idiotypes to the active one get activated, creating the so-called idiotypic network of interactions between B lymphocytes.}
\label{fig:scheme}
\end{figure}

Interestingly, the cells and molecules of the immune system can also recognise, 
react to and regulate each other forming an interacting network. In particular, B clones can bind, 
recognise and inhibit or excite each other forming the so-called {\em idiotypic network} of interactions; indeed, the antigens combining sites, i.e. 
idiotypes, which are expressed on the B cells surface also react to complementary receptors displayed on other B clones \cite{BBRaff}. A schematic representation of the interactions among lymphocytes is shown in fig. \ref{fig:scheme}.
Different B-B interactions mechanisms have been proposed, e.g. via cell receptors or antibodies. 
In this work we do not enter the details of how such interactions arise, we just assume their existence and study their impact on the system's behaviour.
The first formulation of an immune network theory by N. K. Jerne \cite{Jerne} dates back to the early 1970's and was soon after followed by preliminary experimental confirmations \cite{expBB1,expBB2,expBB3} and various mathematical models to capture this interaction mechanism \cite{Perelson}. In particular, linear idiotypic network models \cite{Hoff}, where antibodies  belonging to the $i$-th group could interact with antibodies of the $i-1$ and $i+1$-th group, or a cyclic version of it \cite{Hiern} have been proposed, together with models with random connections between antibodies \cite{Parisi}. In these models the dynamics of the B clones concentration is modelled via large systems of coupled ODEs, mostly solved via numerical simulations.
More recently, experiments also showed a new connection between the existence of an "idiotypic network" and the onset of autoimmune diseases, spotlighting the importance of such interactions and the need for a more careful analysis of its effects \cite{autoimmuneBB1,autoimmuneBB2,autoimmuneBB3}.

To this end, we extend previous models of B-T clones interactions using statistical mechanics tools \cite{jstatba,cavitysollich} to further investigate the effects of 
the idiotypic interactions on the clone activation level and immune response strength.
With our approach, based on belief-propagation techniques \cite{montanari} and tools imported from neural networks models \cite{jphysaas} and spin glasses \cite{MPV}, 
we can derive the expression of experimentally accessible observables such as the distribution of B clone sizes that 
gives important information about the response of the immune system in healthy and pathological situations 
and could help deliver more targeted therapies. 
Our results show that idiotypic interactions increase the stability of B clones activation against biological noise, but they also increase 
cross-talk effects between different clones.
 
The paper is organised as follows. In sec. \ref{sec:model} we introduce the mathematical model, its players and assumptions. 
In sec. \ref{sec:factor} we derive the recursive equations for the marginal distributions of T clones activation (details in \ref{appendix:factor} ) 
and we provide formulae for the distribution of B clone sizes. In particular, we inspect 
their transition from a single peaked to a bimodal distribution, increasing the noise level, in the paramagnetic phase (sec. \ref{sec:para}) and for ferromagnetic 
interactions (sec.  \ref{sec:ferro}), where one can proceed analytically.
In sec. \ref{sec:disorder} we consider the most general case with disordered interactions and 
we derive analytically and via population dynamics simulations the critical line, in the space of model parameters, that separates 
the region where clones act independently from the one where they feel a strong interference due to the presence of other clones. 
\section{The model}\label{sec:model}
In this section, we define a minimal model for the adaptive immune system, which comprises the interactions between B and T clones, 
essential to initiate an immune response, and the idiotypic interactions between B clones, with the aim to quantify their  impact on the immune system activation.
In our model T clones are binary variables or ``spins'' $\sigma_i=\pm 1$, $i=1,\dots,N$, depending on whether they are ($+1$) or not ($-1$) secreting cytokines. 
B clones are described by real variables $b_{\mu}\in\mathbb{R}$, $\mu =1,\dots, P$, 
characterising their log-concentrations with respect to a reference value. 
The interactions between T and B clones are mediated by cytokines, which can be excitatory or inhibitory. 
We model the interaction between a T clone $i$ and a B clone $\mu$, via a variable $\xi_i^{\mu}$
which takes value $1$ if the cytokine is excitatory, $-1$ if it is inhibitory and $0$ if it is absent, due to the specificity of these interactions. 
Both $\sigma$'s and $b$'s are random variables and we expect their joint 
distribution $p(\bsigma,{\bf b}|\bxi)$ for any given realization of the cytokines patterns, 
to be given in equilibrium at inverse noise level $\sqrt{\beta}$, by the Gibbs Boltzmann distribution \cite{PRE}
\bea
p(\bsigma, {\bf b}|\bxi)
=\frac{{\rm e}^{-\sqrt{\beta}H(\bsigma, {\bf b}|\bxi})}{Z}\ ,
\label{eq:joint}
\eea
with Hamiltonian
\be
H(\bsigma,\bb | \bxi)=
-\sum_{i=1}^N\sum_{\mu=1}^P\xi_i^{\mu} \sigma_i b_\mu+
\frac{1}{2\sqrt{\beta}} \sum_{\mu,\nu=1}^P b_\mu A_{\mu\nu}b_\nu \ .
\label{eq:Ham_bi}
\ee 
The matrix $\bA=\{A_{\mu\nu}\}$ models interactions between B clones, 
also called {\em idiotypic interactions} \cite{Jerne,BBRaff}.
In the following, we will often drop the explicit dependence on the B-T interactions $\bxi$ in the Hamiltonian, which we will simply denote by $H(\bsigma,\bb)$, as well as 
in the Boltzmann distribution, $p(\bsigma, {\bf b})$, and in its marginals. 

In this section, we choose a suitable form of the matrix $\bA$ 
to mimic idiotypic interactions, discussing the model's assumptions that motivate our choice. Moreover, we introduce the relevant quantities that we will monitor to probe the system's behaviour. 
We will discuss the graph topology chosen for the B-T interactions in sec. \ref{sec:factor}, where a more convenient (factor graph) representation of the Boltzmann distribution will be 
introduced.

To investigate B-B interactions we choose the simplest configuration of the receptor space which retains the key biological features. We represent B clones receptors as binary strings and assume that complementary strings, 
like e.g. $010\ldots$ and $101\ldots$, excite each other. Noting that one can always label each of $P=2^d$ binary strings of length $d$ by an index $i$, in such a way that the string complementary to $i$ is labeled by $i+P/2$, 
we order the strings on a ring in such a way that each string sits close to similar strings and opposite to complementary ones. 
The reason for this choice is that B-B interactions are believed to be very specific \cite{Jerne,BBman}, with each B clone interacting with a number $\order{(1)}$ of other clones. 
For simplicity, we assume that each B clone interacts only with another clone, but this assumption can be easily relaxed; for example, one may assume that each B clone does not only interact with the cell strictly complementary to it (with Hamming distance $d$), but also with the $d$ clones which are nearly complementary (with a Hamming distance $d-1$).  
We expect that as long as the connectivity of B clones is $\sim d \ll N$ the effect is qualitatively very similar. 
We suppose that the $\mu$-th B clone expansion is triggered by the 
($\mu+P/2$)-th B clone, which is precisely complementary to that B 
clone and each B clone dies at a certain rate.  In the following we will denote by $\hat\mu=\mu+P/2$ the complementary clone of the $\mu$-th B clone. For simplicity, we assume a unitary death rate for all clones.
This leads us to the B-B interactions matrix
\bea
A_{\mu\nu}=\delta_{\mu\nu}-k\delta_{\mu,(\nu + P/2)\ \mathrm{mod}\ P},
\label{eq:A_def}
\eea
with $k\in[0,1)$ representing the strength of idiotypic interactions. 
Assuming that the population dynamics of each clone $\mu$ is given by a gradient descent on $H(\bsigma, {\bf b})$ \cite{jstatba}, 
the non-zero diagonal terms $A_{\mu\mu}$ account for a decay term in the population dynamics of each clone $\mu$, proportional to the population size $b_\mu$ itself. 
Clearly, one could make several other choices for matrix $\bA$. For example, intra-clonal competition can be introduced by allowing non-zero off-diagonal entries in ${\bf A}$. 
These elements would need to scale as $1/P$ to ensure that 
the decay term is overall $\mathcal{O}(1)$. Assuming that off-diagonal terms are all equal i.e. $A_{\mu\nu}\!=\!1/P~\forall~\mu\neq \nu$ and the overall log-concentration of clones $\sum_\mu b_\mu=B$ 
is constant, the loss term in the population dynamics would gain a constant term $-B/P$, which can be absorbed in the definition of $b_\mu$. 
More general matrices can be chosen to account for the system's heterogeneities, here we use (\ref{eq:A_def}) as the simplest choice that retains 
the two key biological ingredients of (i) suppression effects preventing abnormal clonal expansion and (ii) excitatory signalling between complementary clones. 
In addition, (\ref{eq:A_def}) allows for a straightforward matrix inversion, that is useful for later analytical calculations. 
We note that ${\bf A}$ is a Toeplitz matrix, positive definite and symmetric with the properties $A_{\mu\nu}= \bA(\mu-\nu)$, $\bA(n+P)=\bA(n)$ and 
$\bA(n)=\bA(-n)$ and its inverse is
\bea
({\bf A}^{-1})_{\mu\nu}=\frac{1}{1-k^2}\delta_{\mu\nu} + \frac{k}{1-k^2}\delta_{\mu,(\nu + P/2)\ \mathrm{mod}\ P}\ .
\label{eq:inverse}
\eea
In the following we show that many important properties of the system 
described by Hamiltonian (\ref{eq:Ham_bi})
are encoded in the marginal distribution of the $\bsigma$, which is found by 
integrating (\ref{eq:joint}) over the variables $b_\mu$ 
\bea
p(\bsigma)&=&\frac{1}{Z'}\rme^{-\beta H(\bsigma)} \ ,
\label{eq:marginal_eq}
\eea 
leading to a Boltzmann distribution with effective Hamiltonian 
\bea
H(\bsigma)=-\frac{1}{2}\sum_{i,j=1}^N \sigma_i\sigma_j\sum_{\mu,\nu=1}^P\xi^{\mu}_i({\bf A}^{-1})_{\mu\nu}\xi^{\nu}_j \ .
\label{eq:Hsigma1}
\eea
The latter involves only interactions between T clones, 
in the separable form $J_{ij}=\bxi_i {\bf A}^{-1} \bxi_j$ with
$\bxi_i=(\xi_i^1,\ldots,\xi_i^P)$, 
and thus describes an associative network with $P$ diluted patterns 
$\{\bxi^\mu\}_{\mu=1}^P$ \cite{jstatba,PRE,jphysaas}. Here, the patterns 
$\{\bxi^{\mu}\}$ stored in the network represent T clonal strategies for B clonal activations.
We can rewrite the Hamiltonian (\ref{eq:Hsigma1}) as
\bea 
H(\bsigma)=-\frac{1}{2}{\bf m}^T(\bsigma){\bf A}^{-1} {\bf m}(\bsigma)\ ,
\label{eq:Hsigma2}
\eea 
in terms of the vector of magnetizations or order parameters ${\bf m}=(m_1,\ldots,m_P)^T$ and  and its transpose $\bm^T$.
The magnetization defined as
\bea m_{\mu}(\bsigma)=\sum_{i=1}^N\sigma_i\xi^{\mu}_i\ ,
\label{eq:over}
\eea 
gives the overlap between the system configuration $\bsigma$ and pattern $\bxi^\mu$,
 which quantifies the strength of the signal sent by T clones to B clone 
$\mu$, or the activation of B clone $\mu$.

Experimentally, one has access to the B clone size distribution, rather than $p(\bsigma)$. 
Hence, the relevant quantity to probe and compare the model behaviour with experimental data is 
\bea
p(b)=\int {\rm d}\bb \sum_\bsigma p(\bsigma, \bb) \frac{1}{P}\sum_{\mu=1}^P \delta(b-b_\mu)\ .
\label{eq:marg_b}
\eea
To this end, we rewrite (\ref{eq:joint}) as 
\bea
p(\bsigma, \bb)&=&\frac{1}{Z}\rme^{-\half \bb^T \bA \bb +\sqrt{\beta} \bb^T \bm(\bsigma)}
\label{eq:square}
\\
&=& \frac{1}{Z'} p(\bsigma) p(\bb|\sqrt{\beta} \bA^{-1} \bm(\bsigma))\ ,
\eea
where we have completed the square in (\ref{eq:square}) and introduced the joint distribution of the B clone sizes 
\bea
p(\bb|\sqrt{\beta} \bA^{-1} \bm(\bsigma))=\sqrt{\frac{\det \bA}{(2\pi)^N}} 
\rme^{-\half (\bb -\sqrt{\beta} \bA^{-1} \bm(\bsigma))^T \bA (\bb-\sqrt{\beta} \bA^{-1} \bm(\bsigma))}\ ,
\eea
 a multivariate Gaussian distribution with covariance matrix $\bA^{-1}$. The marginal distribution for the $\mu$-th B clone, obtained by integrating the above 
over all $b_\nu$ other than $b_\mu$, will be Gaussian as well with average $\sqrt{\beta} (\bA^{-1}\bm(\bsigma))_\mu$ and variance $({\bf A}^{-1})_{\mu\mu}=1/(1-k^2)~\forall~\mu$ i.e. 
\bea
p(b_\mu|\sqrt{\beta} (\bA^{-1} \bm(\bsigma))_\mu)=\frac{1}{\sqrt{2\pi /(1-k^2)}} 
\rme^{-\frac{(1-k^2)}{2} (b_\mu -\sqrt{\beta} (\bA^{-1} \bm(\bsigma))_\mu)^2 }\ .
\label{eq:bmu}
\eea
Hence, from (\ref{eq:marg_b}) we get
\bea
p(b)=\frac{1}{Z'} \sum_\bsigma p(\bsigma) \frac{1}{P}\sum_{\mu=1}^P p(b|\sqrt{\beta}(\bA^{-1}\bm(\bsigma))_\mu) \ .
\eea
Upon introducing the ``rotated'' overlaps $\bf{\tilde{m}(\bsigma)}=\bA^{-1} \bm(\bsigma)$ we get 
\bea
p(b)&=&\frac{1}{Z'} \sum_{\tilde{m}} \sum_\bsigma p(\bsigma) \frac{1}{P}\sum_{\mu=1}^P \delta(\tilde{m}-\tilde{m}_\mu(\bsigma)) p(b|\sqrt{\beta} \tilde{m})
\nonumber\\
&=&\frac{1}{Z'} \int {\rm d}{\tilde{m}}\, P(\tilde{m}) \,p(b|\sqrt{\beta} \tilde{m}) \ ,
\label{eq:pbi}
\eea
where 
\bea
P(\tilde{m})=\sum_\bsigma p(\bsigma) \frac{1}{P}\sum_{\mu=1}^P\delta({\tilde{m}-\tilde{m}_\mu(\bsigma)})
\label{eq:Pm}
\eea
and 
\bea
p(b|\sqrt{\beta}\tilde{m})=\frac{1}{\sqrt{2\pi/(1-k^2)}} \rme^{-\frac{1-k^2}{2}(b-\sqrt{\beta} \tilde{m})^2}\ .
\label{eq:Gauss_b}
\eea
This shows that the distribution of B clones sizes $p(b)$ is 
readily determined from the distribution $P(\tilde{m})$ of rotated overlaps $\bf{\tilde{m}}(\bsigma)$ in the "marginalised" system 
involving only $\bsigma$, described by (\ref{eq:Hsigma2}), where the $\bb$ have been integrated out.
Hence, the problem of computing B clones distributions, which are experimentally accessible in immunology, reduces to evaluating
the overlap distribution in an associative memory with diluted and coupled patterns. In the absence of idiotypic interactions, i.e. for uncoupled patterns, 
it is seen numerically that for ratios $\alpha=P/N$ not too large and for low noise level $T$, 
the overlap distribution crosses-over from a unimodal distribution peaked in 
zero to a bimodal peaked at large values of the overlap \cite{AABCT_jpa}. In this work we derive 
a tight bound on the noise level $T$ at which this cross-over takes place.
Interestingly, this coincides with the critical temperature at which the 
system undergoes a phase transition, from $\bm=\mathbf{0}$ to $\bm\neq \mathbf{0}$, 
in the regime of extremely diluted B-T interactions and sub-extensive 
number of B clones \cite{jstatba}.
Since the B clones size distribution is a convolution of Gaussian distributions centred 
on the emerging values of the (rotated) overlaps, the phase where $\bm\neq {\bf 0}$, 
can be regarded as the healthy phase of the immune system, where cells numbers are sustained.  
This may provide a theoretical explanation for the need of a {\it basal} activity of the immune system, by which T clones send signals to B clones. This is experimentally
observed, even in the absence of external pathogens, and is believed to be one of the mechanisms to accomplish a {\em homeostatic control} 
of cell numbers \cite{homeostasis}. 

In a similar fashion, we can calculate the equilibrium concentration of any B clone $\mu$ and the activation of any T clone $i$ via
\bea
P_{i\mu}(\sigma,b)&=&\int d\bb \sum_\bsigma p(\bsigma, \bb) 
\delta(b-b_\mu)\delta_{\sigma,\sigma_i}
\nonumber\\
&=&\frac{1}{Z} \sum_\bsigma p(\bsigma) \delta_{\sigma,\sigma_i}
p(b|\sqrt{\beta}\tilde{m}_\mu(\bsigma))
\nonumber\\
&=&\frac{1}{Z'} \int {\rm d}\tilde{m} \,p(b|\sqrt{\beta} \tilde m)
\sum_\bsigma p(\bsigma) \delta_{\sigma,\sigma_i} \delta({\tilde m-\tilde m_\mu(\bsigma)}),
\label{eq:marg2_b}
\eea
showing again that all the information about the physics of the system is 
encoded in the marginalised distribution $p(\bsigma)$.
\section{Factor graph representation}
\label{sec:factor}
We can visualize our system of $N$ T clones and $P$ B clones as a bipartite graph, $\mathcal{G} =(\mathcal{T},\mathcal{B})$ where T clones constitute the $\mathcal{T}$ party and 
B clones represent the $\mathcal{B}$ party, with $N=|\mathcal{T}|$ and $P=|\mathcal{B}|$. We indicate with $\partial \mu=\{ i:\xi_i^{\mu}\neq0 \}$ the set of T clones connected to the B clone $\mu$, 
and with $\partial i=\{ \mu :\xi_i^{\mu}\neq0 \}$ the set of B clones connected to a particular clone $i$, i.e. 
$|\partial\mu|$ is the degree of a B clone $\mu$ and $|\partial i|$ is the degree of a T clone $i$ in the bipartite graph $\mathcal{G}$.  
In our analysis, we will consider random bipartite graph ensembles where the degrees $d_i=|\partial i|$ and $q_\mu=|\partial \mu|$ of the nodes in the two parties are drawn respectively from the 
distributions $P_d(d)=\frac{1}{N}\sum_{i=1}^N\delta_{d,|\partial i|}$, $P_q(q)=\frac{1}{P}\sum_{\mu=1}^P\delta_{q,|\partial\mu|}$, and links are random and independent variables which take values  
$\xi^\mu_i\in\{+1,-1\}$. Conservation of links demands $\sum_{i=1}^Nd_i=\sum_{\mu=1}^Pq_{\mu}$, which gives, for large $N$, 
$N\langle d \rangle = P \langle q \rangle$, where averages are taken over $P_d(d)$ and $P_q(q)$. Different graph topologies have been considered, in the absence of idiotypic interactions, in \cite{cavitysollich}, while here we mostly focus 
on regular graph topologies and on the role of B-B interactions.

In this section, we introduce a factor graph representation of the Boltzmann distribution of the marginalised system described by the Hamiltonian (\ref{eq:Hsigma2}),
which will allow us to derive recursive equations for the marginal distributions of T clones activation and formulae for the distributions 
of the overlaps, which quantify B clones activation.

As a first step, we diagonalise the symmetric matrix ${\bf A}$ by means of the similarity 
transformation $\bD=\bP^{-1}\bA \bP$, 
where $\bD$ is the diagonal matrix constructed 
from the eigenvalues $\{\lambda_\mu\}_{\mu=1}^P$ of $\bA$ and $\bP$ is the orthogonal
matrix of eigenvectors ($\bP^{-1}=\bP^{T}$). Hence, we write the Hamiltonian 
in terms of the transformed vector $\bv(\bsigma)=\bP^{-1}\bM(\bsigma)$ 
\bea 
H(\bsigma)=-\frac{1}{2}{\bf v}^T(\bsigma){\bf D}^{-1} {\bf v}(\bsigma)
=-\frac{1}{2}\sum_\mu v_\mu^2(\bsigma)\frac{1}{\lambda_\mu}\ ,
\eea 
which allows to write $p(\bsigma)$, as defined in \eqref{eq:marginal_eq}, in the factorised form 
\be
p(\bsigma)=\prod_{\mu=1}^P F_\mu(\bsigma) \ ,
\label{eq:factorsmain}
\ee
with factors
\be
F_{\mu}(\bsigma)={\rm e}^{\beta v_\mu^2(\bsigma)/(2\lambda_\mu)}=
\bra 
\rme^{v_\mu(\bsigma)z\sqrt{\beta/\lambda_\mu}}\ket_z 
\label{eq:z_average}
\ee
and 
\begin{equation}
\bra f(z) \ket_z=
\int^{+\infty}_{-\infty} \frac{{\rm d}z}{\sqrt{2\pi}}\rme^{-\frac{1}{2}z^2} f(z) \ .
\label{eq:av_z}
\end{equation}
After simple algebraic manipulations (see \ref{appendix:factor} for details) we can rewrite \eqref{eq:factorsmain} as a product of 
$P/2$ factors
\be
p(\bsigma)=\prod_{\mu=1}^{P/2} f_{\mu\hat{\mu }}(\{\sigma_k, k\in\partial \mu\},\{\sigma_\ell, \ell\in\partial \hat\mu\}) \ ,
\label{eq:psigmaf}
\ee
each involving a pair of complementary clones $\mu,\hat \mu$
\be
f_{\mu\hat{\mu}}(\{\sigma_k, k\in\partial \mu\},\{\sigma_\ell, \ell\in\partial \hat\mu\})= 
\left\langle \exp\left[\sqrt{\beta}\left(y_1
\sum_{k\in \partial\mu}\xi_k^\mu \sigma_k
+y_2
\sum_{\ell\in \partial\hat\mu}\xi_\ell^{\hat\mu} \sigma_\ell
\right)\right]
\right\rangle_{\by},
\label{eq:fmu}
\ee
where 
\bea
\bra (\cdots) \ket_\by=\detx\int \frac{\rmd y_1 \rmd y_2}{2\pi} (\cdots) {\rm e}^{-\frac{1}{2}
\by^T \bC^{-1} \by
}
\label{eq:integral1}
\eea
and 
\bea
\bC^{-1}=\left(
\begin{array}{cc}
1 & -k \\
-k & 1
\end{array}
\right) \ .
\label{eq:Cmatrix1}
\eea
Each factor $f_{\mu\hat{\mu}}$ is a function of the T clones $\{\sigma_k\}$ connected to the $\mu$-th B clone and the $\{\sigma_\ell\}$ connected to the complementary clone $\hat{\mu}$. 
Via this representation we note that the effect of the B-B interactions is equivalent to joining together T clones signalling to complementary B clones.

To compute relevant observables of the system, such as the distribution of overlaps quantifying B clones activation, we introduce the so-called cavity marginals 
$P_{\mu\hat{\mu}}(\sigma_i)$ of $\sigma_i$, when the $i$-th clone is coupled to either the factor $\mu$ or its complementary $\hat{\mu}$ and nothing else, 
and $P_{\setminus \mu\hat{\mu}}(\sigma_i)$ of $\sigma_i$ when coupled to all factors except $\mu$ 
and its complement $\hat{\mu}$. These are often referred to as the {\em messages} from factors $\mu,\hat\mu$ to node $i$ and from node $i$ to factors $\mu,\hat{\mu}$, respectively.
For sparse interactions $\{\xi_i^\mu\}$, our factor graph will be locally tree-like in the thermodynamic limit, with typical loop lengths diverging (logarithmically) with $N$, so we can use the Bethe-Peierls approximation \cite{bethe,montanari} to find the cavity distributions in a recursive fashion
\bea
P_{\setminus \mu\hat{\mu}}(\sigma_i)&=&
\prod_{\nu\in \partial i\setminus\mu\hat{\mu}} P_{\nu\hat{\nu}}(\sigma_i)\ ,
\nonumber\\
P_{\nu\hat{\nu}}(\sigma_i)&=&
\sum_{\{\sigma_{k\in\partial \nu\hat\nu\setminus i}\}}
f_{\nu\hat{\nu}}(\sigma_i, \{\sigma_{k\in \partial \nu\hat\nu \setminus i}\}) 
\prod_{k\in \partial \nu\hat{\nu}\setminus i }
P_{\setminus\nu\hat{\nu}}(\sigma_k)\ ,
\label{eq:iter}
\eea
where $\partial \nu\hat \nu=\partial \nu \cup \partial \hat \nu$.
In \ref{app:BB2} there is a formal derivation of the recursions \eqref{eq:iter} for a model defined on a factor tree and a schematic representation of the graph used to derive the equations. Iterating these equations until convergence, we obtain exact solutions on trees and approximately exact solutions on locally tree-like graphs.

\subsection{Distribution of overlaps} \label{subset:overl}
To study the activation properties of the system, i.e. its ability to retrieve stored patterns of clonal activation, encoded in the pattern overlaps 
$m_\mu=\sum_{k\in \partial\mu}\xi^{\mu}_k\sigma_k$, we look at the joint distribution of complementary clones activation
\bea
P_{\mu\hat{\mu}}(m,\hat{m})=\langle\delta_{m,m_{\mu}(\bsigma)}\delta_{\hat{m},m_{\hat{\mu}}(\bsigma)}\rangle=\sum_{\bsigma}p(\bsigma)\delta_{m,m_{\mu}(\bsigma)}\delta_{\hat{m},m_{\hat{\mu}}(\bsigma)} \ .
\label{eq:pmmhat1}
\eea
Inserting (\ref{eq:psigmaf}) in (\ref{eq:pmmhat1}) and splitting the sums over spins attached to factors $\mu,\hat\mu$ from the sums 
over spins in the rest of the (tree-like) graph (which give the cavity marginals in the absence of the factors $\mu,\hat\mu$) we obtain  
\begin{flalign}P_{\mu\hat{\mu}}(m,\hat{m})=\frac{\sum_{\substack{\{\sigma_k \in\partial\mu\hat\mu\}}}\int {\rm d}y_1 {\rm d}y_2 
{\rm e}^{-\frac{1}{2}\by^T \bC^{-1} \by}{\rm e}^{\sqrt{\beta}(y_1m_\mu(\bsigma)+y_2m_{\hat{\mu}}(\bsigma))}\delta_{m,m_{\mu}(\bsigma)}\delta_{\hat{m},m_{\hat{\mu}}(\bsigma)}
\prod_{k\in\partial\mu\hat{\mu}}P_{\setminus\mu\hat{\mu}}(\sigma_k)}{\sum_{\{\tilde{\sigma}_k\in\partial\mu\hat\mu\}}\int {\rm d}y_1 {\rm d}y_2  
{\rm e}^{-\frac{1}{2}\by^T \bC^{-1} \by}{\rm e}^{\sqrt{\beta}(y_1 m_\mu(\tilde{\bsigma})+y_2m_{\hat{\mu}}(\tilde{\bsigma})}
\prod_{k\in\partial\mu\hat{\mu}}P_{\setminus\mu\hat{\mu}}(\tilde{\sigma}_k)}\ ,\nonumber\\
\label{eq:pmmhatgeneral}
\end{flalign}
where the cavity marginals $P_{\setminus\mu\hat{\mu}}(\sigma_k)$ must be computed recursively from equations \eqref{eq:iter}.
Finally, to compute the B clone size distribution $p(b)$ given in \eqref{eq:pbi}, 
we need the distribution of the rotated overlaps $\tilde{m}_\mu=\frac{m_\mu}{1-k^2}+\frac{k m_{\hat{\mu}}}{1-k^2}$
\bea
P_{\mu\hm}(\tilde{m})= \sum_{ m,\hat{m}} P_{\mu\hat{\mu}}(m,\hat{m})\delta\left(\tilde{m}-\frac{m}{1-k^2}-\frac{k \hat{m}}{1-k^2}\right).
\label{eq:pmtilde}
\eea
In the following sections, we will solve the recursive equations \eqref{eq:iter} in two cases that can be treated analytically, namely the paramagnetic phase (sec. \ref{sec:para}) and the case with ferromagnetic interactions (sec. \ref{sec:ferro}) and we will compute the overlaps and clone sizes distributions in different regimes of the model's parameters. 
In sec. \ref{sec:disorder} we numerically solve the recursive equations \eqref{eq:iter} for the general case with disordered interactions using a population dynamics algorithm. 
\section{Paramagnetic phase}
\label{sec:para}
One can easily see that $P_{\setminus\mu\hat{\mu}}(\sigma_k)=1/2~\forall~\mu,\hat\mu,k$ is always a solution of the recursive equations \eqref{eq:iter}. We refer to this solution as 
the paramagnetic phase, where spins have probability $1/2$ to be $\pm 1$. In this phase,
the distribution of overlaps \eqref{eq:pmmhatgeneral} for the clusters $\mu,\hat\mu$ simplifies to
\bea
\hspace{-2cm}P_{\mu\hat{\mu}}(m,\hat{m})=\frac{\sum_{\{\sigma_k\in\partial\mu\hat\mu\}}\int {\rm d}y_1 {\rm d}y_2  {\rm e}^{-\frac{1}{2}\by^T \bC^{-1} \by}{\rm e}^{\sqrt{\beta}(y_1m_\mu(\bsigma)+y_2m_{\hat{\mu}}(\bsigma))}\delta_{m,m_{\mu}(\bsigma)}\delta_{\hat{m},m_{\hat{\mu}}(\bsigma)}}
{\sum_{\{\tilde{\sigma}_{k\in\partial \mu\hat\mu}\}}\int {\rm d}y_1 {\rm d}y_2  {\rm e}^{-\frac{1}{2}\by^T \bC^{-1} \by}{\rm e}^{\sqrt{\beta}(y_1 m_\mu(\tilde{\bsigma})+y_2m_{\hat{\mu}}(\tilde{\bsigma}))}} \ ,
\label{eq:overpara}
\eea
thus losing the dependence on the interaction with the other clusters in the graph. This means that in the paramagnetic phase 
each cluster in the graph behaves as if it were in isolation.
Let us now analyse and simplify (\ref{eq:overpara}). We use
the delta constraints to remove the spin dependence from the exponential in the numerator, then we perform the integration over $y_1, y_2$ and  
the sum over the spins
\bea
\hspace*{-2cm}
\sum_{\{\sigma_{k\in \partial\mu}, \sigma_{\ell\in\partial\hat\mu}\}}\delta_{m,m_{\mu}(\bsigma)}\delta_{\hat{m},m_{\hat{\mu}}(\bsigma)}
&=&\frac{1}{4\pi^2}\int {\rm d}x_1 {\rm d}x_2 {\rme}^{{\rm i}x_1m+{\rm i}x_2\hat{m}}\sum_{\{\sigma_{k\in\partial\mu}\}}{\rme}^{-{\rm i}x_1\sum_{k\in \partial\mu}\xi_k^\mu\sigma_k} \sum_{\{\sigma_{\ell\in\hat\mu}\}}{\rme}^{-{\rm i}x_2\sum_{\ell\in\partial\hat{\mu}}\xi_\ell^\mu\sigma_\ell}\nonumber\\
&=&\frac{1}{4\pi^2}\int {\rm d}x_1 {\rm d}x_2 {\rme}^{{\rm i}x_1m+{\rm i}x_2\hat{m}}(\cos x_1)^{|\partial\mu|}(\cos x_2)^{|\partial\hat\mu|}
\nonumber\\
&=&\sum_{r=0}^{|\partial\mu|}{|\partial\mu|\choose r}\delta_{m,2r-|\partial\mu|}\sum_{j=0}^{|\partial\hat{\mu}|}{|\partial\hat{\mu}|\choose j}\delta_{\hat{m},2j-|\partial\hat{\mu}|}\ ,
\eea
where we used the Fourier representation of the Kronecker $\delta$'s, the parity of the cosine function to drop the $\xi$'s from its argument,
the binomial expansion of the powers of cosine and finally carried out the integrations. This leads us to the discrete distribution
\bea
P_{\mu\hat{\mu}}(m,\hat{m})=\frac{1}{\mathcal{Z_{\mu\hat\mu}}}{\rm e}^{\frac{\beta}{2(1-k^2)}(m^2+2km\hat{m}+\hat{m}^2)}
{|\partial\mu|\choose \frac{m+|\partial\mu|}{2}}{|\partial\hat{\mu}|\choose \frac{\hat m+|\partial\hat\mu|}{2}}\ ,
\label{eq:pmmhatpara}
\eea
where 
\be
\mathcal{Z_{\mu\hat\mu}}=\sum_{r=0}^{|\partial\mu|}{|\partial\mu|\choose r}\sum_{j=0}^{|\partial\hat{\mu}|}{|\partial\hat{\mu}|\choose j}{\rm e}^{\frac{\beta}{2(1-k^2)}[(2r-|\partial\mu|)^2+2k(2r-|\partial\mu|)(2j-|\partial\hat{\mu}|)+(2j-\partial\hat{\mu})^2]}
\ee
ensures normalization of (\ref{eq:pmmhatpara}) and the binomials are zero for non-integer arguments. 
\begin{figure}[htb!]
\centering
\includegraphics[width=0.48\textwidth]{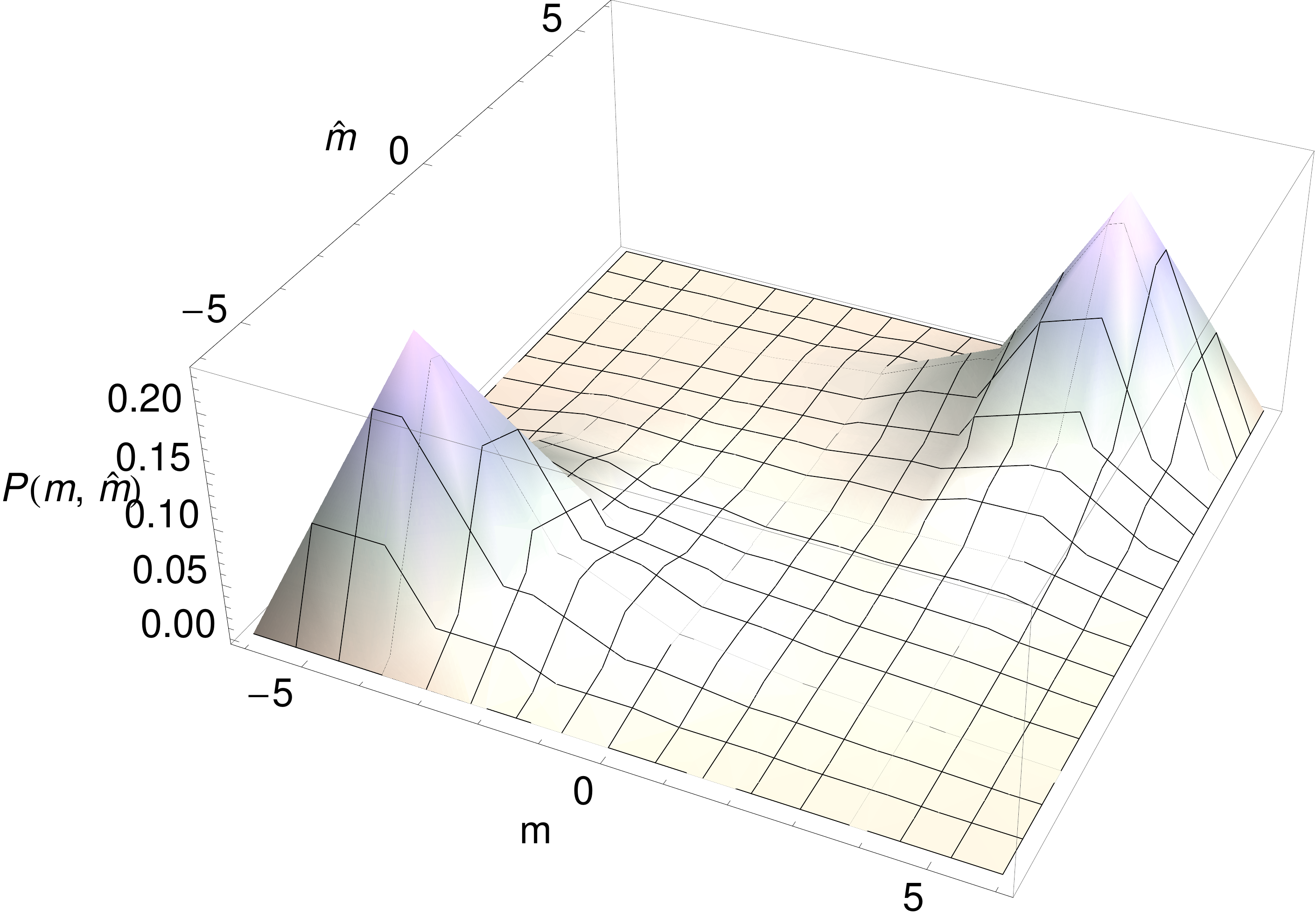}
\includegraphics[width=0.48\textwidth]{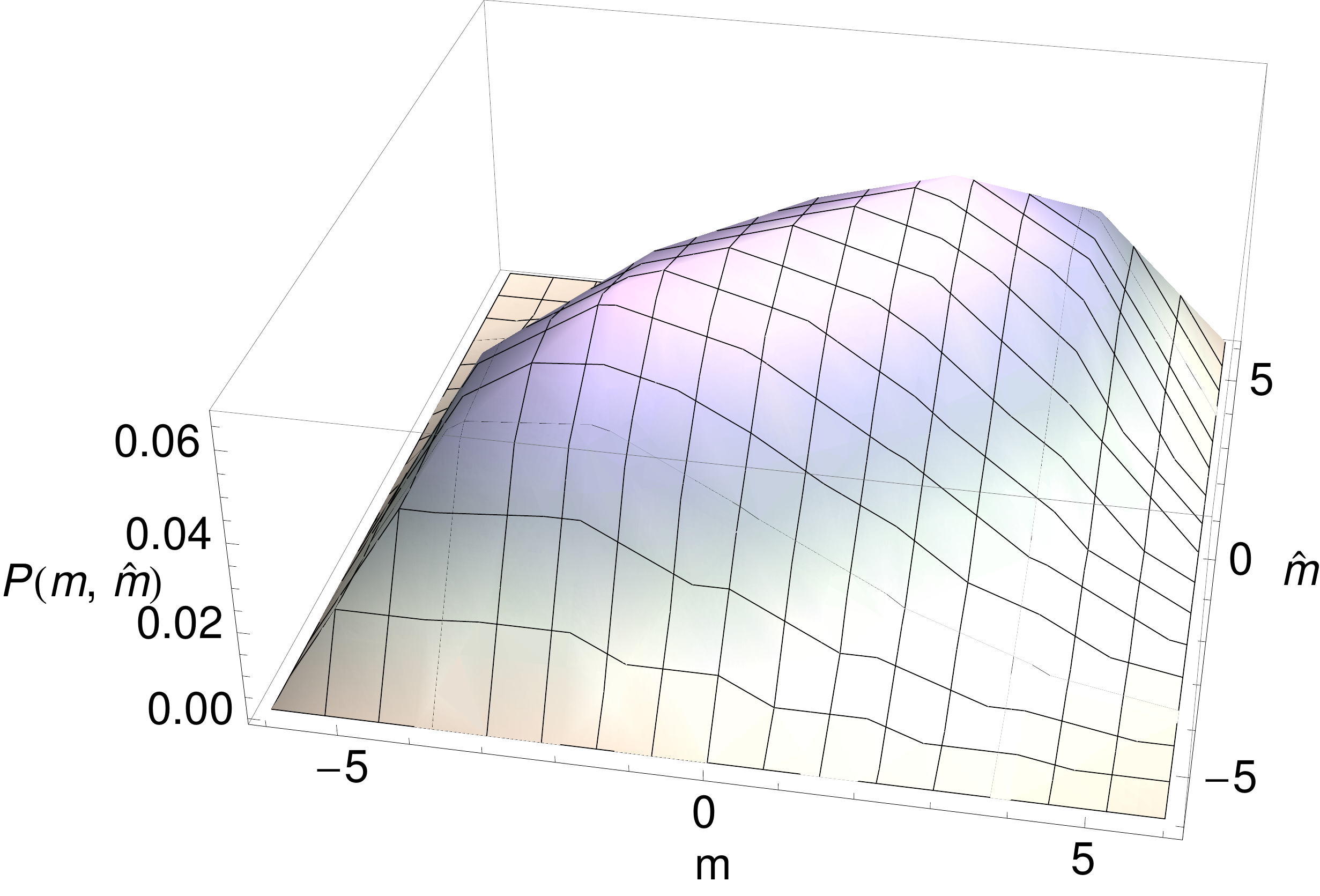}
\caption{3D plot of the joint distribution of complementary B-clones overlaps $P(m,\hat{m}|K)$ \eqref{eq:pmmhav} for $T=5$ (left) and $T=10$ (right) and $k=0.5$. The distribution is computed for a regular graph with degree $K=\hat K=4$ in the paramagnetic phase. Note that here and elsewhere the support of the distribution is discrete, $m, \hat{m}\in \{-4,-2,0,2,4\}$, 
and a continuous interpolating function has been used to guide the eye.} 
\label{fig:pmmhattsamek}
\end{figure}

We note that in the paramagnetic phase the overlap distribution, 
from which our observables of interest can be derived, does not depend on the nature of the interactions $\{\xi_i^\mu\}$, and 
$\mathcal{Z_{\mu\hat\mu}}$ depends on $\mu,\hat\mu$ only through $|\partial\mu|,|\partial\hat\mu|$.
The distribution of rotated overlap follows as
\begin{flalign}
P_{\mu\hm}(\tilde{m})=\sum_{m,\hat{m}}\frac{{\rm e}^{\frac{\beta}{2(1-k^2)}(m^2+2km\hat{m}+\hat{m}^2)}}{\mathcal{Z_{\mu\hat\mu}}}
{|\partial\mu|\choose \frac{m+|\partial\mu|}{2}}{|\partial\hat{\mu}|\choose \frac{\hat m+|\partial\hat\mu|}{2}}
\delta\left(\tilde{m}-\frac{m}{1-k^2}-\frac{k \hat{m}}{1-k^2}\right)\nonumber\ .\\
\label{eq:pmtilde2}
\end{flalign}
Averaging equations \eqref{eq:pmmhatpara} and \eqref{eq:pmtilde2} over graphs with factors degree distribution $P_q$, we obtain respectively
\bea
P(m,\hat{m}|P_q)={\rm e}^{\frac{\beta}{2(1-k^2)}(m^2+2km\hat{m}+\hat{m}^2)}
\sum_{\kappa,\tilde{\kappa}\geq 1}P_q(\kappa)P_{q}(\tilde{\kappa}) \frac{1}{\mathcal{Z_{\kappa,\tilde\kappa}}}
{\kappa\choose \frac{m+\kappa}{2}}{\tilde{\kappa}\choose \frac{\hat{m}+\tilde{\kappa}}{2}} \label{eq:pmmhav}
\eea
and
\bea
P(\tilde{m}|P_q)&=&\sum_{\kappa,\tilde{\kappa}\geq 1}P_q(\kappa)P_{q}(\tilde{\kappa})\frac{1}{\mathcal{Z_{\kappa,\tilde\kappa}}}\sum_{m,\hat{m}}
{\rm e}^{\frac{\beta}{2(1-k^2)}(m^2+2km\hat{m}+\hat{m}^2)}
\nonumber\\
&&\times{\kappa\choose \frac{m+\kappa}{2}}{\tilde{\kappa}\choose \frac{\hat{m}+\tilde{\kappa}}{2}}
\delta\left(\tilde{m}-\frac{m}{1-k^2}-\frac{k \hat{m}}{1-k^2}\right)\ .
\label{eq:pmtildeav}
\eea
In the following we consider the regular graph case, $P_q(\kappa)=\delta_{\kappa,K}$, for which we get
\begin{flalign}
P(m,\hat{m}|K)=\frac{{\rm e}^{\frac{\beta}{2(1-k^2)}(m^2+2km\hat{m}+\hat{m}^2)}}{\mathcal{Z}}
{K\choose \frac{m+K}{2}}{K\choose \frac{\hat m+K}{2}}
\label{eq:pmmhavpara}
\end{flalign}
and
\begin{flalign}
P(\tilde{m}|K)=\frac{1}{\mathcal{Z}}\sum_{m,\hat{m}}{\rm e}^{\frac{\beta}{2(1-k^2)}(m^2+2km\hat{m}+\hat{m}^2)}
{K\choose \frac{m+K}{2}}{K\choose \frac{\hat m+K}{2}}
\delta\left(\tilde{m}-\frac{m}{1-k^2}-\frac{k \hat{m}}{1-k^2}\right)\ .
\label{eq:pmtildeavpara}
\end{flalign}
\begin{figure}[htb!]
\centering
\includegraphics[trim=1.4cm 0cm 0cm 0cm,width=0.43\textwidth]{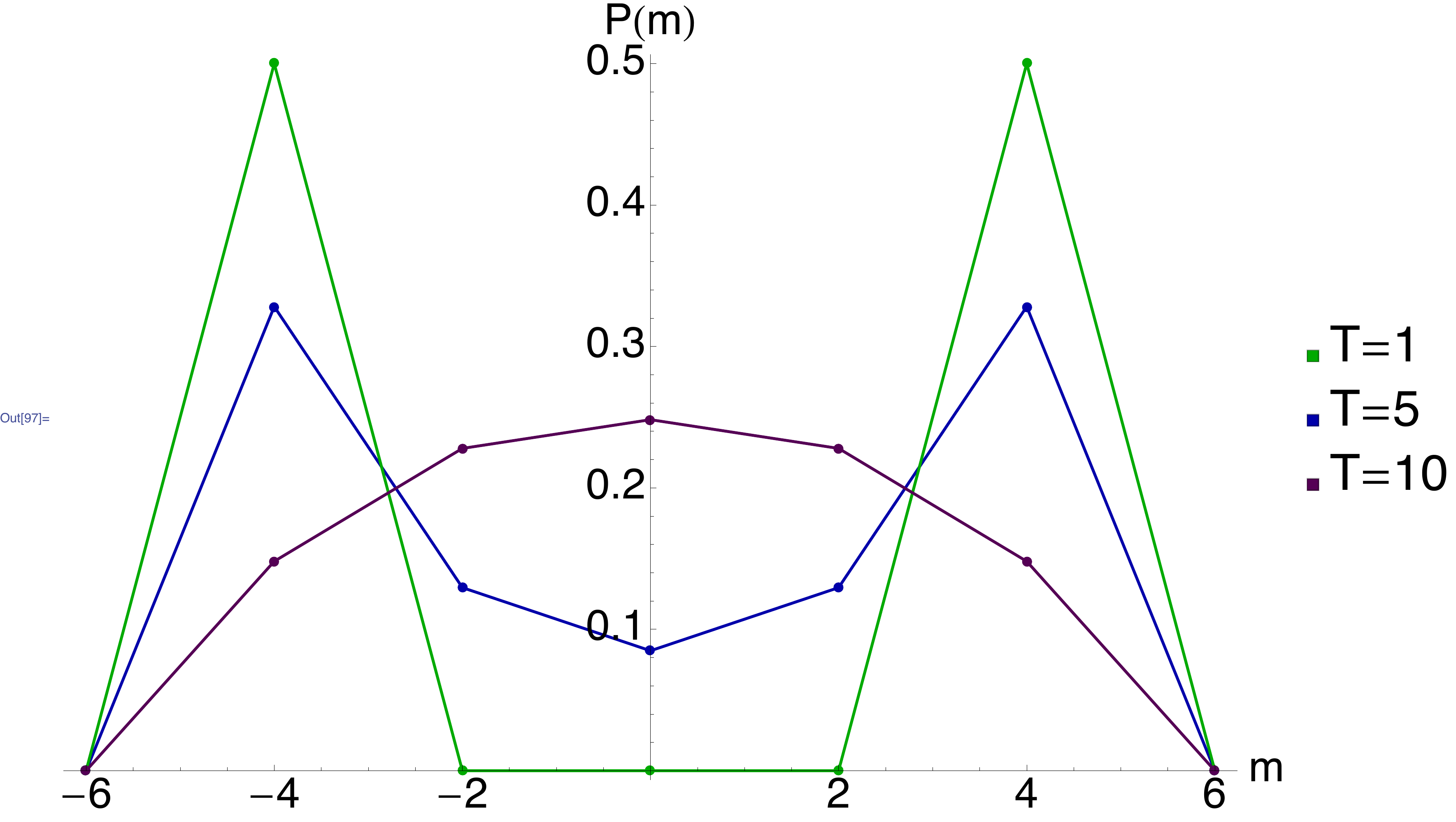}
\includegraphics[trim=1.4cm 0cm 0cm 0cm,width=0.43\textwidth]{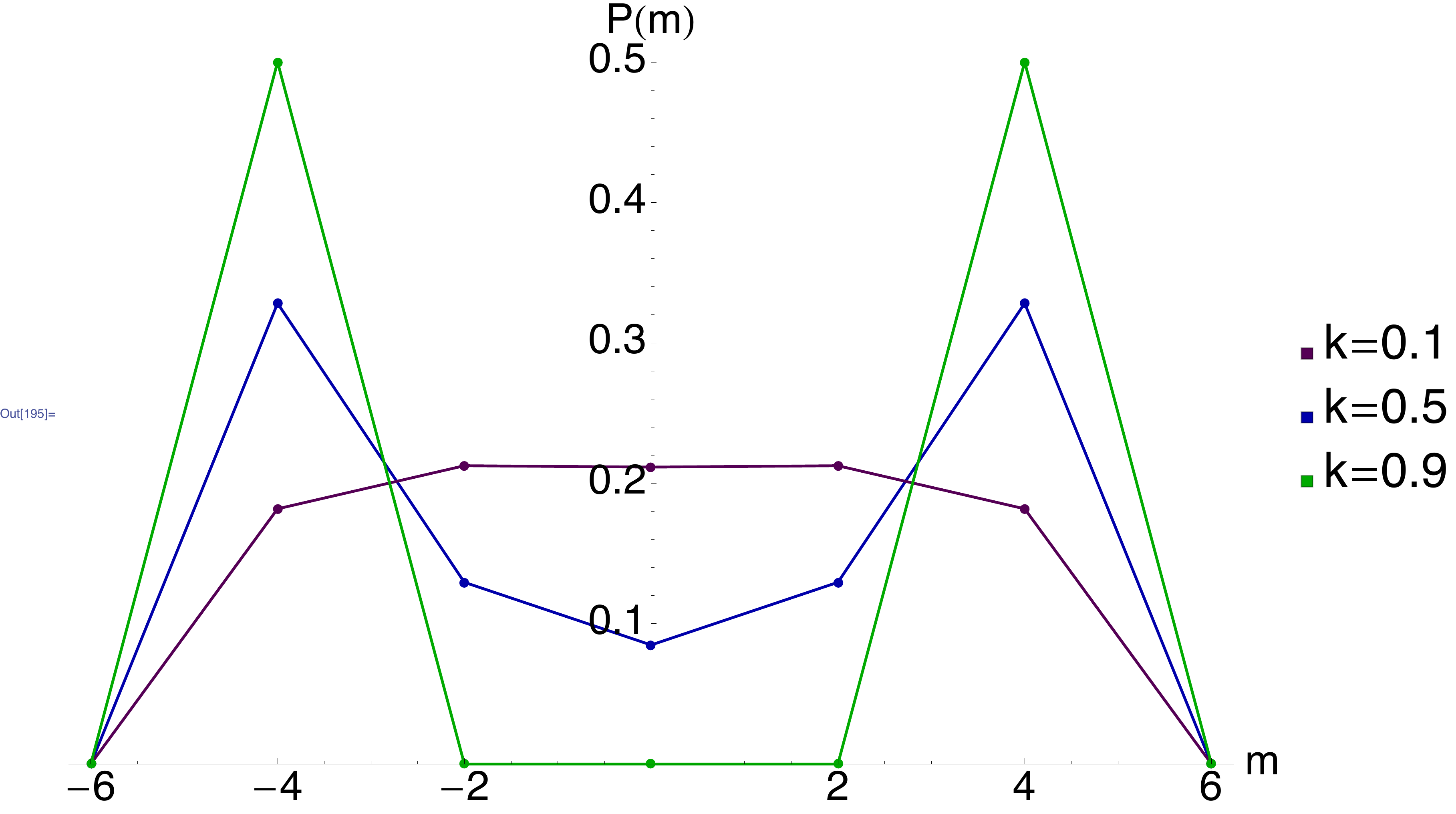}
\caption{Plot of $P(m|K)$ in the paramagnetic phase for a regular graph with degree $K=4$. Left panel: $P(m|K)$ for different temperatures $T=1,5,10$ at fixed $k=0.5$. 
Right panel: $P(m|K)$ for different B-B interaction strengths $k=0.1,0.5,0.9$ at fixed $T=5$. Note that the support of the distribution is discrete (markers), $m\in\{-4,-2,0,2,4\}$. }
\label{fig:pmtk}
\end{figure}
We plot $P(m,\hat{m}|K)$ in figure \ref{fig:pmmhattsamek} for different values of the noise level (or temperature) $T$. 
For high temperature, the distribution is peaked around zero, meaning that in the absence of an antigenic field there is no clonal expansion. 
Decreasing the temperature, the overlap distribution becomes peaked at equal values of $m=\hat m= \pm K$, meaning that pairs of interacting clones 
are likely to receive the same signals, either both excitatory or both inhibitory. The system will 
fluctuate from one peak to the other 
with a timescale $\tau$ proportional to 
the exponential of the free-energy barrier 
between $(m,\hat m)=\pm (K,K)$ and $(0,0)$, hence exponentially large in the finite size $K$ of the two clones $\tau\sim \rme^{\beta K^2/(1-k)}$. 
This means that any clone that is initially expanded, will eventually undergo a contraction in the absence of an antigen, over a typical timescale that increases 
with the size $K$ and the strength $k$ 
of the idiotypic interactions. Hence, one of the roles of 
idiotypic interactions is to prolong the short-term memory of the system.

In fig. \ref{fig:pmtk} we plot $P(m|K)$ obtained by marginalising over $\hat{m}$, for different values of the temperature $T$ and of the B-B interaction strength $k$.
The effect of increasing the B-B interactions strength $k$ is qualitatively similar to 
what we observe when decreasing the temperature. In the next section (\ref{subsec:cross}), we will discuss more in detail the dependence on the parameters 
$T$ and $k$ of the transition from a bimodal to a unimodal overlap distribution.

\begin{figure}[htb!]
\centering
\includegraphics[width=0.48\textwidth]{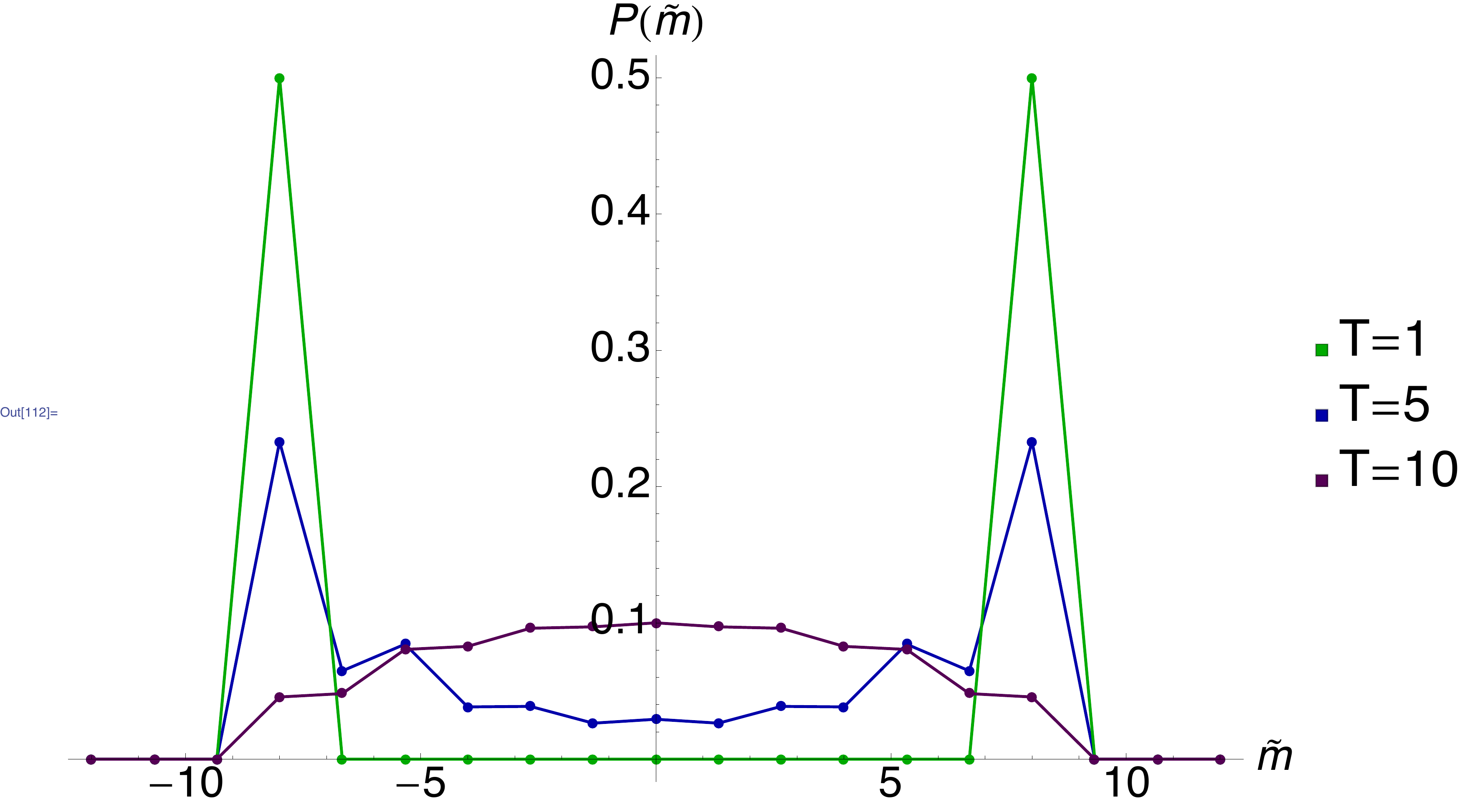}
\includegraphics[width=0.47\textwidth]{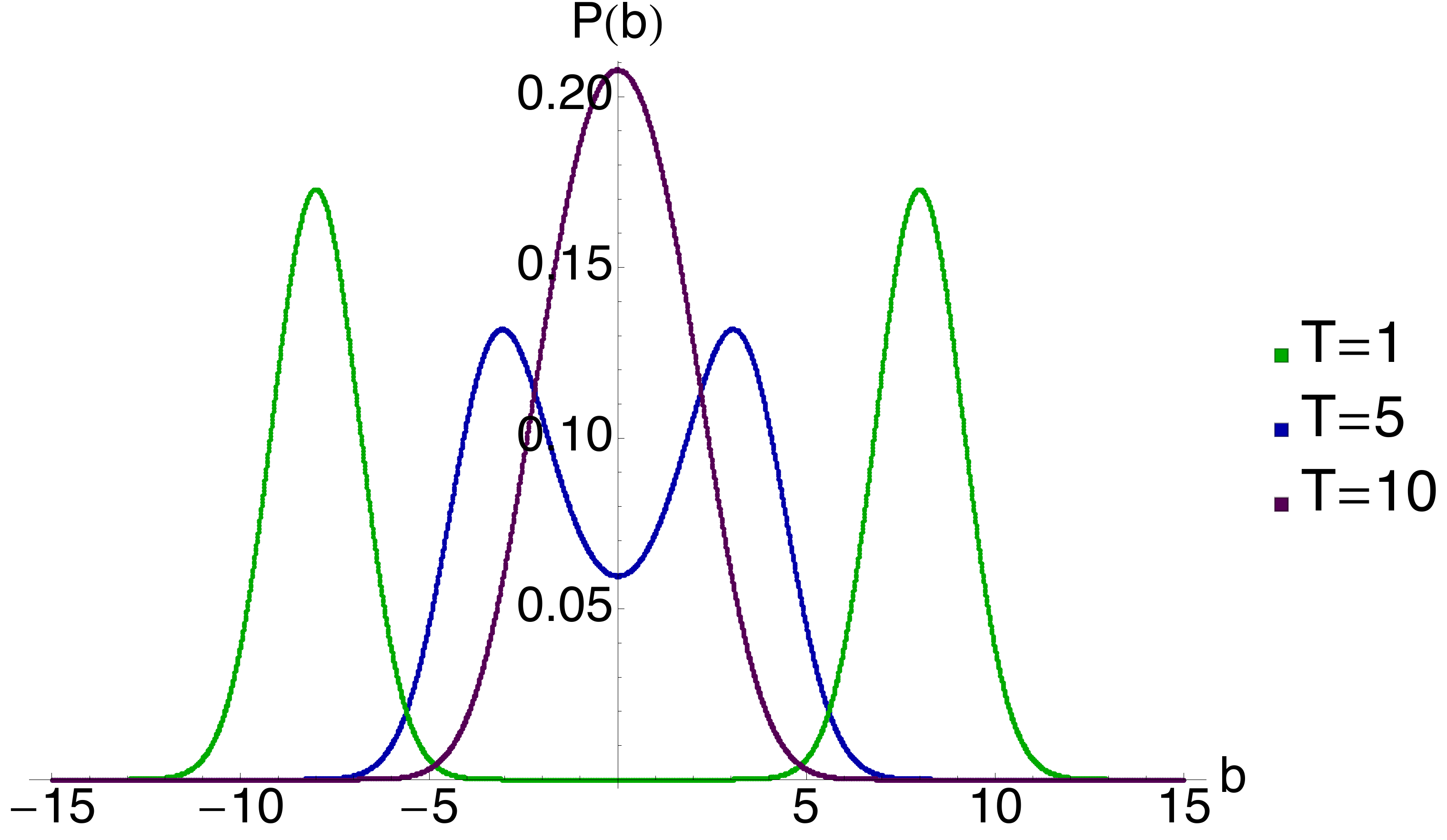}
\caption{Plot of $P(\tilde{m}|K)$ (left) and the associated B clone sizes distribution $p(b)$ (right) for different temperatures $T=1,5,10$, and fixed 
$k=0.5$ for a regular graph with degree $K=4$ in the paramagnetic phase.}
\label{fig:pmtildesamek}
\end{figure}

\begin{figure}[htb!]
\centering
\includegraphics[width=0.48\textwidth]{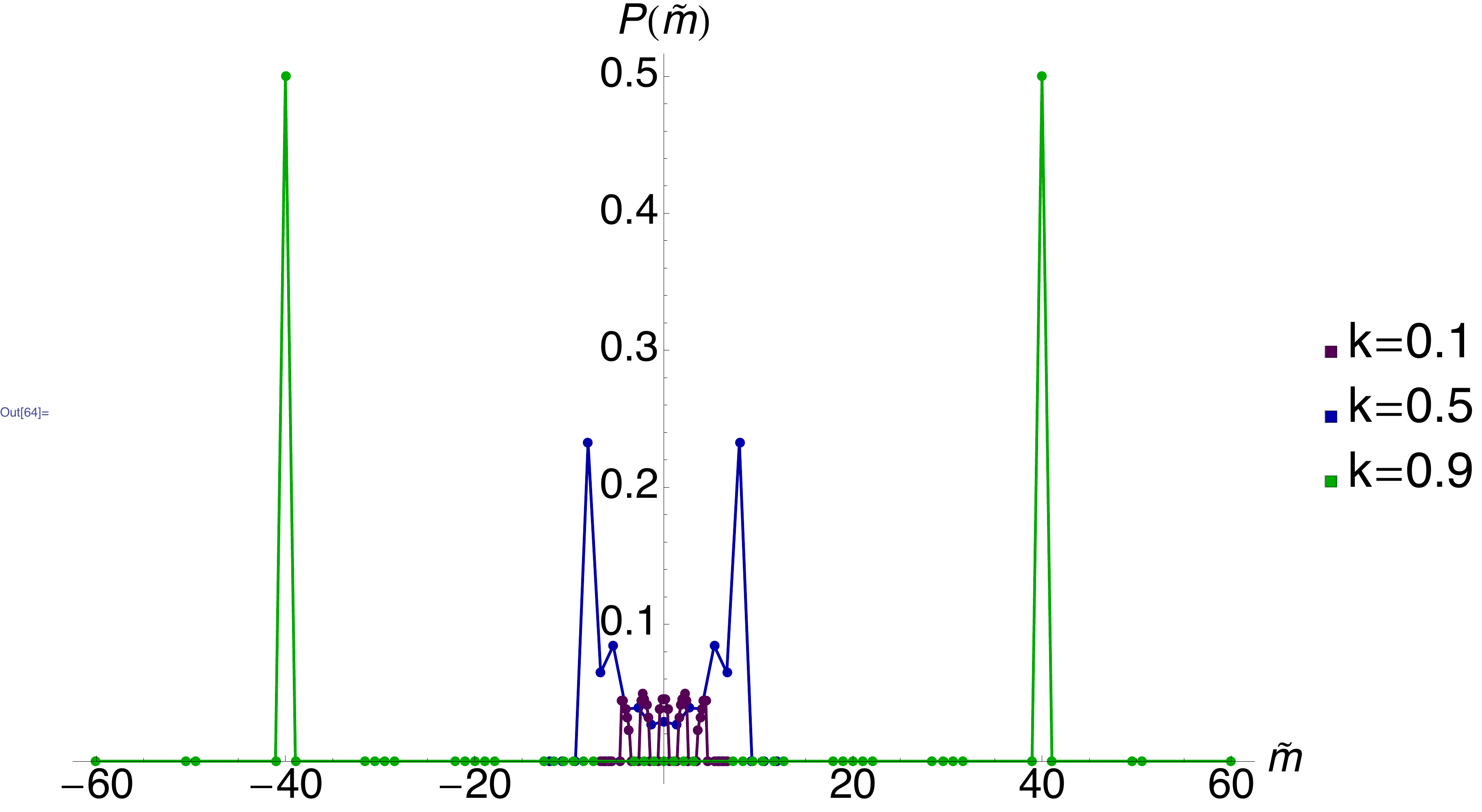}
\includegraphics[width=0.47\textwidth]{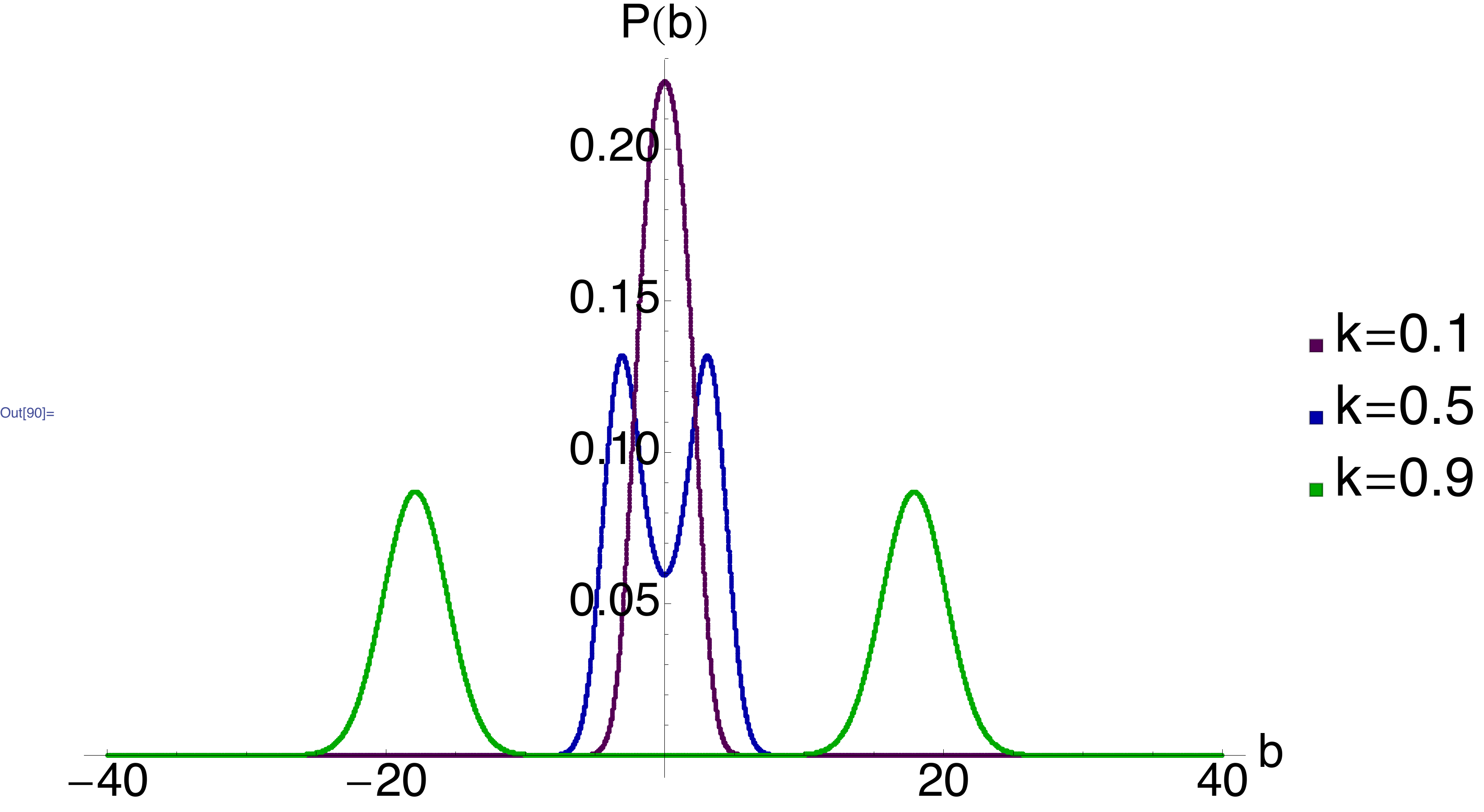}
\caption{Plot of $P(\tilde{m}|K)$ (left) and the associated B clone sizes distribution $p(b)$ (right) for different strength of idiotypic interactions 
$k=0.1,0.5,0.9$, and temperature $T=5$ for a regular graph with degree $K=4$ in the paramagnetic phase.}
\label{fig:pmtildesamet}
\end{figure}

Finally, we plot the distribution of rotated overlaps that is needed to derive the B clone size distribution. 
In fig. \ref{fig:pmtildesamek} (left panel) we show 
the behaviour of $P(\tilde{m}|K)$ when varying the temperature, while in fig. \ref{fig:pmtildesamet} (left panel) we plot it for different values of $k$. 
The behaviour is qualitatively similar to what we discussed for $P(m|K)$
and it directly affects the B clones size distribution. The latter can be computed from $P(\tilde{m}|K)$ 
by using \eqref{eq:pbi}, \eqref{eq:Pm}, \eqref{eq:Gauss_b}, and is shown 
in the right panels of the same figures. 
Fig. \ref{fig:pmtildesamek} (right panel) shows that 
at hight temperature the B clone size distribution 
is peaked in zero, meaning that no clonal expansion or 
contraction occurs in the system. 
Lowering the temperature, the distribution develops two peaks 
leading the system to oscillate, in the absence of an antigenic field, 
between "memorised" (expanded or
contracted) states, where B clones are boosted or suppressed, respectively. 
An antigenic field will force the system to remain in the expanded state. 
In fig. \ref{fig:pmtildesamet} (right panel) one sees 
the peaks shifting at larger clonal sizes when $k$ is increased, 
showing that idiotypic interactions may help
boosting the proliferation of B cells. Furthermore, when 
decreasing $k$ at fixed $T$ 
the distribution crosses over from a bimodal to a unimodal distribution 
peaked in zero, meaning that clonal expansion and contraction 
are more resilient to noise in the presence of idiotypic interactions.

\subsection{Crossover transition}\label{subsec:cross}
In this section, we study the dependence on the noise level
$T$ and on the idiotypic interaction strength
$k$ of the crossover transition of the overlap distribution 
from unimodal to bimodal (see fig. \ref{fig:pmmhattsamek} and \ref{fig:pmtk}).

We note that this crossover transition is {\em not} 
a phase transition, due to the effective finite size $K$ of the system. In contrast with what happens in 
the case $K\sim N^{1-\gamma}$, with $\gamma<1$ (and $P\sim N^\gamma$) analysed in \cite{jphysaas},   
there is no order parameter that becomes non-zero at the crossover. Before the crossover,
one has a broad distribution of the order parameter, peaked in zero, while after the crossover 
one has a typical timescale 
for the system to make a transition between one peak and the other, rather than a
full ergodicity breaking. 
We will refer to the line in the $(T, k)$ plane where the crossover 
takes place, as the {\em single cluster activation line}, as it represents the onset of 
B clonal activation in the paramagnetic phase. 
In order to derive the activation line, 
we consider the distribution of the magnetizations $m\in\{-K,\ldots,K\}$ and $\hat{m}\in\{-\hat{K},\ldots,\hat{K}\}$ in the cluster of size $K+\hat K$
\be
P(m,\hat m|K,\hat K)=\frac{1}{Z}{\rme}^{\frac{\beta}{2(1-k^2)}\left(m^2+2km\hat{m}+\hat{m}^2\right)} {K\choose \frac{K+m}{2}} {\hat{K}\choose \frac{\hat K + \hat m}{2}}\ ,
\label{eq:Pclust}
\ee
where the normalising constant 
\begin{equation}
Z=\sum_{m, \hat{m}} {\rme}^{\frac{\beta}{2(1-k^2)}\left(m^2+2km\hat{m}+\hat{m}^2\right)} {K\choose \frac{K+m}{2}} {\hat{K}\choose \frac{\hat K + \hat m}{2}}
\label{eq:Zclust}
\end{equation}
gives the partition function. From it one can derive the free energy $F=-\log Z$,
which is expected to give information on the ``critical behaviour'' of the cluster.
We note that $F$ cannot be directly computed but we can obtain bounds on this function by considering bounds for $Z$. In particular, 
we can exploit the inequalities 
\be
\frac{1}{K+1}\rme^{K\mathcal{S}\left(\frac{r}{K}\right)}
\leq{K\choose r }
\leq\rme^{K\mathcal{S}\left(\frac{r}{K}\right)},
\label{eq:inequal}
\ee
where $\mathcal{S}(p)=-p\log p - (1-p)\log(1-p)$. 
The second inequality, after the change of variables $m=\frac{m}{K}$ and $\hat{m}=\frac{\hat m}{\hat K}$, gives us the upper bound
\begin{eqnarray}
\hspace*{-2cm}Z&\leq&\sum_{m, \hat{m}} {\rme}^{\frac{\beta}{2(1-k^2)}\left(K^2 m^2+2K\hat K km\hat{m}\hat K^2+\hat{m}^2\right)   + K\mathcal{S}\left(\frac{1+m}{2}\right)   +\hat{K}\mathcal{S}\left(\frac{1+\hat{m}}{2}\right) } \leq(2K\!+\!1)\!(2\hat{K}\!+\!1) {\rme}^{\sup_{m,\hat{m}\in[-1,1]}\!\phi_1(m,\hat{m})}\label{eq:Zub},
\end{eqnarray}
where we have defined the function
\begin{eqnarray} 
\hspace*{-2cm}\phi_1(m,\hat{m})&=&  \frac{\beta}{2(1-k^2)}\left(K^2m^2+2kK\hat{K}m\hat{m}+\hat{K}^2\hat{m}^2\right) 
+ K\mathcal{S}\left(\frac{1+m}{2}\right)   +\hat{K}\mathcal{S}\left(\frac{1+\hat{m}}{2}\right) \ ,
\label{def:phi_1}
\end{eqnarray}
from which the lower bound on the free energy follows
\begin{eqnarray}
F\geq - \sup_{m,\hat{m}\in[-1,1]} \phi_1(m,\hat{m}) -
\log\left((2K\!+\!1)(2\hat{K}\!+\!1)\right)\label{eq:Flb}. 
\end{eqnarray}
Let us now consider the function $\phi_1(m,\hat{m})$.  The stationary points of this function satisfy the ``mean-field'' equations
\begin{eqnarray}
m&=&\tanh\left(  \frac{\beta}{1-k^2}\left(   Km+\hat{K}k\hat{m}\right)   \right)\ ,\label{eq:MF}\\
%
\hat{m}&=&\tanh\left(  \frac{\beta}{1-k^2}\left(   \hat{K}\hat{m}+Kk m\right)   \right)\ .
\end{eqnarray}
We note that the ``paramagnetic'' point $(m, \hat{m})=(0,0)$ is always a solution to the above equations. However, the point $(0,0)$ becomes unstable when 
the largest eigenvalue of the Jacobian $J$ of the system (\ref{eq:MF}) 
evaluated in $(0,0)$
\begin{equation}
J=\frac{\beta}{1-k^2}\left(
\begin{array}{cc}
K & \hat K k
\\
Kk & \hat K
\end{array}
\right)
\end{equation}
becomes greater than one. This happens for 
\begin{eqnarray} 
\beta&\geq&  \frac{    2(1-k^2)      }{  K + \hat{K}     +\sqrt{K^2  +2K\hat{K} \left(2k^2-1\right) +\hat{K}^2     }   }\ , \label{eq:beta-ineq}
\end{eqnarray}
which for $K=\hat{K}$ reduces to $\beta\geq\frac{1-k}{K}$. In this regime, the point $(0,0)$ ceases to be a maximum and becomes a 
saddle-point of the function $\phi_1(m,\hat{m})$.  
Interestingly, the line $T=K/(1-k)$ coincides with the critical line of the real phase transition occurring in the system with a sub-extensive number $P\sim N^{\gamma}$ of extremely diluted patterns, 
with $\gamma<1$, analysed in \cite{jstatba}, where the typical size of a cluster is $K \sim N^{1-\gamma}$.

This connection is understood by looking at the first inequality in (\ref{eq:inequal}), which gives an upper bound on $F$. 
Let us define $\alpha$ such that $\hat K =\alpha K$ and set $\beta=\frac{\tilde\beta}{K}$ then  
\begin{eqnarray}
Z&\geq&\frac{1}{(K\!+\!1)\!(\hat{K}\!+\!1)}\sum_{m, \hat{m}}\rme^{\frac{\beta}{2(1-k^2)}\left(K^2m^2+2kK\hat{K}m\hat{m}+\hat{K}^2\hat{m}^2\right) + K\mathcal{S}\left(\frac{1+m}{2}\right)   +\hat{K}\mathcal{S}\left(\frac{1+\hat{m}}{2}\right)}  \nonumber\\
&&~~~~ =\frac{1}{(K\!+\!1)\!(\hat{K}\!+\!1)}\sum_{m, \hat{m}}\rme^{K\phi_2(m,\hat{m})}\nonumber \\
 &&~~~~~=\frac{\rme^{K\sup_{m_1,m_2\in[-1,1]}\phi_2(m_1,m_2  )}}{(K\!+\!1)\!(\hat{K}\!+\!1)}\sum_{m, \hat{m}}\rme^{-K\left(\sup_{m_1,m_2\in[-1,1]}\phi_2(m_1,m_2)-\phi_2(m,\hat{m})\right)}\label{eq:Zlb},
\end{eqnarray}
 where we have defined the function
 \begin{eqnarray} 
\phi_2(m,\hat{m})&=&  \frac{\tilde\beta}{2(1-k^2)}\left(m^2+2k\alpha m\hat{m}+\alpha^2\hat{m}^2\right)  + \mathcal{S}\left(\frac{1+m}{2}\right)   +\alpha\mathcal{S}\left(\frac{1+\hat{m}}{2}\right)\label{def:phi_2}.
\end{eqnarray}
From the above lower bound on $Z$ we get the upper bound 
\begin{eqnarray}
F&\leq&   -K\sup_{m,\hat m\in[-1,1]}\phi_2(m,\hat m)  -\log \left(\sum_{m, \hat{m}}\rme^{-K\left(\sup_{m_1,m_2\in[-1,1]}\phi_2(m_1,m_2)-\phi_2(m,\hat{m})\right)}\right )\nonumber\label{eq:Fub}\\&&~~~~~~~~~~~~~~~~~~+\log\left((K\!+\!1)(\alpha K\!+\!1)\right)\ .
\end{eqnarray}
The stationary points of the function $\phi_2(m,\hat m)$ satisfy the equations 
\begin{eqnarray}
m&=&\tanh\left(  \frac{\tilde\beta}{1-k^2}\left(   m+\alpha k\hat{m}\right)   \right)\ ,\\
%
\hat{m}&=&\tanh\left(  \frac{\tilde\beta}{1-k^2}\left(   \alpha\hat{m}+k m\right)   \right)\ ,
\end{eqnarray}
which,  if we reverse the transformations $\hat K =\alpha K$ and $\beta=\frac{\tilde\beta}{K}$, gives us again the mean-field equations (\ref{eq:MF}).  We note that 
$\phi_1(m,\hat m)=K\phi_2(m,\hat m)$ when $\hat K =\alpha K$ and $\beta=\frac{\tilde\beta}{K}$, hence in the limit of large $K$ and 
assuming $\alpha<\infty$, the upper bound (\ref{eq:Fub}) and the lower bound (\ref{eq:Flb}) give for the free energy density 
$\frac{F}{K}\rightarrow -\sup_{m,\hat m\in[-1,1]}\phi_2(m,\hat m)$ as 
$K\rightarrow\infty$. This suggests that the result (\ref{eq:MF}) can be interpreted as the infinite size 
approximation of the finite size system (\ref{eq:Pclust}), which gives the connection with results derived in \cite{jstatba}. This system becomes ``critical'', 
i.e. develops bi-stability,  when the equality in (\ref{eq:beta-ineq}) is satisfied.  

Numerically, it is possible to identify a crossover temperature by 
looking at the height difference between the peaks in $0$ and $K$. 
This method is exact in the case $K=2$, where the support consists of three points, and will 
provide a lower bound for the crossover temperature for $K>2$ (an exact numerical method would consist in locating the temperature at which the peak in $0$ stops being the maximum over the support of the distribution). 

\begin{figure}[htb!]
\centering
\includegraphics[width=0.52\textwidth]{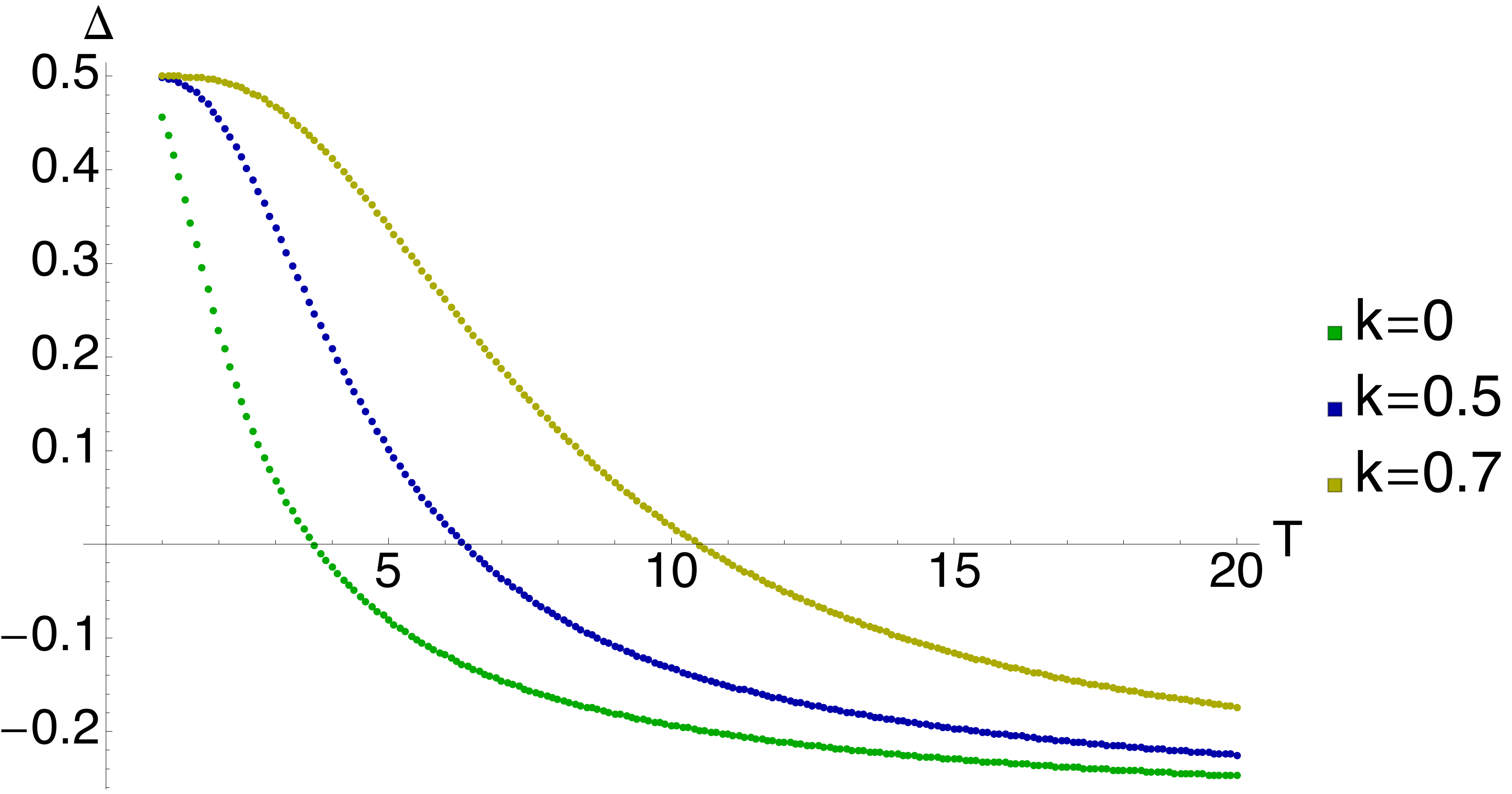}
\includegraphics[width=0.45\textwidth]{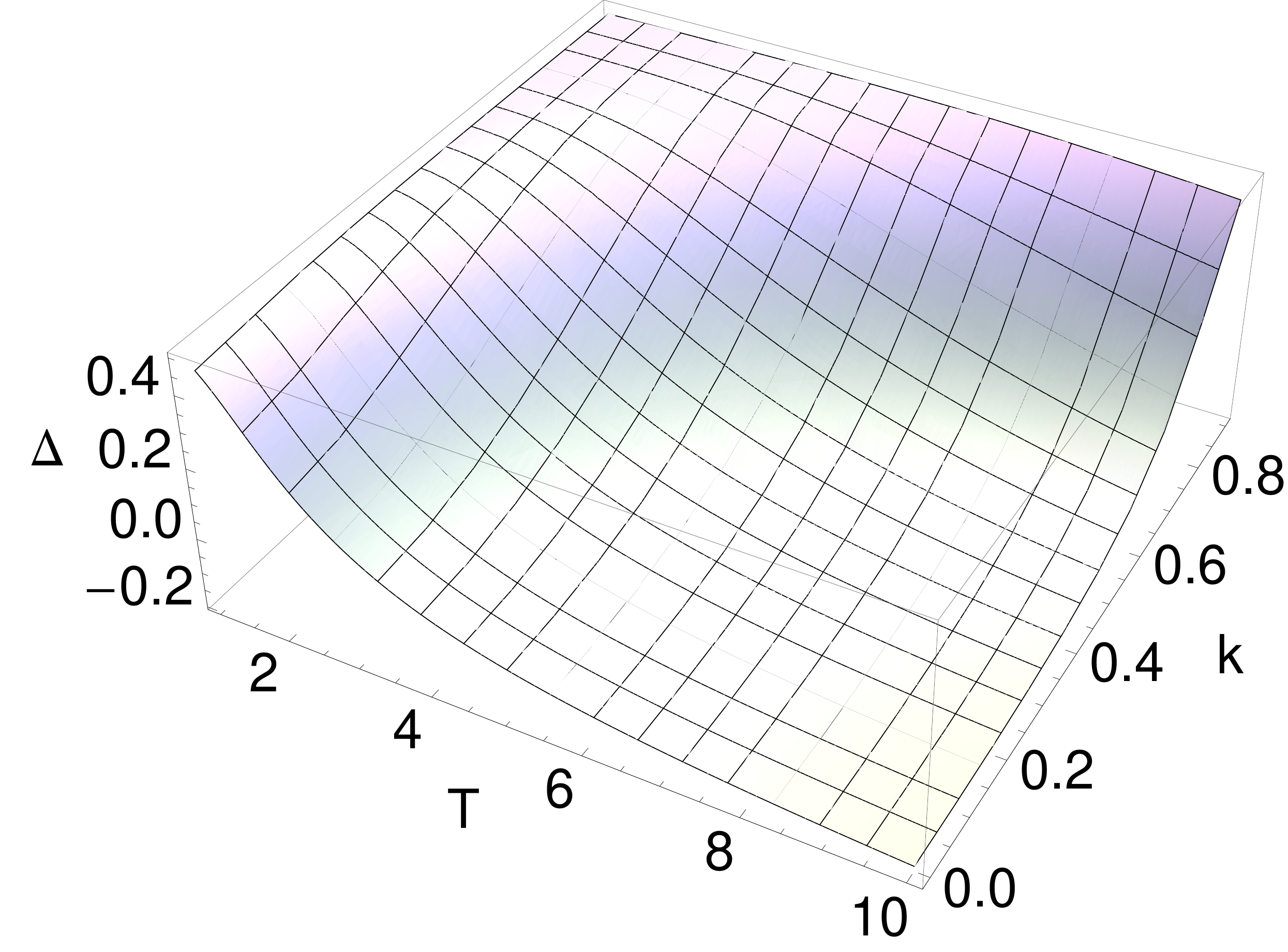}
\caption{Plot of the peak difference $\Delta=P(K|K)-P(0|K)$ as a function of the temperature (left) and 3D plot (right) with $K=3$. }
\label{fig:peakd}
\end{figure}

\begin{figure}
\centering
\includegraphics[width=0.3\textwidth]{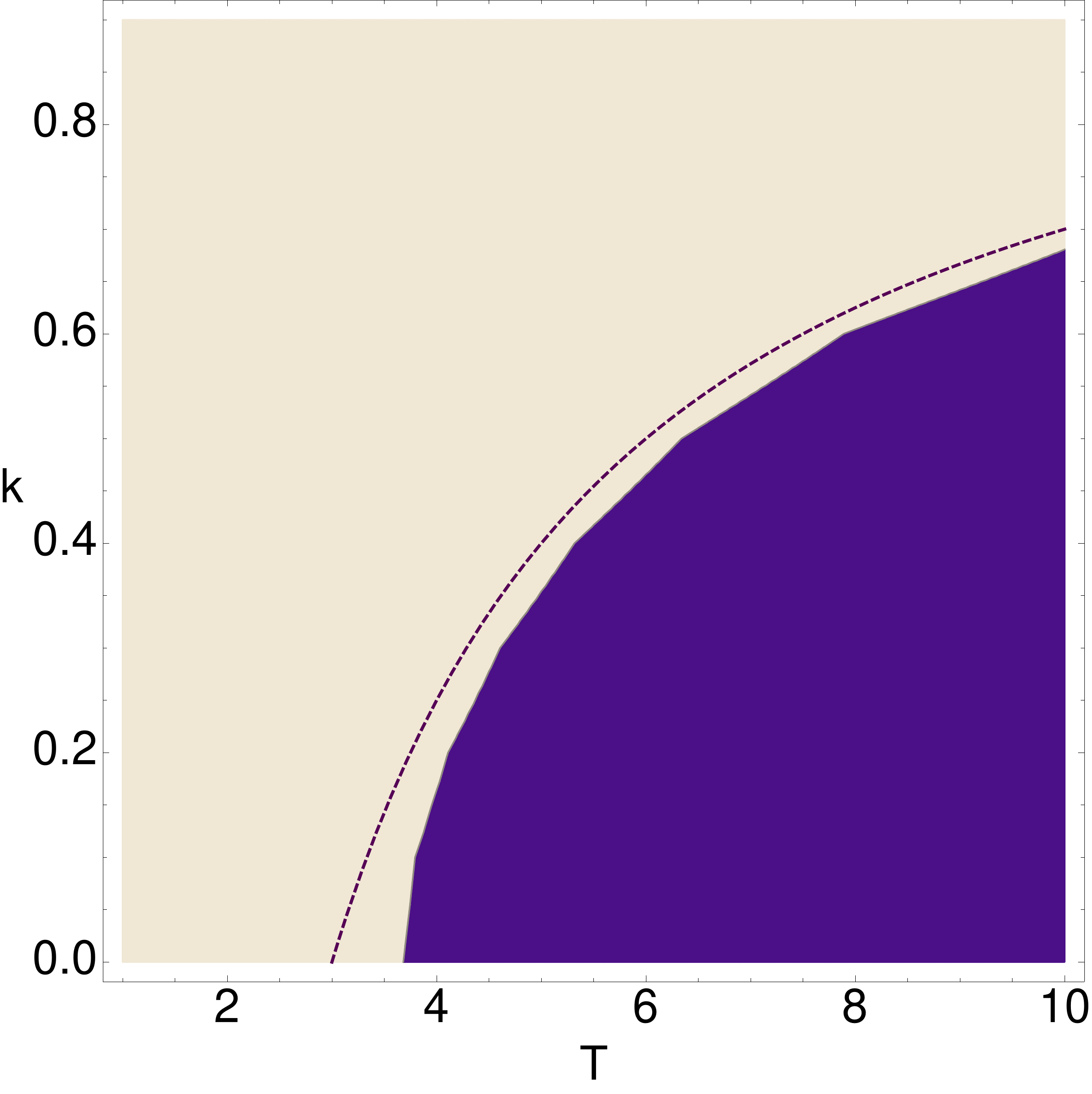}
\includegraphics[width=0.3\textwidth]{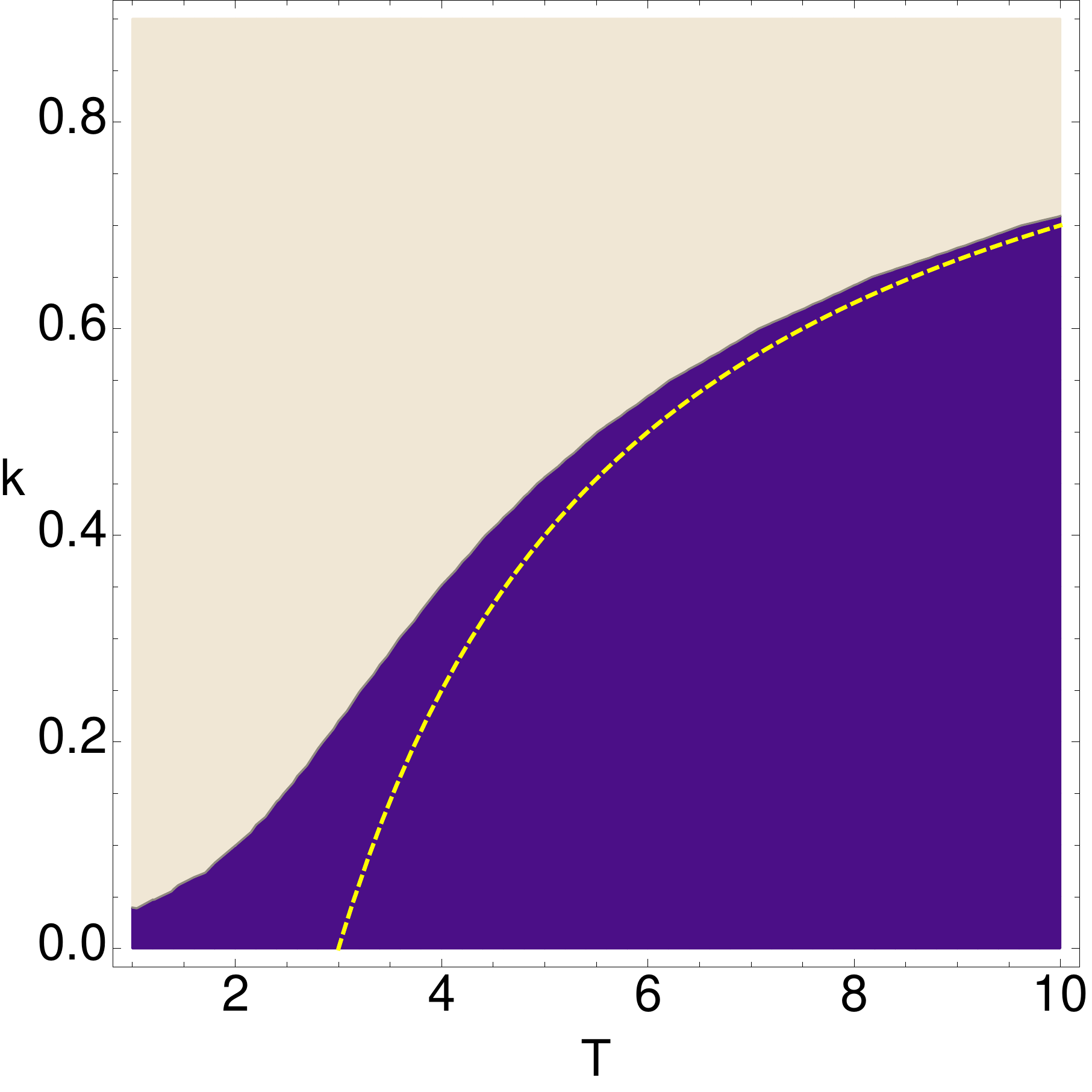}
\caption{Transition line in the plane $(T,k)$ from unimodal to bimodal distribution.  Left: Zero-contour plot of the peak difference $\Delta$ in the plane $(T,k)$ monitoring $P(m |\mathcal{P})$ for a regular graph with degree $K=3$ when coupled to a factor with Poisson distribution (average degree $K=3$). Right: Zero-contour plot of the peak difference monitoring $P(m_{tot}|K)$ with $K=3$. 
The dashed lines represent the activation line $T_a=\frac{K}{1-k}$ predicted for large $K$, consistently with results in \cite{jstatba}.}
\label{fig:peakcontour}
\end{figure}

We consider $P(m|K,P_q)=\sum_{\hat{K}\geq0}P_q(\hat{K})\sum_{\hat{m}}P(m,\hat{m}|K, P_q)$, which can be written as 
\begin{equation}
P(m|K, P_q)=\sum_{\hat{K}\geq 0}P_q(\hat{K}) \left(\frac{\sum_{j=0}^{\hat{K}}{\hat{K}\choose j} {\rm e}^{\frac{\beta}{2(1-k^2)}[m^2+(2j-\hat{K})^2+2km(2j-\hat{K})]}
{K\choose \frac{m+K}{2}}}{\sum_{r=0}^{K}{K \choose r}\sum_{s=0}^{\hat{K}}{\hat{K} \choose s}{\rm e}^{\frac{\beta}{2(1-k^2)}[(2r-K)^2+2k(2r-K)(2s-\hat{K})+(2s-\hat{K})]}}  \right),
\end{equation}
where we have fixed the size of one clone to $K$ and marginalised over the overlap $\hat{m}$ of the other clone. Its size $\hat K$ is assumed, in what 
follows, to be drawn from a
Poisson distribution $\pi_K$ with average $K$. We plot in fig. \ref{fig:peakd} the peak difference $\Delta=P(K|K, \pi_K)-P(0|K, \pi_K)$ 
as a function of $T$ and $k$ and we show 
in fig. \ref{fig:peakcontour} (left panel) the line, in the plane $(T,k)$, where $\Delta$ first becomes positive. For low values of $K$, 
this is expected to give a good approximation for the activation line where the crossover transition occurs, and is seen to be in good agreement 
with the critical line 
$T_a=K/(1-k)$, theoretically predicted for regular graph topology\footnote{This provides the main node of the Poissonian distribution used here.} 
and large size $K$ (dashed line).
Alternatively, for a regular topology with degree $K$ one can monitor the total cluster magnetization $m_t=m+\hat m$ via 
$P(m_{t}|K)=\sum_{m,\hat{m}}P(m,\hat{m}|K)\delta_{m_{t},m+\hat{m}}$, although we expect here the peak 
difference $P(2K|K)-P(0|K)$ to give a worse estimate of the transition line, given to the larger support of the distribution.
The latter can be written as
\bea
\hspace*{-2cm}
P(m_{t}|K)=\sum_{m}\sum_{\hat{m}}\frac{{\rm e}^{\frac{\beta}{2(1-k^2)}(m^2+2km\hat{m}+\hat{m}^2)}
{K\choose \frac{m+K}{2}}{K\choose \frac{\hat m+K}{2}}}
{\sum_{\ell=0}^{K}{K\choose \ell}\sum_{j=0}^{K}{K\choose j}{\rm e}^{\frac{\beta}{2(1-k^2)}((2\ell-K)^2+2k(2\ell-K)(2j-K)+(2j-K)^2}}\delta_{m_{tot},m+\hat{m}}
\eea
and the line where the peak difference first becomes positive is plotted in fig. \ref{fig:peakcontour} (right panel). This indeed provides 
a lower bound on the activation temperature $T_a=K/(1-k)$ shown in the same plot, for guidance, as a dashed line.

\section{Ferromagnetic interactions}\label{sec:ferro}
We can study cross-talks effects between clusters in the case of ferromagnetic interactions $\xi_i^\mu=\xi_i^{\hat{\mu}}=1$, $\forall~i,\mu,\hat{\mu}$, 
where we can progress analytically even away from the paramagnetic phase. To this purpose we consider spins interacting on a random regular factor-graph with 
$|\partial i|=L$ and  $|\partial \mu|=|\partial \hat{\mu}|=K$. In the thermodynamic limit $N\to\infty$, the graph is tree-like 
and due to the ferromagnetic nature of the interactions all factors (\ref{eq:fmu}) are equivalent. 
It follows that all the cavity distributions are equivalent $P_{\setminus\mu\hat{\mu}}(\sigma)=P^c(\sigma)$, $\forall~\mu,\hat\mu$. 
Parametrizing them as $P^c(\sigma_i)\propto{\rm e}^{\beta\phi\sigma_i}$ and denoting 
$\mathcal{D}{\bf y}= \frac{{\rm d}y_1{\rm d}y_2}{2\pi}{\rme}^{\frac{-1}{2}{{\bf y}^T {\bf C}^{-1} {\bf y}}}$,
we can write the recursive equation \eqref{eq:iter} as
\begin{flalign}
P^c(\sigma)=\frac{1}{\mathcal{Z}}\left[\int_{-\infty}^{+\infty}\mathcal{D}{\bf y} \sum_{\{\sigma_k\},\{\tau_k\}}{\rm e}^{\sqrt{\beta}\sigma y_1+y_1\left(\sqrt{\beta}\sum_{k=1}^{K-1}\sigma_k\right)+y_2\left(\sqrt{\beta}\sum_{k=1}^{K}\tau_k\right)+\beta\phi(\sum_{k=1}^{K-1}\sigma_k+\sum_{k=1}^{K}\tau_k)}\right]^{L-1} \ ,
\end{flalign}
where we denoted with $\sigma$ the spins attached to factor $\mu$, with $\tau$ those attached to 
$\hat\mu$, and assumed that, due to sparsity 
of interactions, $\sigma_i$ is attached to factor $\mu$ only. 

\begin{figure}[htb!]
\centering
\includegraphics[width=0.41\textwidth]{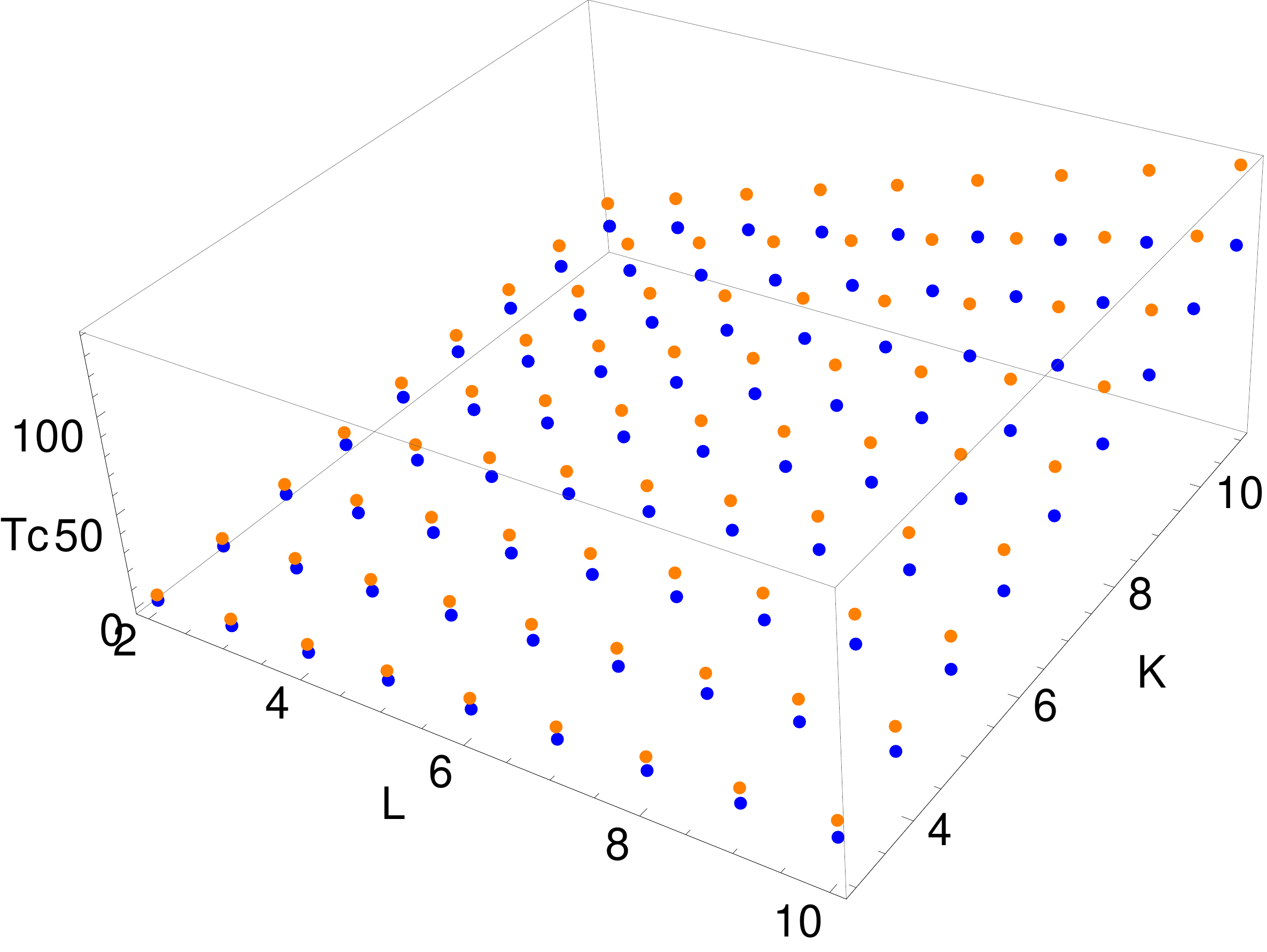}
\hfill
\includegraphics[trim= 0cm 5.5cm 0.2cm 0.6cm,clip, width=0.47\textwidth]{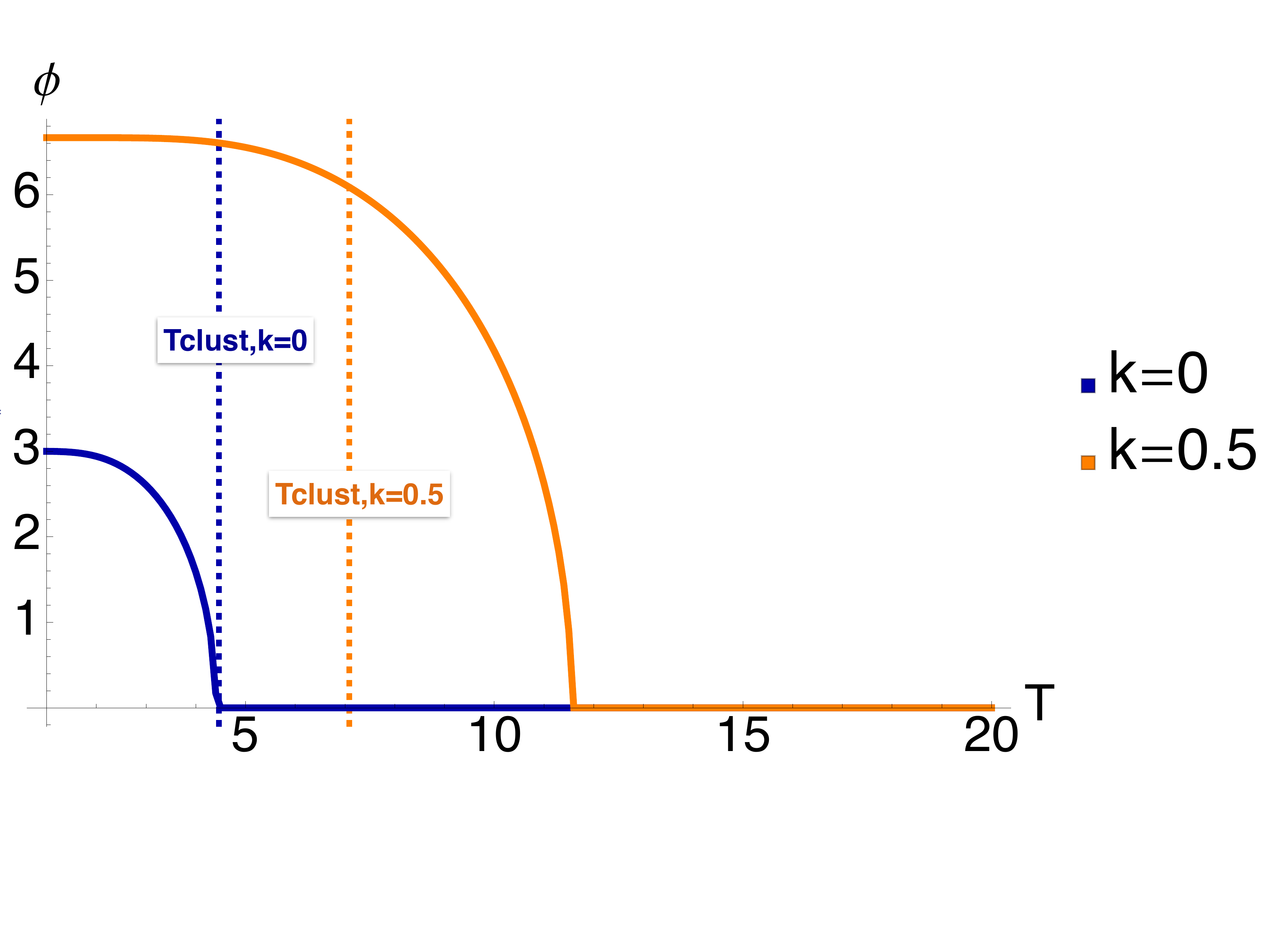}
\caption{Left: Critical temperature $T_c$ for the transition $\phi=0\to\phi\neq0$  for a regular graph with degrees $K,L$ 
for the values of the B-B interaction strength $k=0.5$ (orange) and $k=0$ (blue). Right: Plot of $\phi$ as a function of the temperature $T$ for $K=4$ and $L=2$. 
Dashed lines represent the single cluster activation temperature 
(sec. \ref{sec:para}).}
\label{fig:ferrotrans}
\end{figure}

For $\phi=0$ we retrieve the paramagnetic phase analysed in sec. \ref{sec:para} with $P^c(\sigma_i)=1/2~\forall~\sigma_i=\pm 1$, 
while for $\phi\neq 0$ each cluster will receive a signal from the others, acting as a field.
Summing over $\{\sigma_k\},\left\{\tau_k\right\}$ we obtain
\begin{flalign}
P^c(\sigma)=\frac{\left[\int_{-\infty}^{+\infty}\frac{{\rm d}{\bf y}}{2\pi}{\rm e}^{\frac{-{\bf y}^T {\bf C}^{-1}{\bf y}}{2}}{\rm e}^{\sqrt{\beta}\sigma y_1}\left[\cosh(\sqrt{\beta}y_1+\beta\phi)\right]^{K-1}\left[\cosh(\sqrt{\beta}y_2+\beta\phi)\right]^{K}\right]^{L-1}}{\sum_{\tilde{\sigma}}\left[\int_{-\infty}^{+\infty}\frac{{\rm d}{\bf y}}{2\pi}{\rm e}^{\frac{-{\bf y}^T {\bf C}^{-1}{\bf y}}{2}}{\rm e}^{\sqrt{\beta}\tilde{\sigma} y_1}\left[\cosh(\sqrt{\beta}y_1+\beta\phi)\right]^{K-1}\left[\cosh(\sqrt{\beta}y_2+\beta\phi)\right]^{K}\right]^{L-1}} 
\label{eq:ferropc}
\end{flalign}
and using $P^c(1)=a{\rm e}^{\beta\phi}$ and $P^c(-1)= a{\rm e}^{-\beta\phi}$, we get
\begin{eqnarray}
\phi=\frac{1}{2\beta}\log\frac{P^c(1)}{P^c(-1)}\ .
\end{eqnarray}
This leads to the following self-consistency equation for $\phi$
\begin{flalign}
\phi=\frac{L-1}{2\beta}\log\left(\frac{\int_{-\infty}^{+\infty}\frac{{\rm d}{\bf y}}{2\pi}{\rm e}^{\frac{-{\bf y}^T {\bf C}^{-1}{\bf y}}{2}}{\rm e}^{\sqrt{\beta} y_1}\left[\cosh(\sqrt{\beta}y_1+\beta\phi)\right]^{K-1}\left[\cosh(\sqrt{\beta}y_2+\beta\phi)\right]^{K}}{\int_{-\infty}^{+\infty}\frac{{\rm d}{\bf y}}{2\pi}{\rm e}^{\frac{-{\bf y}^T {\bf C}^{-1}{\bf y}}{2}}{\rm e}^{-\sqrt{\beta} y_1}\left[\cosh(\sqrt{\beta}y_1+\beta\phi)\right]^{K-1}\left[\cosh(\sqrt{\beta}y_2+\beta\phi)\right]^{K}} \right) \ .
\label{eq:phiferro1}
\end{flalign}
Clearly $\phi=0$ is always a solution, but we expect it to become unstable at low temperature. With simple manipulations we can write \eqref{eq:phiferro1} in the following form
\begin{flalign}
\phi=\frac{L-1}{2}\log\left(\frac{\sum_{\ell=0}^{K-1} {K-1 \choose \ell}\sum_{p=0}^K  {K \choose p} \int_{-\infty}^{+\infty}\frac{{\rm d}{\bf y}}{2\pi}{\rm e}^{\frac{-{\bf y}^T {\bf C}^{-1}{\bf y}}{2}}{\rm e}^{{\bf J}^T {\bf y} +\beta\phi(2\ell+2p-2K+1)}}{\sum_{\tilde{\ell}=0}^{K-1} {K-1 \choose \tilde{\ell}}\sum_{\tilde{p}=0}^K  {K \choose \tilde{p}} \int_{-\infty}^{+\infty}\frac{{\rm d}{\bf \tilde{y}}}{2\pi}{\rm e}^{\frac{-{\bf \tilde{y}}^T {\bf C}^{-1}{\bf \tilde{y}}}{2}}{\rm e}^{{\bf M}^T{\bf \tilde{y}}+\beta\phi(2\tilde{\ell} + 2\tilde{p}-2K+1)}}\right)\ ,\nonumber\\
\end{flalign}
where ${\bf J}^T = (\sqrt{\beta}(2+2\ell-K),\sqrt{\beta}(2p-K))$ and  
${\bf M}^T = (\sqrt{\beta}(2\tilde{\ell}-K),\sqrt{\beta}(2\tilde{p}-K))$.
Solving the integral in eq. \eqref{eq:integral1} and using the matrix ${\bf C}$ defined in \eqref{eq:Cmatrix1} we obtain
\begin{flalign} \phi=\frac{L-1}{2}\log\left(\frac{\sum_{\ell=0}^{K-1} {K-1 \choose \ell} \sum_{p=0}^K  {K \choose p}{\rm e}^{\frac{\beta((2\ell-K+2)^2+ (2p-K)^2+2k(2\ell-K+2)(2p-K))}{1-k^2}+\beta\phi(2\ell+2p-2K+1)}}{\sum_{\tilde{\ell}=0}^{K-1} {K-1 \choose \tilde{\ell}}\sum_{\tilde{p}=0}^K  {K \choose \tilde{p}}{\rm e}^{\frac{\beta((2\tilde{\ell}-K)^2+(2\tilde{p}-K)^2+2k (2\tilde{\ell}-K)(2\tilde{p}-K))}{1-k^2}+\beta\phi(2\tilde{\ell} + 2\tilde{p}-2K+1)}}\right)\ .
\label{eq:phic}
\end{flalign}
Via bifurcation analysis we can determine the critical temperature $T_c$ 
at which cavity fields bifurcate to a non-zero value (we will provide full details 
for general types of interactions in sec. \ref{sec:disorder}). 
In fig. \ref{fig:ferrotrans} (left),
we plot the critical temperature $T_c$ 
as a function of the vertex and factor degree
for two different values of the B-B interaction 
strength $k$: increasing $k$ widens the region where the cavity fields are 
non-zero, the so-called {\em interference region}. 
Given that in the ferromagnetic case the cavity fields are homogeneous, 
the interference between cytokine patterns is constructive and does not 
disrupt the system's parallel retrieval of information. 
For a comparison between the critical temperature $T_c$ and the single 
cluster activation temperature $T_a$, see the right panel of fig. 
\ref{fig:ferrotrans}, where we plot the cavity field versus the temperature 
for fixed values of $K=4$ and $L=2$.

\begin{figure}
\centering
\includegraphics[width=0.45\textwidth]{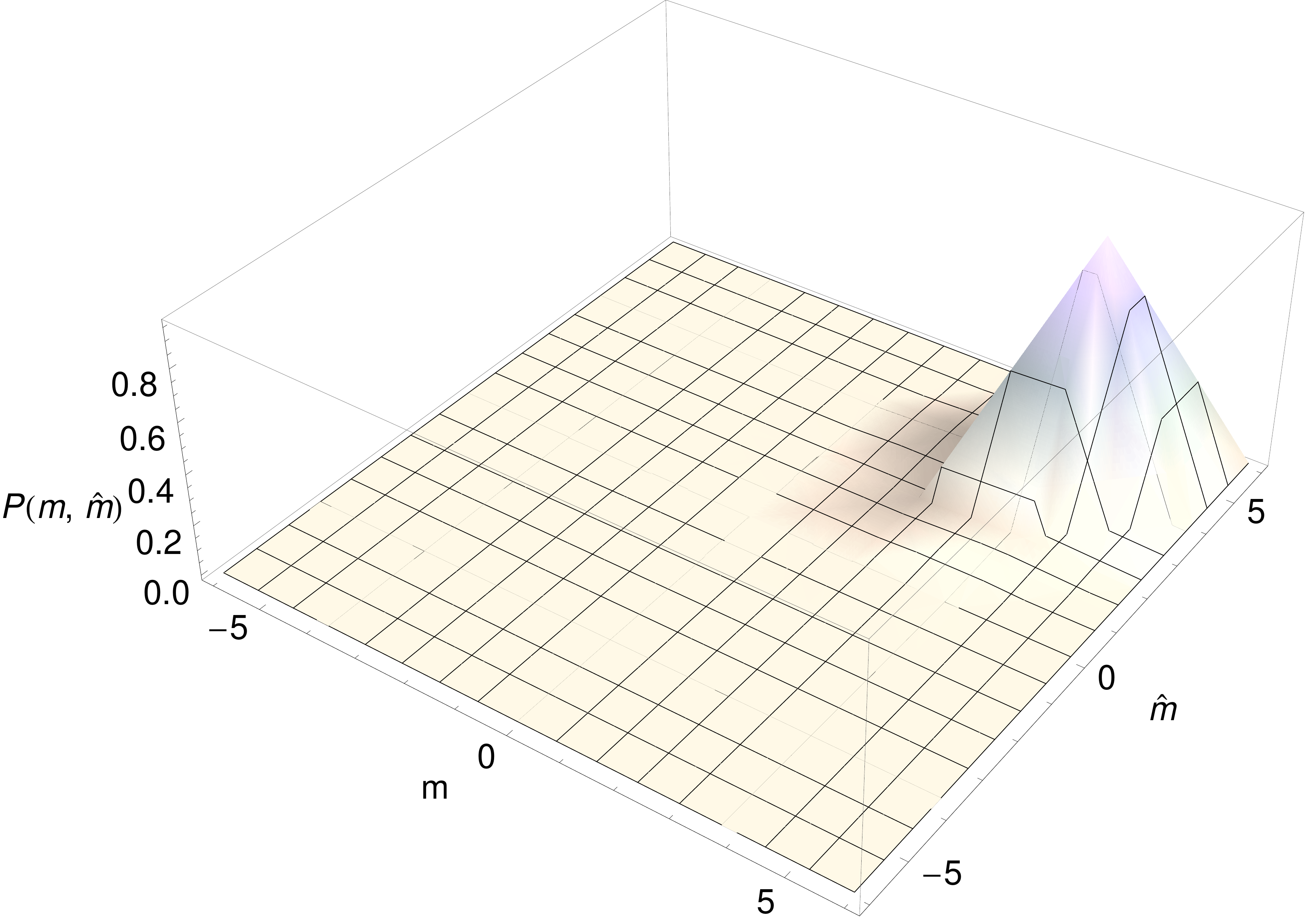}
\includegraphics[width=0.45\textwidth]{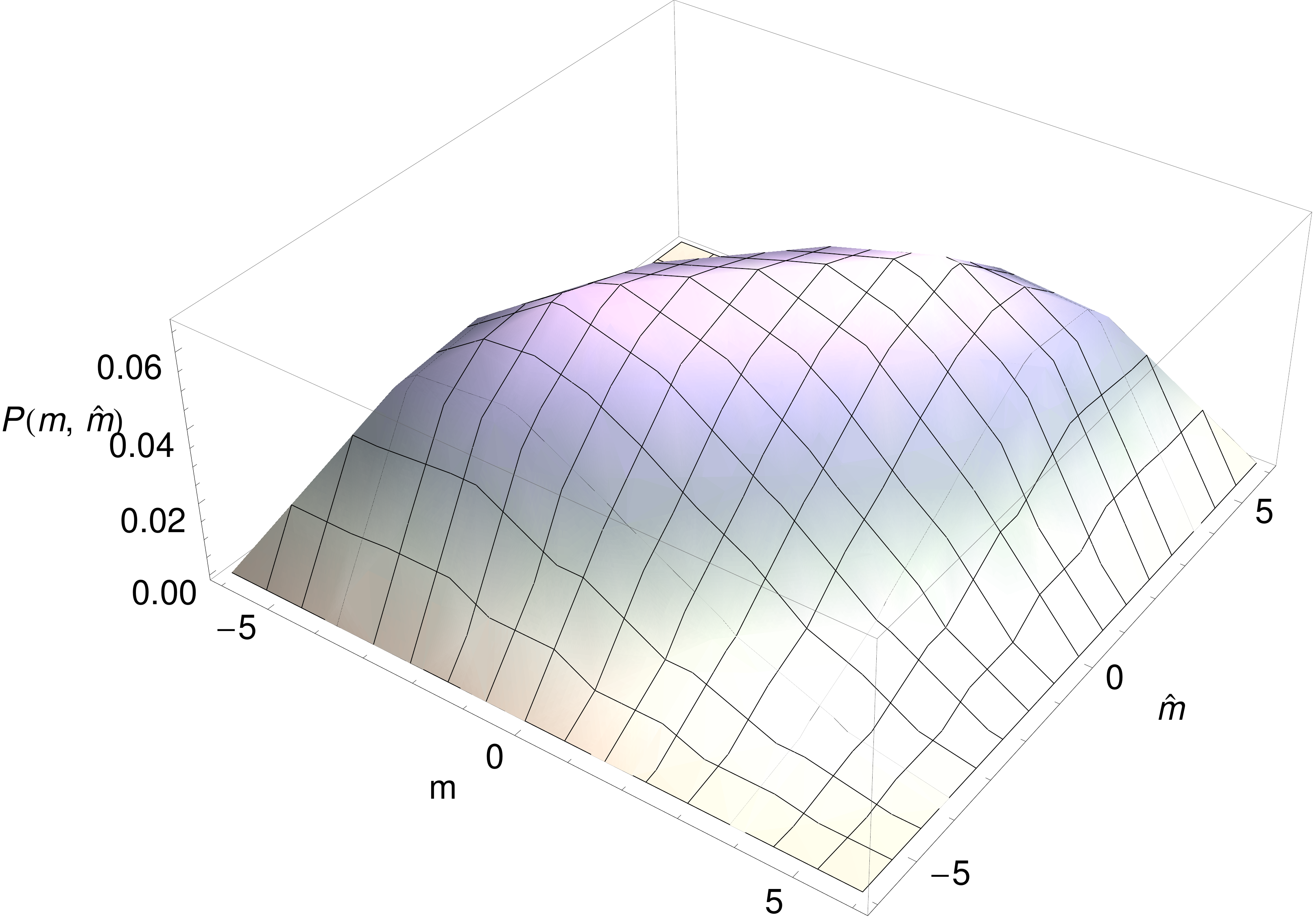}
\caption{Plot of $P(m, \hat{m}|K)$ \eqref{eq:pmmhferro} for $T=5$ (left panel) and $T=15$ (right panel) fixing $k=0.5$, for a regular graph with  $K=4,L=2$ with ferromagnetic interactions.}
\label{fig:pmmhferro}
\end{figure}

As in the paramagnetic case, we compute the overlap distribution defined in 
\eqref{eq:pmmhat1} for regular factor graphs with degree $K$ (generalizations 
to non-regular topologies are straightforward) from
\begin{flalign}
 P(m,\hat{m}|K)=\frac{{\rm e}^{\frac{\beta}{2(1-k^2)}(m^2+\hat{m}^2+2k m \hat{m})+\beta \phi (m+\hat{m})}\sum_{\ell=0}^{K}{K \choose \ell}\sum_{p=0}^K {K \choose p}\delta_{m,2\ell-K}\delta_{\hat{m},2 p-K}}{\sum_{\tilde{\ell}=0}^{K} {K \choose \tilde{\ell}}\sum_{\tilde{p}=0}^K  {K \choose \tilde{p}}{\rm e}^{\frac{\beta((2\tilde{\ell}-K)^2+(2\tilde{p}-K)^2+2k(2\tilde{\ell}-K)(2\tilde{p}-K)^2)}{2(1-k^2)}+\beta\phi(2\tilde{\ell} + 2\tilde{p}-2K+1)}}\ ,
\label{eq:pmmhferro}
\end{flalign}
where $\phi$ is self-consistently obtained from eq. \eqref{eq:phic}. In fig. \ref{fig:pmmhferro} we plot the distribution \eqref{eq:pmmhferro} 
for different values of the temperature and fixed degrees $K=4$, $L=2$. For $T>T_c$ where $\phi=0$ (see fig. \ref{fig:ferrotrans}, right), 
the overlap distribution \eqref{eq:pmmhferro} reduces to (\ref{eq:pmmhavpara}), 
which describes the behaviour of a single cluster in the paramagnetic phase and does not depend on the nature (e.g. ferromagnetic or disordered) of 
the interactions. 
Since $T_c$ is larger than the single cluster activation temperature (see fig. \ref{fig:ferrotrans}, right), the system is found in the high-temperature regime of the paramagnetic phase and displays an overlap distribution peaked in zero. Lowering the temperature, 
$\phi$ becomes non-zero, hence each clone receives signals from other clones in the system. Therefore, we expect that the predictions 
obtained in the paramagnetic phase become inaccurate in this regime. Numerical evaluation of \eqref{eq:pmmhferro} 
shows that at low temperature one has again a crossover of the overlap distribution 
to an activated regime, where it displays a single peak at either positive or 
negative values of the magnetization, depending on the initial conditions (see fig. \ref{fig:pmmhferro}, left). Indeed, 
clonal cross-talk manifests itself, in this regime, as a field that pins the system in one peak or the other, and the system develops a non-zero global magnetization (as for ferromagnetic interactions the field is homogeneous).
\begin{figure}
\centering
\includegraphics[width=0.45\textwidth]{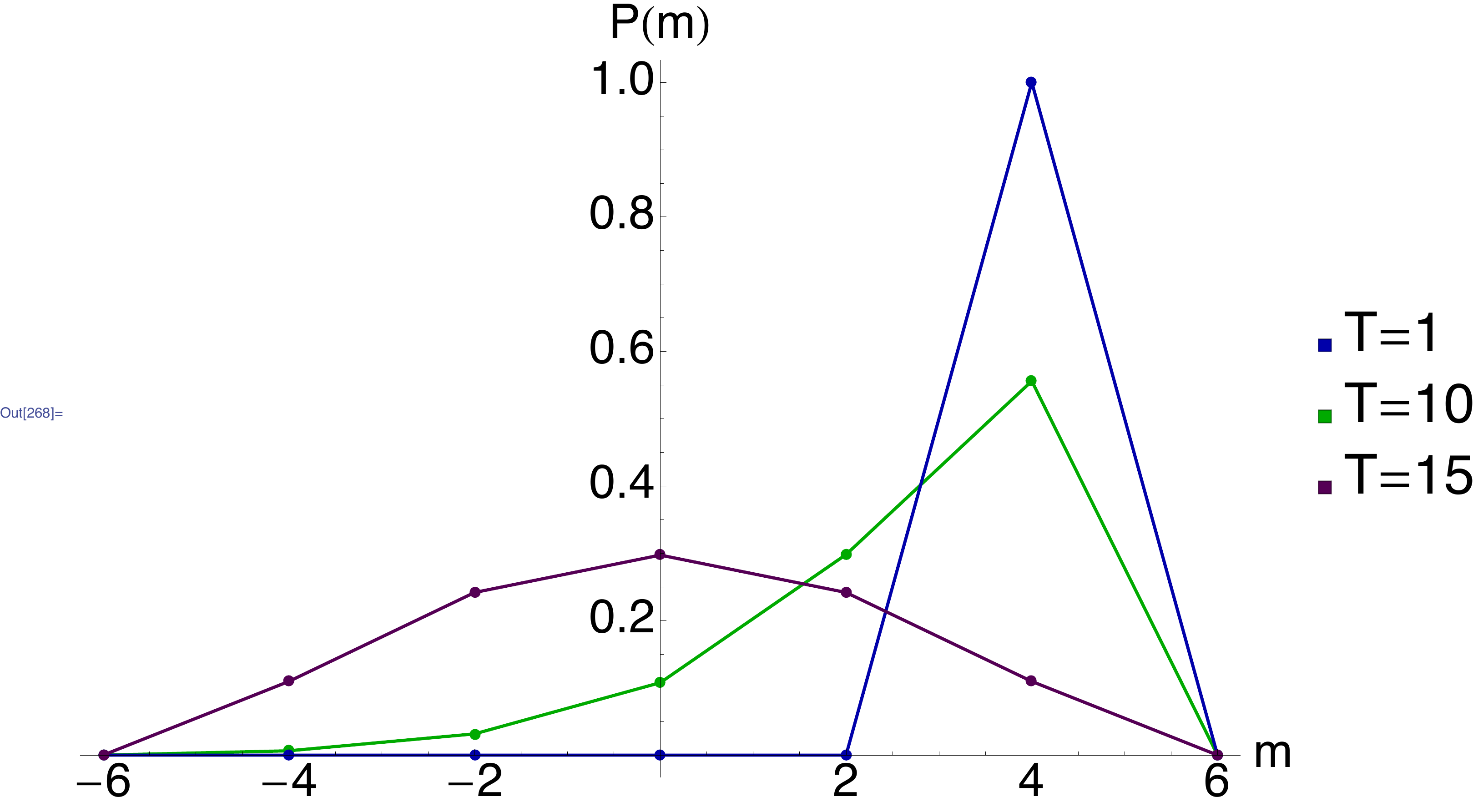}
\includegraphics[width=0.45\textwidth]{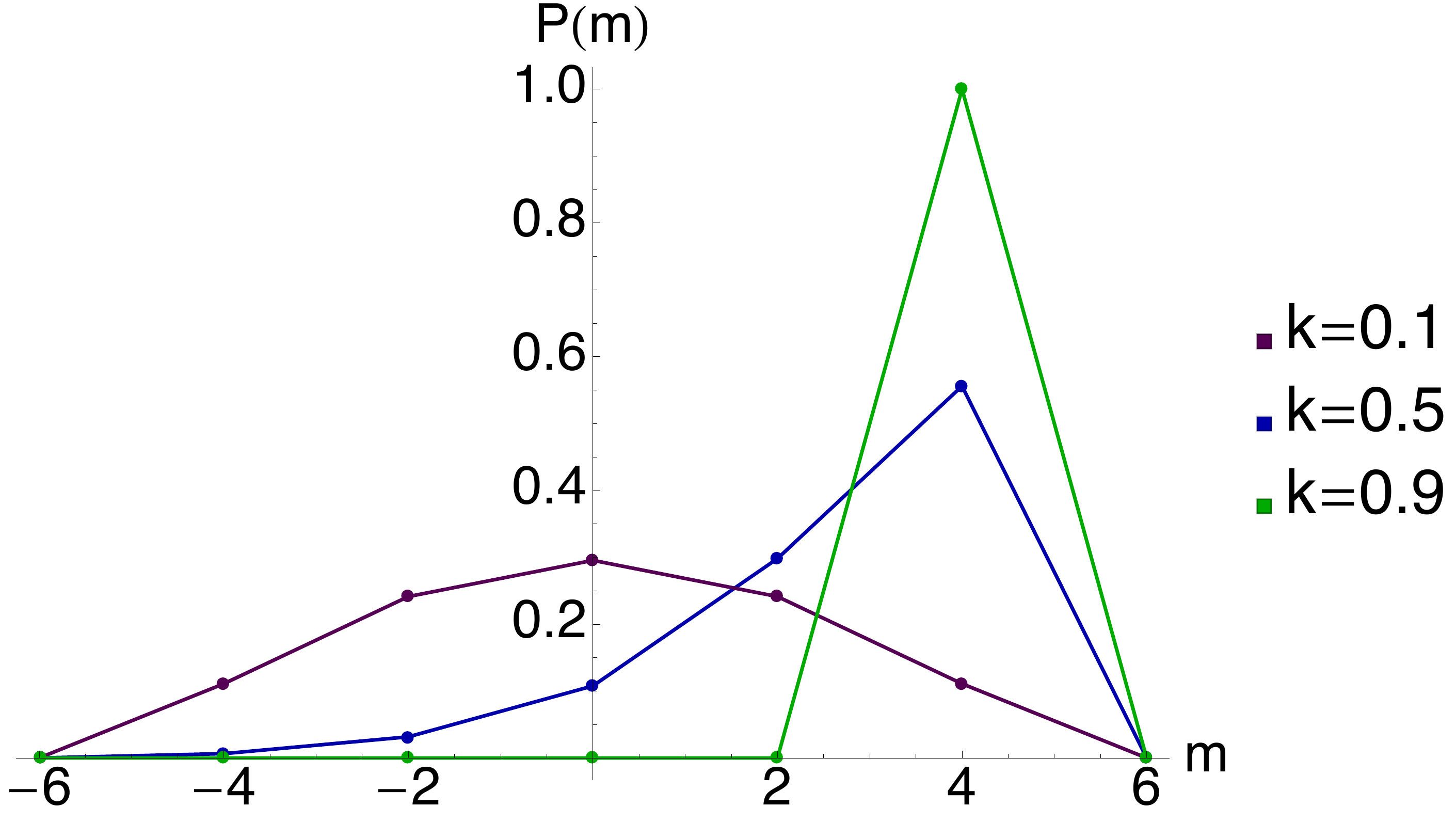}
\caption{Plot of $P(m)$ for a regular graph with $K=4,L=2$ with ferromagnetic interactions. Left: $P(m)$ varying the temperature $T=1,10,15$ with $k=0.5$. Right:  $P(m)$ varying B-B interaction strength $k=0.1,0.5,0.9$ for $T=10$. }
\label{fig:ferropm}
\end{figure}

\begin{figure}
\centering
\includegraphics[trim=1.4cm 0cm 0cm 0cm, clip=true,width=0.48\textwidth]{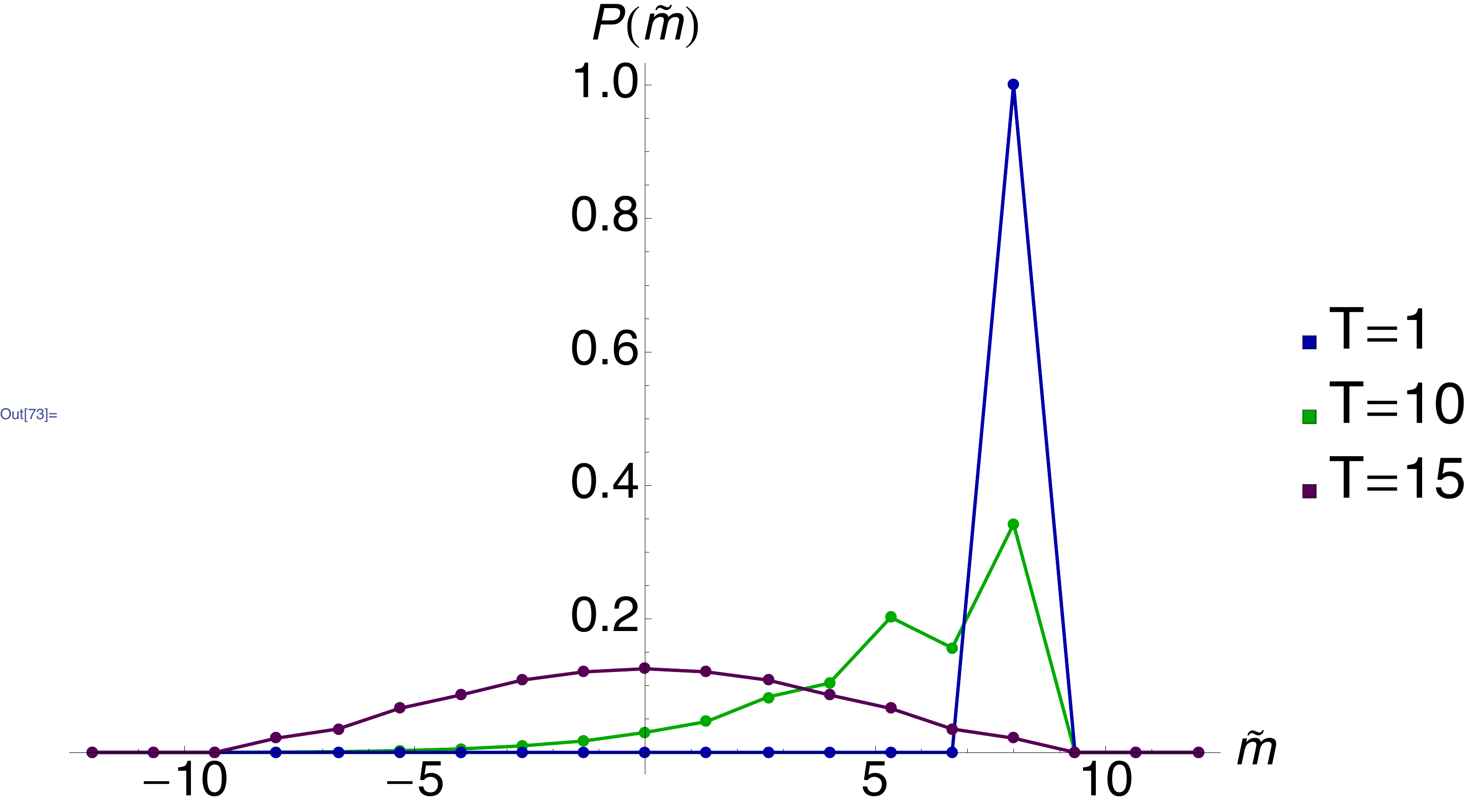}
\includegraphics[trim=1.4cm 0cm 0cm 0cm, clip=true, width=0.48\textwidth]{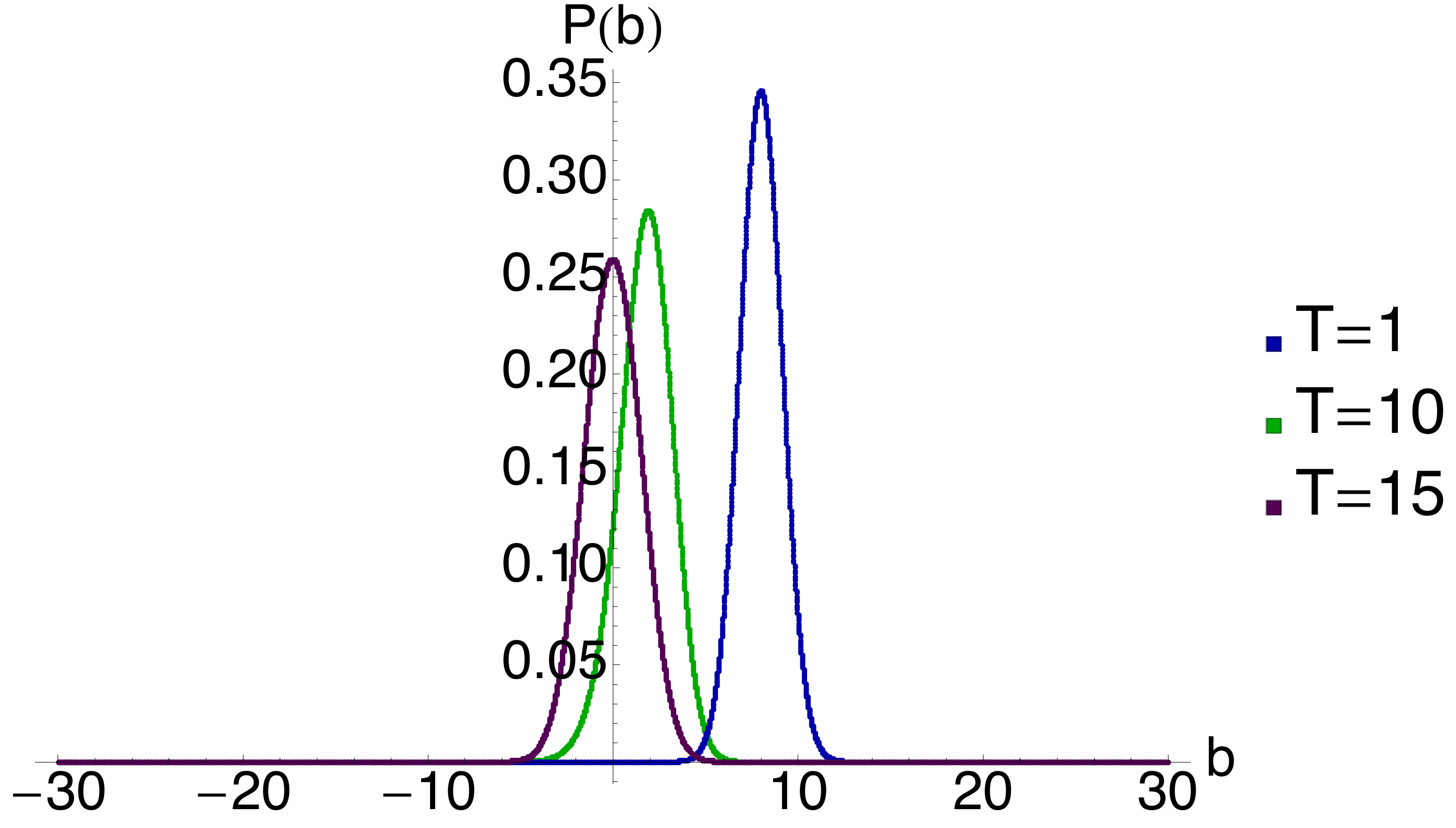}
\caption{Plot of $P(\tilde{m}|K)$ (left) and the corresponding B clones distribution $P(b)$ (right) for different temperatures $T=1,5,10$ for a regular graph with $K=4,L=2$ and $k=0.5$ with ferromagnetic interactions.}
\label{fig:pmtildeferrot}
\end{figure}

\begin{figure}
\centering
\includegraphics[trim=1.4cm 0cm 0cm 0cm, clip=true,width=0.48\textwidth]{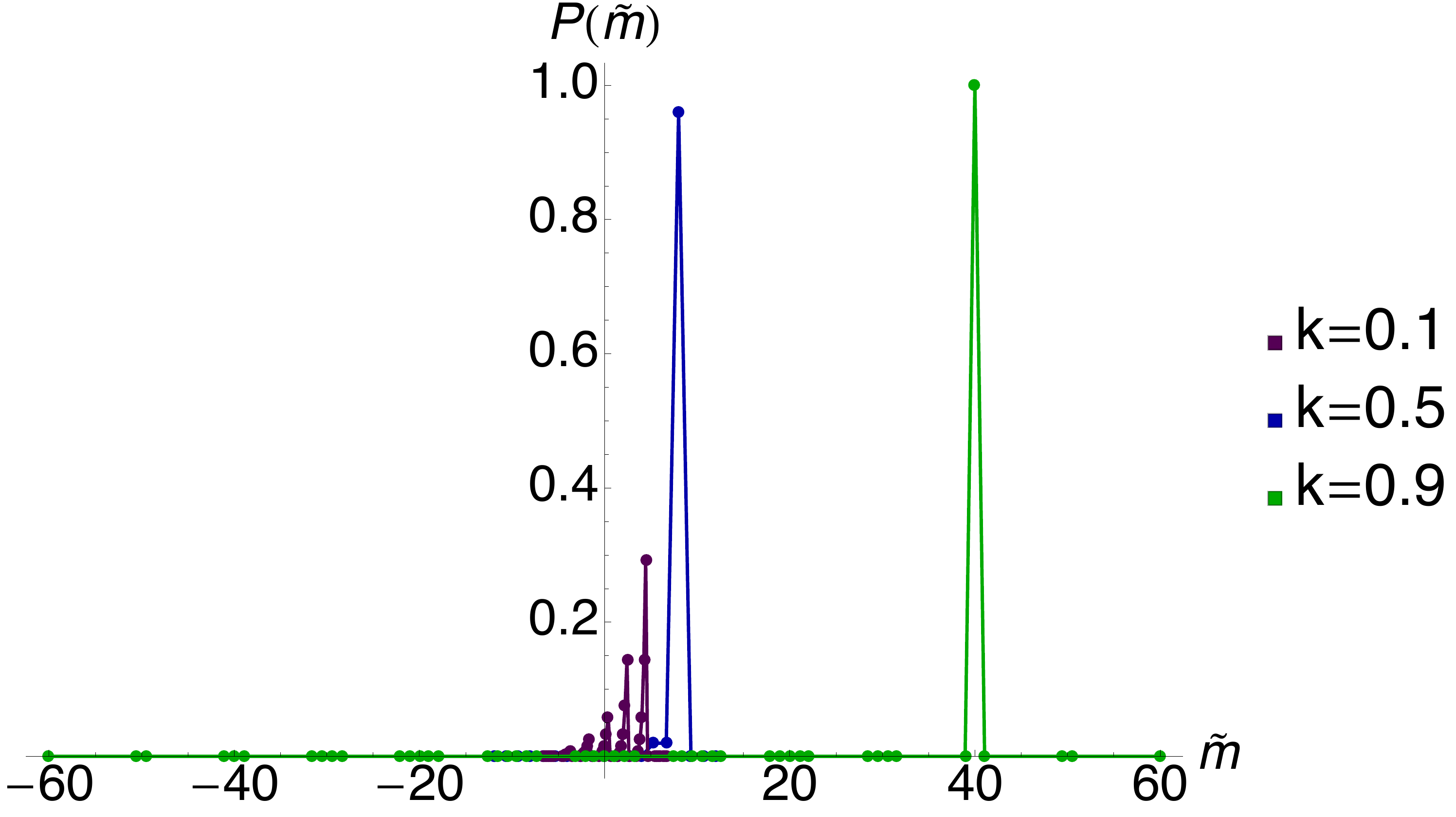}
\includegraphics[trim=1.4cm 0cm 0cm 0cm, clip=true, width=0.48\textwidth]{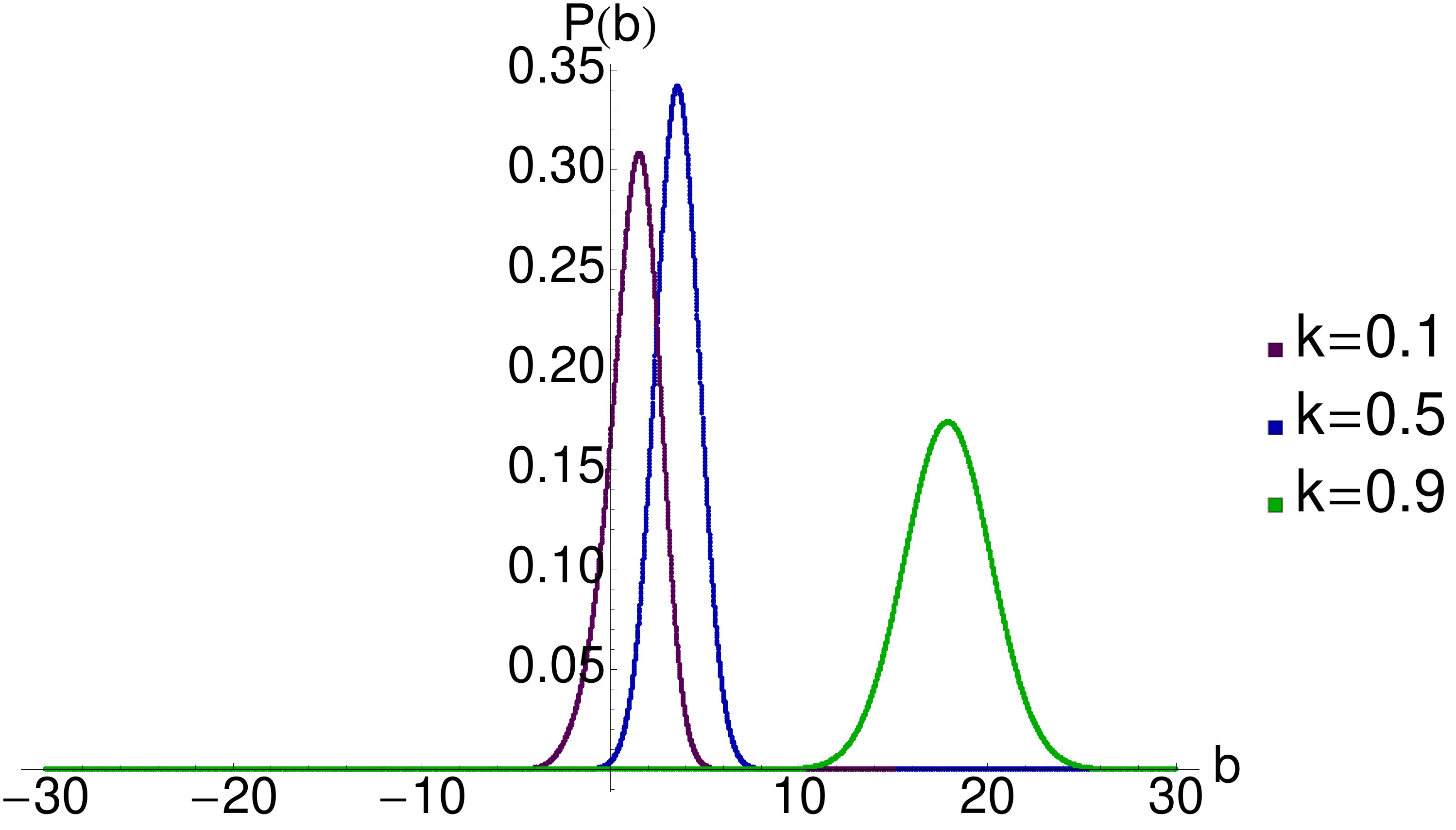}
\caption{Plot of $P(\tilde{m}|K)$ (left) and the corresponding B clones distribution $P(b)$ (right) for different B-B interaction strength $k=0.1, 0.5, 0.9$ for a regular graph with $K=4,L=2$ and $T=5$ with ferromagnetic interactions.}
\label{fig:pmtildeferrok}
\end{figure}

Marginalising over $\hat{m}$, we can get information about the overlap distribution of the $\mu$-th B clone given that it interacts with his complement $\hat{\mu}$: in fig. \ref{fig:ferropm} we plot it for different values of the temperature and the B-B interaction strength, and observe qualitatively the same transition described above. 
From the joint distribution $P(m,\hat{m}|K)$, we can compute the distribution $P(\tilde{m}|K)$
of rotated overlaps $\tilde{m}_\mu=\frac{m_\mu}{1-k^2}+\frac{k m_{\hat{\mu}}}{1-k^2}$, which is plotted in fig. \ref{fig:pmtildeferrot} (left panel)
for different temperatures and in fig. \ref{fig:pmtildeferrok} (left panel) for different values of $k$. 

\begin{figure}[htb]
\centering
\includegraphics[width=0.48\textwidth]{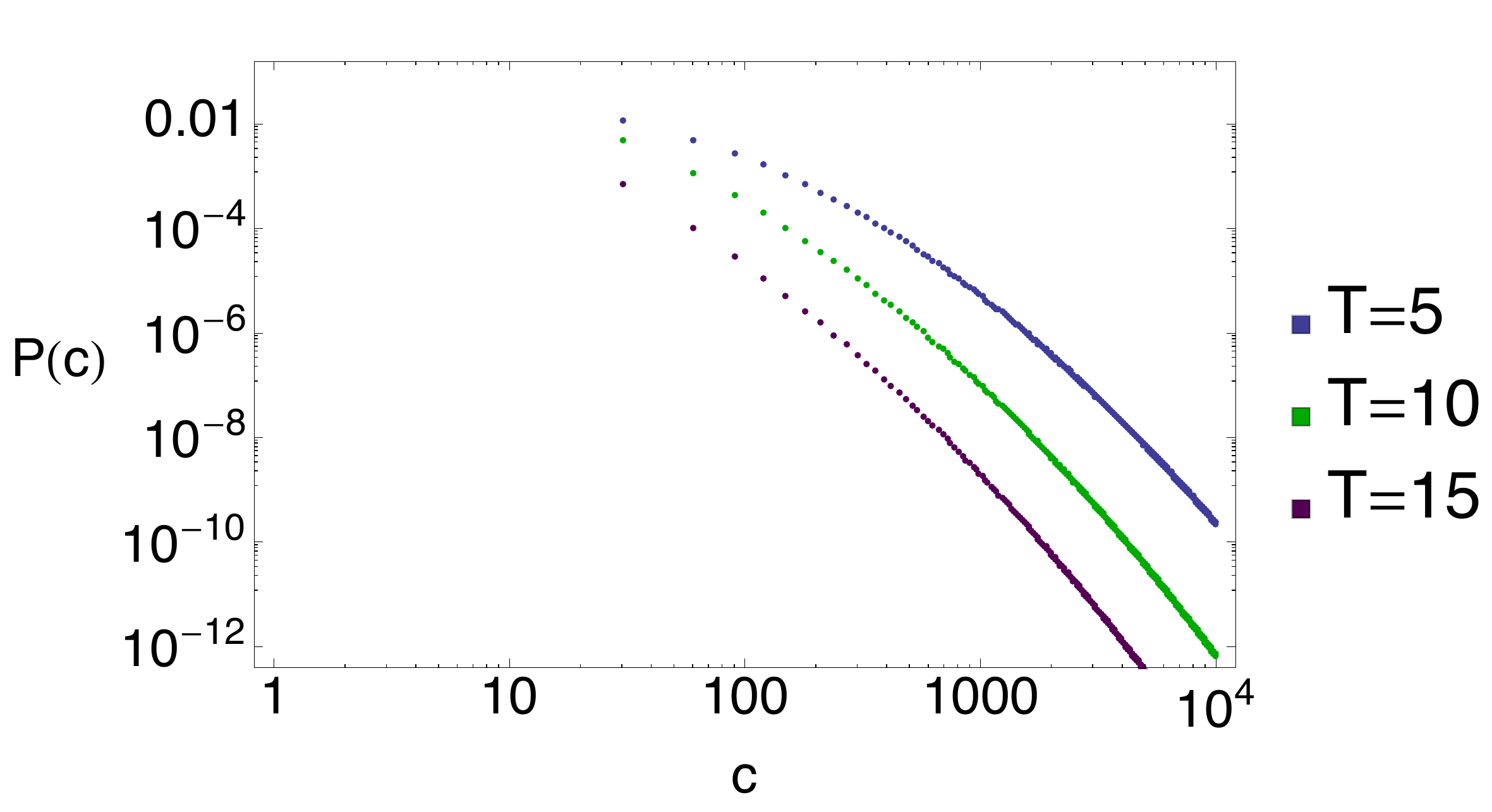}
\includegraphics[width=0.48\textwidth]{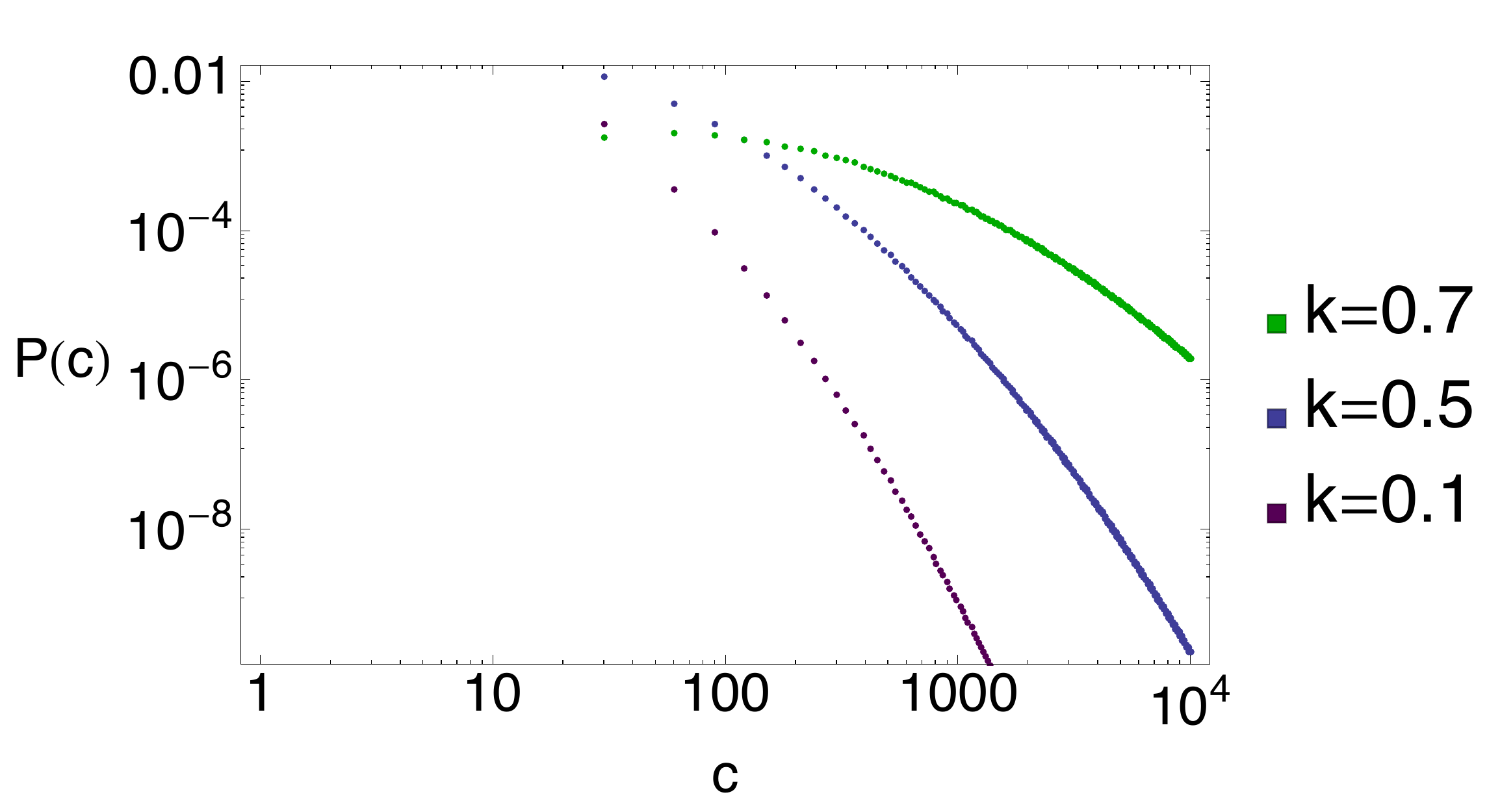}
\caption{Log-Log plot of $\mathcal{P}(c)$ for a regular graph with $K=4,L=2$ with ferromagnetic interactions.
Left: $\mathcal{P}(c)$ varying the temperature fixing $k=0.5$. Right: $\mathcal{P}(c)$ varying $k$ for $T=5$. }
\label{fig:pconc}
\end{figure}

We can finally use the rotated overlap distribution 
$P(\tilde{m})$ to derive the B clones size distribution, as shown in sec. \ref{sec:model}. In fig. 
\ref{fig:pmtildeferrot} and \ref{fig:pmtildeferrok} (right panels) we show respectively its behaviour in temperature and in $k$. 
At low temperature, the B clone size distribution is peaked around non-zero values, meaning that B clones are expanding. 
Increasing $k$, increases the probability of having strong clonal expansions even at high noise levels.
B clonal sizes are often experimentally measured as concentrations and are known to 
follow a Zipf's law \cite{Bdist1}. Using our definition of clonal sizes as (relative) log-concentrations $b=\log c/c_0$ we can 
get concentration distributions 
as $\mathcal{P}(c)= \int P(b)\delta(c-{\rm e}^b){\rm d}b =\frac{1}{c} P(\log(c))$. 
In fig. \ref{fig:pconc} we show the log-log plot of 
$\mathcal{P}(c)$ for different values of $T$ (left) and $k$ (right). 
In particular, we see that increasing $k$ and decreasing $T$, increases the probability of having clones in high concentrations. 
More in general, our model enables us to determine the single most important parameters that affect the tail behaviour of these distributions and might be useful to infer the network connectivity and operational noise in health and disease situations. 
\section{Disordered interactions}\label{sec:disorder}
\begin{figure}
\centering
\includegraphics[width=0.5\textwidth]{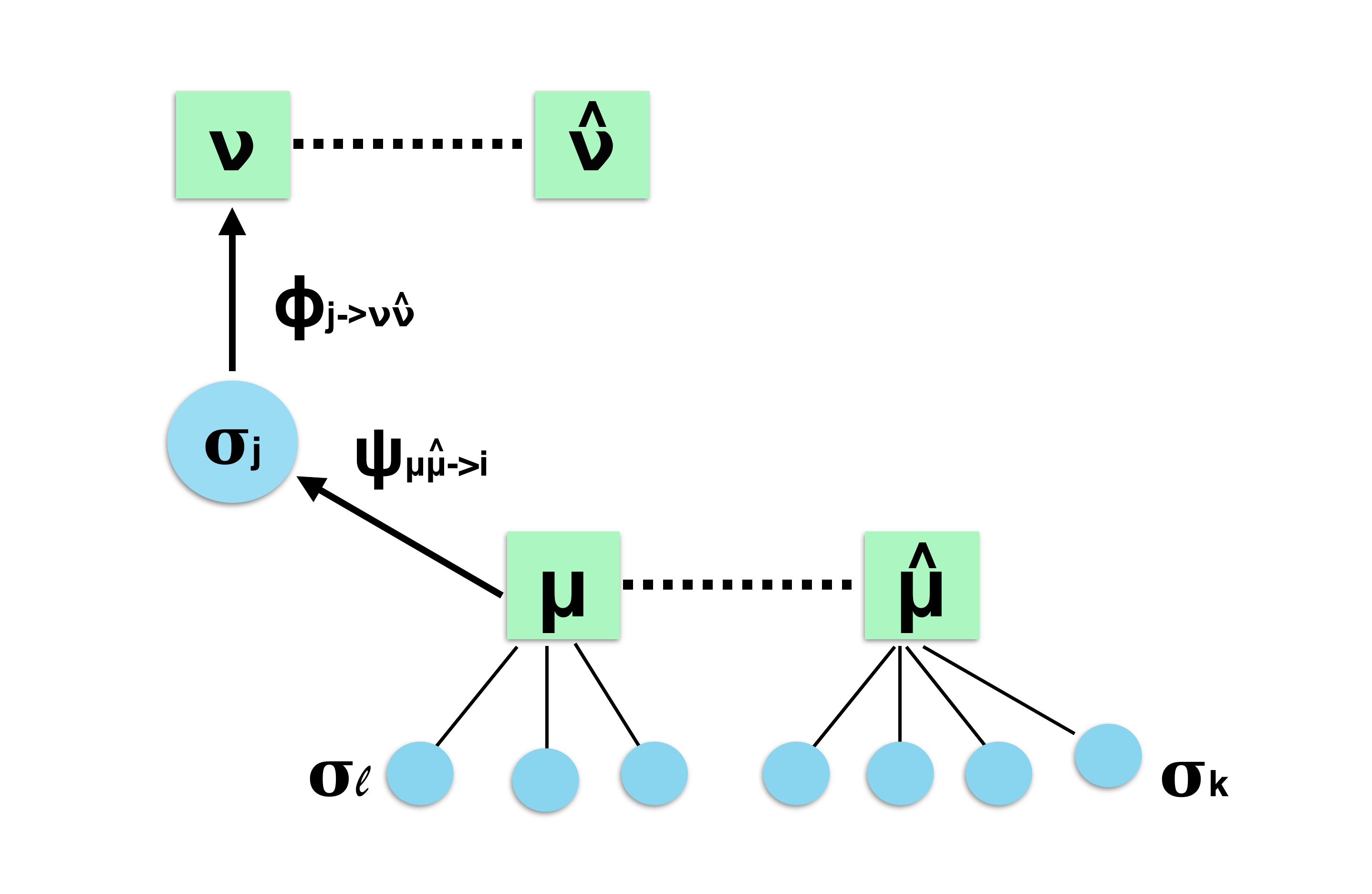}
\caption{Schematic representation of the messages $\phi_{j\to\nu\hat{\nu}}$ from node $j$ to factors $\nu\hat{\nu}$, and 
$\psi_{\mu\hm\to j}$, from the factors $\mu\hat{\mu}$ to node $j$ in the factor tree used to derive equations \eqref{eq:updatepsi} and 
\eqref{eq:updatephi}.}
\label{fig:factor2}
\end{figure}

In order to discuss more general cases of interactions it is useful to parametrise $P_{\mu\hat{\mu}}(\sigma_j)$, the {\em message} from factor $\mu$ and $\hat{\mu}$ to node $i$, by an effective field $\psi_{\mu\hat{\mu} \to j}$ and $P_{\setminus \nu\hat{\nu}}(\sigma_j)$, the {\em message} from node $j$ to factors $\mu,\hat{\mu}$, by the effective field $\phi_{j \to \nu\hat{\nu}}$ as
\bea
P_{\mu\hat{\mu}}(\sigma_j)&\propto&\rme^{\beta \sigma_j \psi_{\mu\hat{\mu} \to j}}\ ,
\label{eq:psi}\\
P_{\setminus \nu\hat{\nu}}(\sigma_j)&\propto&\rme^{\beta \sigma_j \phi_{j \to \nu\hat{\nu}}} \ .
\label{eq:phi}
\eea
A schematic representation of these messages in a factor graph can be seen in fig. \ref{fig:factor2}.
Using the relation $\bra \sigma_j\ket = \frac{ \sum_{\sigma_j} P_{\mu\hat{\mu}}(\sigma_j)\sigma_j}{\sum_{\sigma_j} P_{\mu\hat{\mu}}(\sigma_j)}=\tanh (\beta \psi_{\mu\hat{\mu}\to j})$ we can derive an expression for $ \psi_{\mu\hat{\mu}\to j}$. From \eqref{eq:iter} and the definition of $f_{\mu\hm}(\{\sigma_{k\in\partial \mu}\},\{\sigma_{\ell\in\partial \hat{\mu}}\})$ in \eqref{eq:fmu}, we have
\bea
P_{\mu\hm}(\sigma_j)&=&\frac{1}{\mathcal{Z}_{\mu\hat{\mu}}}\sum_{\{\sigma_{k\in\partial \mu}\},\{\sigma_{\ell\in\partial\hat\mu}\}}
\int \frac{\rmd y_1 \rmd y_2}{2\pi/\detx}  {\rm e}^{-\frac{1}{2}\by^T \bC^{-1} \by}{\rm e}^{\sqrt{\beta}(y_1\sum_{k\in\partial\mu}\xi^\mu_k\sigma_k+y_2\sum_{\ell\in\partial\hat{\mu}}\xi^{\hat{\mu}}_\ell\sigma_\ell)}
\nonumber\\
&&\times
\left[\prod_{k\in\partial\mu\hat{\mu}\setminus j}P_{\setminus\mu\hat{\mu}}(\sigma_k)\right]\ , 
\label{eq:pmusigma}
\eea
where $\mathcal{Z}_{\mu\hat{\mu}}$ is the normalization. Manipulating \eqref{eq:pmusigma} by using the parametrisation \eqref{eq:phi}, yields 
\bea
P_{\mu\hm}(\sigma_j)&=&\frac{\detx}{\mathcal{Z}_{\mu\hat{\mu}}}\sum_{\{\sigma_k\},\{\sigma_\ell\}}\int \frac{\rmd y_1 \rmd y_2}{2\pi}  {\rm e}^{-\frac{1}{2}\by^T \bC^{-1} \by}{\rm e}^{\sqrt{\beta}y_1\xi^\mu_j\sigma_j}
\left[\prod_{k\in\partial\mu\setminus j}{\rm e}^{(\sqrt{\beta}y_1\xi^\mu_k+\beta\phi_{k\to\mu\hm})\sigma_k}\right]
\nonumber\\
&&\times
\left[\prod_{\ell\in\partial\hm}{\rm e}^{(\sqrt{\beta} y_2\xi^{\hat{\mu}}_\ell+\beta\phi_{\ell\to\mu\hm})\sigma_\ell}\right].
\eea
Note that due to the sparsity of the links, we assume that each spin is connected to either $\mu$ or its complementary $\hat{\mu}$ factor. Summing over  $\{\sigma_k\},\{\sigma_\ell\}$, we obtain
\bea
\hspace{-2.5cm}P_{\mu\hm}(\sigma_j)=\frac{\detx}{\mathcal{Z}_{\mu\hat{\mu}}}\int \frac{\rmd y_1 \rmd y_2}{2\pi}  {\rm e}^{-\frac{1}{2}\by^T \bC^{-1} \by}{\rm e}^{\sqrt{\beta}y_1\xi^\mu_j\sigma_j}
\left[\prod_{k\in\partial\mu\setminus j}2\cosh{(\sqrt{\beta}y_1\xi^\mu_k+\beta\phi_{k\to\mu\hm})}\right]\nonumber\\ \times\left[\prod_{\ell\in\partial\hm}2\cosh{(\sqrt{\beta}y_2\xi^{\hat{\mu}}_\ell+\beta\phi_{\ell\to\mu\hm})}\right]\ .
\label{eq:pmusigmamod}
\eea
Hence the cavity fields in \eqref{eq:psi} are found as 
\begin{equation}
\hspace{-0.2cm} \psi_{\mu\hm\to j} =\atanh\left(\frac{\langle\sinh{\sqrt{\beta}(y_1\xi^\mu_j)}
\prod_{k\in\partial\mu\setminus j}\cosh{(\sqrt{\beta}y_1\xi^\mu_k+\beta\phi_{k\to\mu\hm})}\prod_{
\ell\in\partial\hat{\mu}}\cosh{(\sqrt{\beta}y_2\xi^{\hat{\mu}}_\ell+\beta\phi_{\ell\to\mu\hm}))}\rangle_{\by}}
{\langle\cosh{\sqrt{\beta}(y_1\xi^\mu_j)}
\prod_{k\in\partial\mu\setminus j}\cosh{(\sqrt{\beta}y_1\xi^\mu_k+\beta\phi_{k\to\mu\hm})}\prod_{
\ell\in\partial\hat{\mu}}\cosh{(\sqrt{\beta}y_2\xi^{\hat{\mu}}_\ell+\beta\phi_{\ell\to\mu\hm})}\rangle_{\by}}\right) \ .\\ \\
\label{eq:updatepsi}
\end{equation}
Finally, using the recursive equation \eqref{eq:iter} and the parametrisation \eqref{eq:phi}, \eqref{eq:psi} it follows also that 
\begin{equation} 
\hspace{-2cm} \phi_{j\to\nu\hat{\nu}}=\sum_{\mu\hat{\mu}}\psi_{\mu\hm\to j} \ .
\label{eq:updatephi}
\end{equation}
The cavity fields equations \eqref{eq:updatepsi},\eqref{eq:updatephi} can be iterated until convergence \cite{montanari}. 
We can show analytically and check numerically that $\psi_{\mu\hm \to j}=0, \forall \mu,\hat\mu,j$ is a fixed point of \eqref{eq:updatepsi} for any 
value of $k$, simply using symmetries of the integral. 
In the following we will study the transition from zero to non-zero cavity fields and we will discuss how it affects the retrieval properties and the 
functioning of the system.

\subsection{Distributions of cavity fields }
In the large $N$ limit the solution of the cavity equations  \eqref{eq:updatephi}, \eqref{eq:updatepsi} can be characterised via the distribution of messages or fields, $W_{\psi}(\psi)$ and
$W_{\phi}(\phi)$.
The field distributions can be computed as follows, denoting by $\Psi(\{\xi^\mu_i\},\{\xi^{\hat{\mu}}_i\},\{\phi_{k\to\mu\hm}\},\{\phi_{\ell\to\mu\hm}\})$ the r.h.s.\ of (\ref{eq:updatepsi}),
\be
W_\psi(\psi)=\sum_e\frac{Q(e)e}{\langle e \rangle}\langle\langle\delta(\psi-\Psi(\phi_1,...,\phi_{e-1},\{\xi^1,\dots,\xi^e\})\rangle\rangle_{\bxi,\bphi} \ ,
\label{eq:wpsi}
\ee 
and taking the average over i.i.d. values of the (non-zero) $\{\xi^e\}$ and over i.i.d. fields 
$\phi_1,...,\phi_{e-1}$ drawn from $W_{\phi}(\phi)$.  In the above expression, $Q(e)e/\langle e \rangle$ is the probability of picking an edge 
connected to a cluster of degree $e$. Since the cluster is composed by the union of the nodes signalling to factor 
$\mu$ and $\hat{\mu}$, its degree distribution $Q(e)$ follows from the degree distribution of the disjoint factors $\mu, \hat{\mu}$ as  
\be
Q(e)=\sum_{q,\hat{q}}P_q(q)P_q(\hat{q})\delta(e-q-\hat{q})\ ,
\ee
and
\bea 
\hspace*{-1cm}\langle e \rangle =\sum_e Q(e)e =
\sum_e \sum_{q,\hat{q}}e P_q(q)P_q(\hat{q})\delta(e-q-\hat{q})=\sum_{q,\hat{q}}P_q(q)P_q(\hat{q})(q+\hat{q})=2\langle q \rangle \ .
\eea 
Similarly we have 
\be
W_{\phi}(\phi)= 
\sum_d \frac{d P_d(d)}{\langle d \rangle} \bigg\langle\bigg\langle \delta\left( \phi - \sum_{\mu=1}^{d-1}\psi_{\mu} \right) \bigg\rangle\bigg\rangle_{\bxi,\bpsi} \ ,
\label{eq:wphi}\ee
where the average is over i.i.d. values of the (non-zero) $\{\xi^e\}$ and i.i.d. fields $\psi_1,...,\psi_{d-1}$ drawn from $W_{\psi}(\psi)$. $P_d(d)$ represents the 
probability of picking a node of degree $d$. Field distributions can then be obtained numerically by a population dynamics (PD) algorithm 
\cite{bethe}, details of which are provided in \ref{appendix:algorithm}.
\subsection{Small fields expansion and bifurcation lines}
Transitions from zero to non-zero cavity fields can be located by monitoring bifurcations away from zero of the moments of the field distribution.
Depending on the model's parameters, either the first or the second moment will bifurcate away from zero first \cite{cavitysollich}. 
To this end, we Taylor expand for small fields the rhs of equation \eqref{eq:updatepsi} to the first order and set ${\bphi}=\mathbf{0}$, which gives
\begin{align}
&\Psi(\xi^\mu_j,\{\xi^\mu_k\},\{\xi^{\hat{\mu}}_\ell\},\{\phi_{k\to\mu\hm}\},\{\phi_{\ell\to\mu\hm}\})\simeq\nonumber\\ &\simeq
\sum_{k\in\partial\mu\setminus j}\phi_{k\to\mu\hm}
 \frac{\langle \sinh(\sqrt{\beta}y_1\xi_j^{\mu}) \sinh(\sqrt{\beta} y_1\xi_k^{\mu})\prod_{r \in \partial\mu\setminus \{k,j\}} \cosh(\sqrt{\beta}y_1\xi_r^{\mu})\prod_{\ell \in \partial\hm} \cosh(\sqrt{\beta}y_2\xi_\ell^{\hm})
\rangle_{{\bf y}}}{\langle\prod_{k \in \partial\mu} \cosh(\sqrt{\beta} y_1\xi_k^{\mu}) \prod_{\ell \in \partial\hm} \cosh(\sqrt{\beta}y_1\xi_\ell^{\hm})
\rangle_{{\bf y}}}\nonumber+\\
&+\sum_{\ell\in\partial\hat\mu}\phi_{\ell\to\mu\hm}
 \frac{\langle \sinh(\sqrt{\beta}y_1\xi_j^{\mu}) \sinh(\sqrt{\beta} y_2\xi_\ell^{\hm})\prod_{r \in \partial\mu\setminus j} \cosh(\sqrt{\beta}y_1\xi_r^{\mu})\prod_{p \in \partial\hm \setminus \ell} \cosh(\sqrt{\beta}y_2\xi_p^{\hm})
\rangle_{{\bf y}}}{\langle\prod_{k \in \partial\mu} \cosh(\sqrt{\beta} y_1\xi_k^{\mu}) \prod_{\ell \in \partial\hm} \cosh(\sqrt{\beta}y_1\xi_\ell^{\hm})
\rangle_{{\bf y}}}\nonumber=\\
&=\sum_{k\in\partial\mu\setminus j}\phi_{k\to\mu\hm}
\Omega^a(\xi^\mu_k,\xi^\mu_j,\{\xi^\mu_r\},\{\xi^{\hat{\mu}}_\ell\})+\sum_{\ell\in\partial\hat\mu}\phi_{\ell\to\mu\hm}
\Omega^b(\xi^{\hm}_\ell,\xi^\mu_j,\{\xi^\mu_r\},\{\xi^{\hat{\mu}}_p\})\ ,
\label{eq:update}
\end{align}
where 
\begin{align}
&\Omega^a(\xi^1,\dots,\xi^e)=\frac{\langle \sinh (\sqrt{\beta}y_1\xi^1)\sinh (\sqrt{\beta}y_1\xi^2)\prod_{r=3}^{q-1}\cosh{(\sqrt{\beta}y_1\xi^r)}\prod_{\ell=q}^{e}\cosh(\sqrt{\beta}y_2\xi^{\ell})\rangle_{{\bf y}}}{\langle \prod_{r=1}^{q-1}\cosh(\sqrt{\beta}y_1\xi^r)\prod_{\ell=q}^{e}\cosh(\sqrt{\beta}y_2\xi^{\ell}) \rangle_{{\bf y}}}\ ,\\ 
&\Omega^b(\xi^1,\dots,\xi^e)=\frac{\langle \sinh (\sqrt{\beta}y_1\xi^1)\sinh (\sqrt{\beta}y_2\xi^2)\prod_{r=3}^{q}\cosh{(\sqrt{\beta}y_1\xi^r)}\prod_{\ell=q+1}^{e}\cosh(\sqrt{\beta}y_2\xi^{\ell})\rangle_{{\bf y}}}{\langle \prod_{r=1}^{q-1}\cosh(\sqrt{\beta}y_1\xi^r)\prod_{\ell=q}^{e}\cosh(\sqrt{\beta}y_2\xi^{\ell}) \rangle_{{\bf y}}}\ .
\end{align}
Moments can be obtained using the fields distribution \eqref{eq:wpsi},\eqref{eq:wphi} and averaging the small field expansion \eqref{eq:update} over $\phi,\psi,\xi$ .
For the mean bifurcation we get
\begin{flalign}
&\langle \psi\rangle_{\psi}= \langle \phi\rangle_{\phi}\sum_{e,q} \frac{P_q(q)P_q(e-q)e}{2\langle q\rangle}\left[(q-1)\langle \Omega^a(\xi^1,\dots,\xi^e)\rangle_{\bxi}+(e-q)\langle \Omega^b(\xi^1,\dots,\xi^e)\rangle_{\bxi} \right]\ ,\\ 
&\langle\phi\rangle_{\phi}=\langle \psi\rangle_{\psi}\sum_d\frac{P_d(d)d (d-1)}{\langle d\rangle}\ .
\label{eq:fieldmean}
\end{flalign}
Combining them we obtain
\begin{flalign}
\langle \psi\rangle_{\psi}= \langle \psi\rangle_{\psi}\sum_{d,e,q}\frac{P_d(d)d (d-1)}{\langle d\rangle}\frac{P_q(q)P_q(e-q)e}{2\langle q\rangle}\left[(q-1)\langle \Omega^a(\xi^1,\dots,\xi^e)\rangle_{\bxi}+(e-q)\langle \Omega^b(\xi^1,\dots,\xi^e)\rangle_{\bxi} \right]\ ,
\label{eq:fieldmeancomb}
\end{flalign}
with solutions $\langle \psi \rangle_\psi=0$ or 
\be
1=\sum_d\frac{P_d(d)d (d-1)}{\langle d\rangle}\sum_{e,q} \frac{P_q(q)P_q(e-q)e}{2\langle q\rangle}\left[(q-1)\langle \Omega^a(\xi^1,\dots,\xi^e)\rangle_{\bxi}+
(e-q)\langle \Omega^b(\xi^1,\dots,\xi^e)\rangle_{\bxi} \right].
\ee
For the variances we obtain
\begin{flalign}
&\langle \psi^2\rangle_{\psi}=\langle \phi^2\rangle_{\phi}\sum_{e,q}\frac{P_q(q)P_q(e-q)e}{2\langle q\rangle}\bigg[(q-1)\langle
(\Omega^a(\xi^1,\dots,\xi^e)^2\rangle_{\bxi}+(e-q) \langle (\Omega^b(\xi^1,\dots,\xi^e))^2\rangle_{\bxi} 
\bigg]\ ,
\label{eq:fieldvar_psi}  \\
&\langle\phi^2\rangle_{\phi}=\langle \psi^2\rangle_{\psi}\sum_d\frac{P_d(d)d (d-1)}{\langle d\rangle}\ .
\label{eq:fieldvar1}
\end{flalign}
Combining equations \eqref{eq:fieldvar_psi} and \eqref{eq:fieldvar1} we have
\bea
\langle \psi^2\rangle_{\psi}&=&\langle \psi^2\rangle_{\psi}\sum_{d,e,q}\frac{P_d(d)d (d-1)}{\langle d\rangle}\frac{P_q(q)P_q(e-q)e}{2\langle q\rangle}
\nonumber\\
&&\times\bigg[(q-1)\langle(\Omega^a(\xi^1,\dots,\xi^e))^2\rangle_{\bxi}+(e-q) \langle (\Omega^b(\xi^1,\dots,\xi^e))^2\rangle_{\bxi} 
\bigg]\ , \label{eq:fieldvarcomb}
\eea
with solutions $ \langle \psi^2\rangle_{\psi}=0$ or 
\be
1=\sum_d\frac{P_d(d)d (d-1)}{\langle d\rangle}\sum_{e,q} \frac{P_q(q)P_q(e-q)e}{2\langle q\rangle}\bigg[(q-1)\langle(\Omega^a(\xi^1,\dots,\xi^e))^2\rangle_{\bxi}+(e-q) \langle (\Omega^b(\xi^1,\dots,\xi^e))^2\rangle_{\bxi}\bigg]\ .
\ee
Both eq. \eqref{eq:fieldmeancomb} and \eqref{eq:fieldvarcomb} have a trivial solution with zero moments or a more complicated one, which gives us the line 
where moments become different from zero. This will depend on the system's temperature, 
the B-B interaction strength, the graph topology encoded in the distributions $P_d(d), Q(e)$ and the distribution of the disordered interactions $\xi$'s.
In the next subsection, we will study the bifurcations for different choices of the disorder. We will focus on the regular graph topology with vertex degree $L$ and factor degree $K$; hence, we choose $P_d(d)=\delta_{d,L}$ and $P_q(q)=\delta_{q,K}$. We will obtain the critical line numerically via population dynamics simulations and analytically from \eqref{eq:fieldmeancomb}, \eqref{eq:fieldvarcomb}. 

\subsubsection{Symmetric pattern distributions}
\begin{figure}[htb!]
\centering
\includegraphics[width=0.7\textwidth]{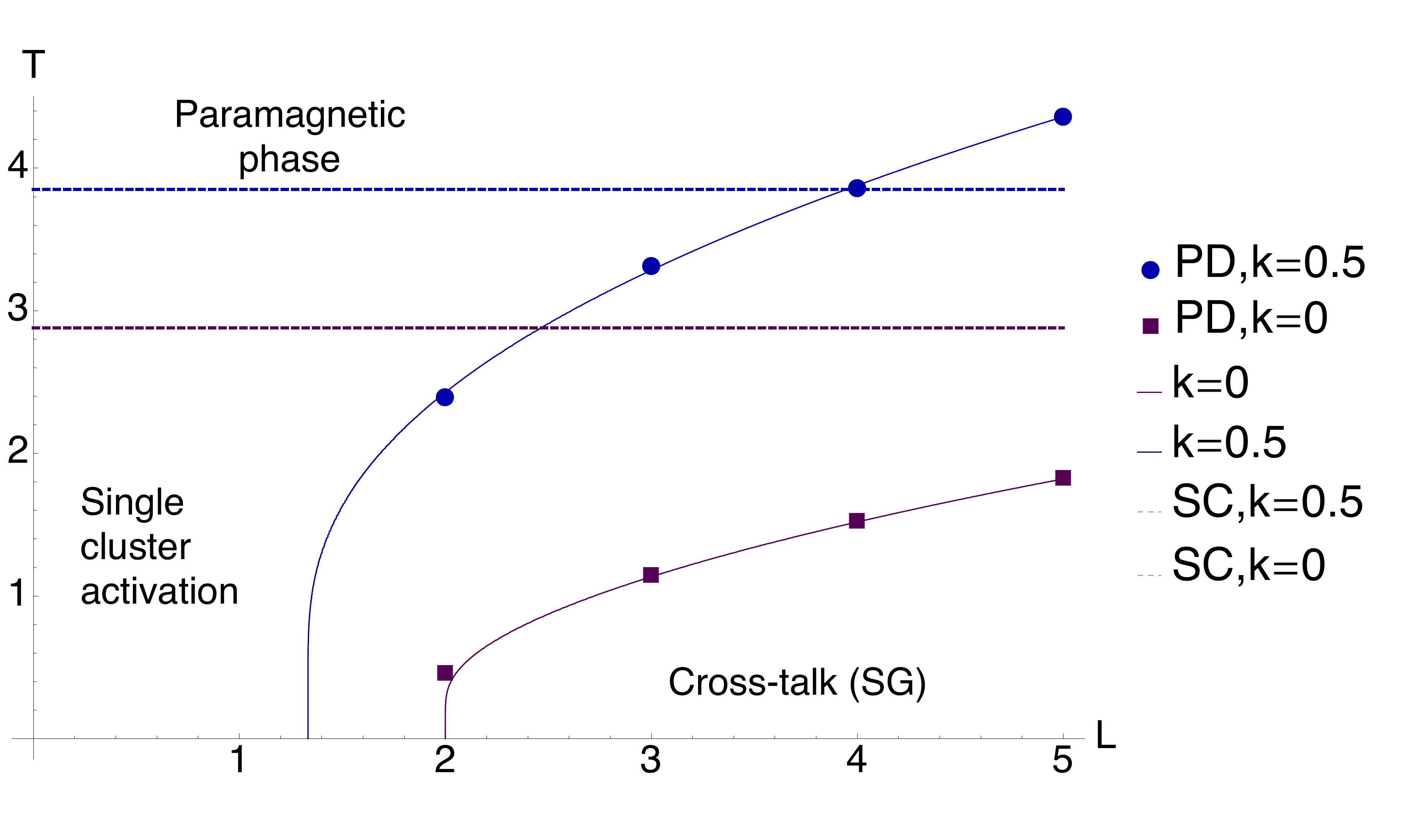}
\caption{Critical line for symmetric patterns distribution (bifurcation in variance) in the plane $(T,L)$ for different values of B-B interaction strength $k=0, 0.5$. We consider a regular graph with factor degree $K=2$. At high temperature the system is in the {\em paramagnetic phase}, where 
clusters are independent of each other. Crossing the (solid) lines, cavity fields become non-zero and the clonal interference increases entering the {\em cross-talk} or spin glass (SG) region \cite{bethe}. Markers (PD) represent numerical results obtained via population dynamics simulations (population size $M=10^4$, see 
\ref{appendix:algorithm} for details). Dashed lines represent the single cluster (SC) activation temperatures, derived in sec. \ref{subsec:cross}.}
\label{fig:clinemreg1}
\end{figure}

We first consider the case of symmetrically distributed $\xi$'s, i.e. $\mathbb{P}(\xi)=\frac{1}{2}\delta_{\xi,+1}+\frac{1}{2}\delta_{\xi,-1}$.
In this case there is no instability from growing means as the field distribution is always symmetric and, indeed, the r.h.s of \eqref{eq:fieldmeancomb} 
averages to zero.
The bifurcation is, therefore, detectable from the instability of growing variances, while the mean remains zero. 
Specialising \eqref{eq:fieldmeancomb} to the regular graph case with $P_d(d)=\delta_{d,L}$ and $P_q(q)= \delta_{q,K}$ we get
\begin{flalign}
1 = (L-1)\bigg[(K-1)\left(\frac{\langle \sinh ^2(\sqrt{\beta}y_1)\cosh^{K-2}{(\sqrt{\beta}y_1)}\cosh^{K}(\sqrt{\beta}y_2)\rangle_{{\bf y}}}{\langle \cosh^K(\sqrt{\beta}y_1)\cosh^{K}(\sqrt{\beta}y_2)\rangle_{{\bf y}}} \right)^2 
\nonumber\\ + K \left(\frac{\langle\sinh(\sqrt{\beta}y_1)\sinh(\sqrt{\beta}y_2)\cosh^{K-1}{(\sqrt{\beta}y_1)}\cosh^{K-1}(\sqrt{\beta}y_2)\rangle_{{\bf y}}}{\langle \cosh^K(\sqrt{\beta}y_1)\cosh^{K}(\sqrt{\beta}y_2)\rangle_{{\bf y}}}\right)^2\bigg] \ .
\label{eq:simmbif1}
\end{flalign}
We can easily compute the large $T$-limit ($\beta\to0$) in \eqref{eq:simmbif1}: the R.H.S clearly tends to zero, hence no solutions with non-zero fields exists at high temperature. For $\beta\to\infty$ averages are dominated by large values of $y_1,y_2$ yielding $\sinh^2(\sqrt{\beta}y_1)\sim \cosh^2(\sqrt{\beta}y_1)$ and $1=(L-1) (2K-1)$. Hence, for a regular graph there is no bifurcation as long as $L<\frac{2K}{2K-1}$.
For the general $\beta$-dependence, we compute the Gaussian averages over $\bf{y}$ explicitly, obtaining
\begin{flalign}
&\frac{1}{L-1} = (K-1)\left(\frac{\sum_{p=0}^2\sum_{r=0}^{K-2}\sum_{s=0}^{K}(-1)^p{2\choose p}{K-2 \choose r}{K\choose s}{\rm e}^{\frac{\beta}{2(1-k^2)}((K-2r-2p)^2 + (K-2s)^2+2k(K-2r-2p)(K-2s))}}{\sum_{\ell,f=0}^K{K\choose \ell}{K\choose f}{\rm e}^{\frac{\beta}{2(1-k^2)}((K-2\ell)^2 + (K-2f)^2+2k(K-2\ell)(K-2f))}} \right)^2 
\nonumber \\ &+ K \left(\frac{\sum_{n,h=0}^1\sum_{j,v=0}^{K-1}(-1)^{n+h}{1\choose n}{K-1 \choose j}{K-1\choose v}{1\choose h}{\rm e}^{\frac{\beta}{2(1-k^2)}((K-2j-2h)^2 + (K-2n-2v)^2+2k(K-2j-2h)(K-2n-2v))}}{\sum_{\tilde{\ell},\tilde{f}=0}^K{K\choose\tilde{\ell}}{K\choose\tilde{f}}{\rm e}^{\frac{\beta}{2(1-k^2)}((K-2\tilde{\ell})^2 + (K-2\tilde{f})^2+2k(K-2\tilde{\ell})(K-2\tilde{f}))}}\right)^2\ .\nonumber \\
\label{eq:clinemreg}
\end{flalign}
The latter condition gives the critical line for the variances bifurcation in $T=1/\beta$ as a function of $L$. 
We plot this line in fig. \ref{fig:clinemreg1}, in the presence and in the absence of idiotypic interactions. 
We also compare the theoretical bifurcation line in \eqref{eq:clinemreg} with the data obtained from population dynamics simulations (markers). 
For $k=0$ the line 
is in agreement with results obtained in \cite{cavitysollich}.
Increasing the B-B interactions strength, widens the interference region as one of the effects is to merge clusters (in this case of equal sizes) together. 
As in the case with ferromagnetic interactions, the role of the critical line consists in separating the region of clonal cross-talk from the region with no interference. However with disordered interactions the interference manifests itself as inhomogeneous fields 
and will therefore represent an additional source of noise in the system (besides the thermal noise).  
For $T>T_{c}(K,L)$, the cavity fields are zero and each clique is signalling to a particular B clone, without feeling the interference of the others. In the neural networks jargon, the system works, in this region, as a parallel processor able 
to retrieve multiple cytokine patterns $\bxi^\mu$ simultaneously. For $T<T_{c}(K, L)$ the system is in the {\em clonal cross-talk region} or so-called
{\em spin glass phase} (SG) \cite{cavitysollich} where random fields act on each clique, reducing the parallel processing capabilities and 
making the signalling process to B clones less effective. The dashed lines highlight the temperature at which single clusters become active 
(see sec. \ref{subsec:cross}) for different $k$. Lowering the temperature the system is subjected to both an increased clonal interference 
(crossing the solid lines) and to a unimodal-bimodal transition in the overlap distributions for single clusters.  Increasing $k$, the temperature at which 
the cross-over transition happens increases, meaning that the system increases its tolerance to high noise levels.

\subsubsection{Non-Symmetric pattern distributions}
In this subsection we add a degree of asymmetry $a\in[-1,+1]$  
to study the bifurcations in the first moment. Hence, we consider   the $\xi$'s entries to be distributed according to $\mathbb{P}(\xi)=\frac{1+a}{2}\delta_{\xi,+1}+\frac{1-a}{2}\delta_{\xi,-1}$.
Computing the $\xi$-averages in \eqref{eq:fieldmean}  and specialising the equations for the regular graph case with $P_d(d)=\delta_{d,L}$ and $P_q(q)= \delta_{q,K}$ we get
\begin{flalign}
a^{-2}&=(L-1)\bigg[(K-1)\frac{\langle \sinh ^2(\sqrt{\beta}y_1)\cosh^{K-2}{(\sqrt{\beta}y_1)}\cosh^{K}(\sqrt{\beta}y_2)\rangle_{{\bf y}}}{\langle \cosh^K(\sqrt{\beta}y_1)\cosh^{K}(\sqrt{\beta}y_2)\rangle_{{\bf y}}}
\nonumber\\&+K\frac{\langle \sinh(\sqrt{\beta}y_1)\sinh(\sqrt{\beta}y_2)\cosh^{K-1}{(\sqrt{\beta}y_1)}\cosh^{K-1}(\sqrt{\beta}y_2) \rangle_{{\bf y}}}{\langle \cosh^K(\sqrt{\beta}y_1)\cosh^{K}(\sqrt{\beta}y_2) \rangle_{{\bf y}}}\bigg]\ .
\label{eq:bifmeanbeta}
\end{flalign}
Note that when $a\to 0$ the transition point diverges and we recover the symmetric case.
First, we analyse the limits for high and zero temperature: for $\beta\to0$ eq. \eqref{eq:bifmeanbeta} does not admit any solution, hence the first moment is zero. At zero temperature, i.e. $\beta \to \infty$, we have a transition to non-zero mean for  $a^{-2}=(L-1)(2K-1)$.
For a regular graph with vertex degree $L$ and factor degree $K$, the general $\beta$-dependence can be obtained from 
\begin{flalign}
&\frac{a^{-2}}{(L-1)}=(K-1)\left(\frac{\sum_{p=0}^2\sum_{r=0}^{K-2}\sum_{s=0}^{K}(-1)^p{2\choose p}{K-2 \choose r}{K\choose s}{\rm e}^{\frac{\beta}{2(1-k^2)}((K-2r-2p)^2 + (K-2s)^2+2k(K-2r-2p)(K-2s))}}{\sum_{\ell,f=0}^K {K\choose \ell}{K\choose f}{\rm e}^{\frac{\beta}{2(1-k^2)}((K-2\ell)^2 + (K-2f)^2+2k(K-2\ell)(K-2f))}} \right) 
\nonumber \\ &+ K \left(\frac{\sum_{n,h=0}^1\sum_{j,v=0}^{K-1}(-1)^{n+h}{1\choose n}{K-1 \choose j}{K-1\choose v}{1\choose h}{\rm e}^{\frac{\beta}{2(1-k^2)}((K-2j-2h)^2 + (K-2n-2v)^2+2k(K-2j-2h)(K-2n-2v))}}{\sum_{\tilde{\ell}=0}^K\sum_{\tilde{f}=0}^{K}{K\choose\tilde{\ell}}{K\choose\tilde{f}}{\rm e}^{\frac{\beta}{2(1-k^2)}((K-2\tilde{\ell})^2 + (K-2\tilde{f})^2+2k(K-2\tilde{\ell})(K-2\tilde{f}))}}\right)\ .
\nonumber\\ 
\end{flalign}
Note that, even in presence of a non-symmetric patterns distribution, there may be a bifurcation in the field variances at zero means, whose critical line will still be given by eq. \eqref{eq:clinemreg}. In this situation, the physical bifurcation is the first one taking place when lowering the temperature.

\begin{figure}[htb!]
\centering
\includegraphics[width=0.7\textwidth]{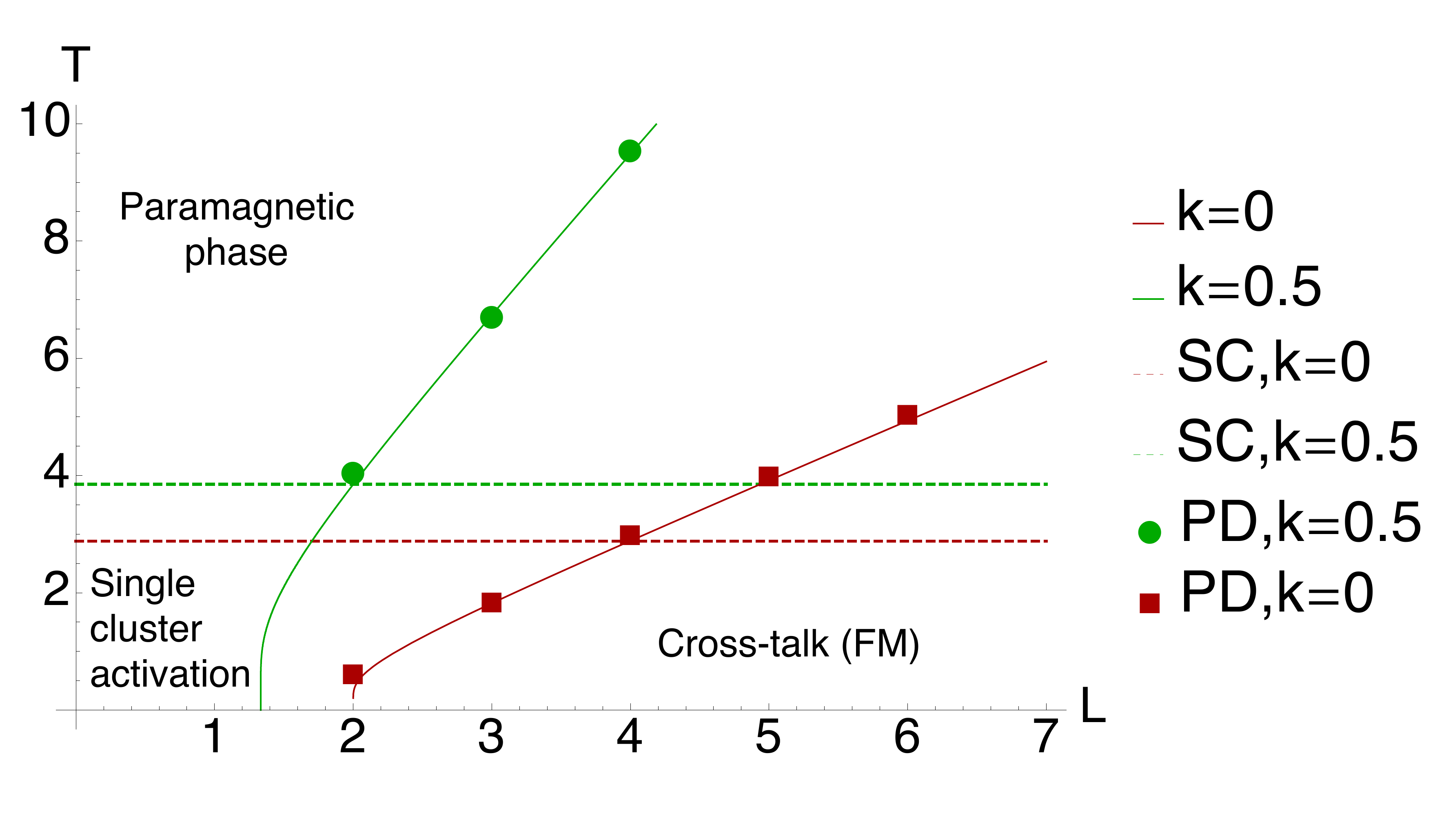}
\caption{Critical line for the bifurcation of means for a regular graph with ferromagnetic interactions, i.e. $a=1$. 
and factor degree $K=2$.
At high temperature the system is in the {\em paramagnetic phase}, where each cluster within the graph acts independently from the others. Crossing the (solid) lines for different $k$'s, cavity fields become non-zero and the clonal interference increases entering the 
{\em cross-talk} or ferromagnetic (FM) region. Markers (PD) represent numerical results obtained via population dynamics simulations 
(population size $M=10^4$, see \ref{appendix:algorithm} for details). Dashed lines represent the single cluster (SC) activation temperatures, 
derived in sec. \ref{subsec:cross}.}
\label{fig:clineferro}
\end{figure}

In fig. \ref{fig:clineferro} we plot the critical line with $k=0.5$ and $k=0$ for $a=1$, which retrieves the case 
of ferromagnetic interactions $\xi_i^\mu=+1~\forall~i,\mu$ analysed in sec. \ref{sec:ferro}. Here the critical line is given by bifurcation of the means. 
The cross-talk region is enlarged by the presence of B-B interactions. Here however, 
interference is constructive, due to the ferromagnetic nature of interactions.

\section{Conclusions and outlooks}
In this work, we modelled the interacting system of B and T clones, main constituents of the adaptive immune system. In particular, we investigated the 
effect of idiotypic interactions among B clones, using belief propagation techniques and extending a previous model studied in the sub-extensive regime \cite{jstatba}.
We derived cavity equations for the factor graph associated with the system and we discussed preliminary simple cases, which can be solved analytically. 
In particular, we considered the paramagnetic phase and ferromagnetic interactions and monitored the behaviour of the overlap distributions. The latter is useful to calculate the B clonal size distributions varying the temperature and the 
strength of the idiotypic interactions. We also derived the activation line where the overlap distribution shows a crossover behaviour from a phase of inactive to a phase of active B clones.   
We find, in particular, that B-B interactions increase the activation temperature making the 
system more resilient to noise. Indeed, one of the effects of the idiotypic 
network is to merge groups of T clones signalling to complementary B clones together, producing more stable signals and prolonging memory in the system.
Having a model that predicts what are the most important parameters affecting  the B clonal distribution is particularly welcome as this observable has recently become experimentally accessible and is important to understand the immune system's state \cite{Bdist2}. 

We then generalised our analysis to the case of disordered interactions. Here we calculated the 
critical line marking the onset of {\em clonal cross-talk}, both analytically via bifurcation analysis and numerically via population dynamics simulations, 
and compare the critical line with the activation line. B-B interactions have again the effect of enhancing B clonal activation at high noise levels,
while making their interference more pronounced at low noise level. This may result in a 
destructive interference compromising the retrieval of cytokines patterns. 

This work paves the way for an investigation of the immune system which is both theoretically sound and able to connect to 
experimental work. Pathways for future research may include the modelling of T-T interactions, which are known to play an important role in the 
self/non-self discrimination process \cite{quorum}, and the relaxation of the assumption that B clones and T clones 
evolve in the same thermal noise and on the same timescales, which could shed light on the effect of cellular environment on the immune response 
as shown recently in \cite{mozeika}. Finally, the assumption that each T clone is able to secrete both excitatory and inhibitory cytokines 
may be replaced with the more realistic scenario of two populations of T clones, each responsible for sending one type of signal only. 
On the experimental side, our study suggests that oscillations of single clone populations may be observed on typical timescales.
The availability of experimental data on clonal expansions and contraction patterns in time, in the absence of antigens, may thus feed important 
information into the model.

\appendix
\section{Factor representation}\label{appendix:factor}
In this section we work out the distribution
\be
p(\bsigma)=\prod_{\mu=1}^P F_\mu(\bsigma)\ ,
\label{eq:factors}
\ee
where
\be
F_{\mu}(\bsigma)=
\bra 
\rme^{v_\mu(\bsigma)z\sqrt{\beta/\lambda_\mu}}\ket_z \ ,
\label{eq:z_average}
\ee
with $\bra \cdot \ket_z$ denoting the average (\ref{eq:av_z}) over a Gaussian distribution with zero mean and unit variance.
We recall that $\bv(\bsigma)=\bP^{-1}\bM(\bsigma)$ and $\bP$ is the orthogonal 
matrix whose columns are the eigenvectors $\{\bn^\mu\}_{\mu=1}^P$ of the matrix $\bA$ defined in (\ref{eq:A_def}). We can write
\bea
v_\mu(\bsigma)=\sum_\nu (\bP^{T})_{\mu\nu}M_\nu(\bsigma)=
\sum_\nu n_\nu^\mu M_\nu(\bsigma)
=\sum_\nu n_\nu^\mu \sum_{i\in\partial \nu} \xi_i^\nu \sigma_i \ ,
\label{eq:rotated}
\eea
where $\partial\nu=\{i:\xi^{\nu}_i\neq 0\}$ and 
$n_\nu^\mu$ is the $\nu$-component of the eigenvector ${\bf n}^\mu$ 
of $\bA$, associated to the eigenvalue $\lambda_\mu$.
The matrix $\bA$ has two eigenvalues, each with degeneracy $P/2$
\bea
\lambda_1=1-k \quad\quad {\rm and}\quad\quad 
\lambda_2=1+k \ .
\eea
The $P/2$ eigenvectors $\bn^\mu$, $\mu=1,\ldots,P/2$,
associated to $\lambda_1$, have components
\be
n_\nu^\mu=\frac{1}{\sqrt{2}}(\delta_{\nu,\mu}+\delta_{\nu,\mu+P/2})\ ,\quad 
\mu=1,\ldots,P/2 \ ,
\ee
whereas the $P/2$ eigenvectors $\bn^\mu$, $\mu=P/2+1,\ldots,P$, associated to $\lambda_2$, have components
\be
n_\nu^\mu=\frac{1}{\sqrt{2}}(\delta_{\nu,\mu}-\delta_{\nu,\mu+P/2})
\ ,\quad 
\mu=P/2+1,\ldots,P \ .
\ee
Hence there are only two contributions to the sum over $\nu$ in 
(\ref{eq:rotated})
\bea
v_\mu(\bsigma)&=&\frac{1}{\sqrt{2}}\left(
n_\mu^\mu \sum_{k\in \partial\mu}\xi_k^\mu \sigma_k+
n_{\mu+P/2}^\mu \sum_{\ell\in \partial(\mu+P/2)}\xi_\ell^{\mu+P/2} \sigma_\ell\right).
\eea
Defining the scaled eigenvector ${\bf x}^\mu=\frac{{\bf n}^\mu}{\sqrt{2\lambda_\mu}}$, we obtain
\bea
F_\mu(\bsigma)=\left\langle 
\exp\left[\sqrt{\beta}z
x_\mu^\mu \sum_{k\in \partial\mu}\xi_k^\mu \sigma_k+ \sqrt{\beta}z
x_{\mu+P/2}^\mu \sum_{\ell\in \partial(\mu+P/2)}\xi_\ell^{\mu+P/2} \sigma_\ell
\right]
\right\rangle_z \ .
\eea
Next, we split the product over $\mu$ in 
(\ref{eq:factors}), to separate
the contribution from $\mu\leq P/2$ and $\mu>P/2$
\bea
\hspace*{-1cm}p(\bsigma)&=&\prod_{\mu=1}^{P/2}\bra 
\rme^{v_\mu(\bsigma)z\sqrt{\beta/\lambda_\mu}}\ket_z
\prod_{\nu=P/2+1}^{P}\bra 
\rme^{v_\nu(\bsigma)z\sqrt{\beta/\lambda_\nu}}\ket_z
\nonumber\\
&=&
\prod_{\mu=1}^{P/2}\left\langle \exp\left[\sqrt{\beta}z\left(
x_\mu^\mu\sum_{k\in \partial\mu}\xi_k^\mu \sigma_k+
x_{\mu+P/2}^\mu\sum_{\ell\in \partial(\mu+P/2)}\xi_\ell^{\mu+P/2} \sigma_\ell
\right)
\right]
\right\rangle
\nonumber\\
&&\times
\prod_{\nu=P/2+1}^{P}\left\langle
\exp\left[\sqrt{\beta}z
\left(x_\nu^\nu\sum_{k\in \partial\nu}\xi_k^\nu \sigma_k+ 
x_{\nu+P/2}^\nu\sum_{\ell\in \partial(\nu+P/2)}\xi_\ell^{\nu+P/2} \sigma_\ell
\right)
\right]
\right\rangle \ .
\nonumber
\eea
Setting $\nu=(\mu+P/2)_{mod P}$ we have
\bea 
\hspace*{-1cm}p(\bsigma)&=&\prod_{\mu=1}^{P/2}\left\langle \exp\left[\sqrt{\beta}z\left(
x_\mu^\mu\sum_{k\in \partial\mu}\xi_k^\mu \sigma_k+
x_{\mu+P/2}^\mu\sum_{\ell\in \partial(\mu+P/2)}\xi_\ell^{\mu+P/2} \sigma_\ell
\right)
\right]
\right\rangle
\nonumber\\
&&\times \prod_{\mu=1}^{P/2}\left\langle \exp\left[\sqrt{\beta}z\left(
x_\mu^{\mu+P/2}\sum_{k\in \partial\mu}\xi_k^\mu \sigma_k+
x_{\mu+P/2}^{\mu+P/2}\sum_{\ell\in \partial(\mu+P/2)}\xi_\ell^{\mu+P/2} \sigma_\ell
\right)
\right]
\right\rangle \ .
\nonumber\\
\label{eq:combined_av}
\eea
and combining the averages yields
\bea
\hspace*{-2cm}p(\bsigma)= \prod_{\mu=1}^{P/2}\left\langle \exp\left[\sqrt{\beta}
\left(z_1 x_\mu^\mu+z_2 x_\mu^{\hat{\mu}}
\right)
\sum_{k\in \partial\mu}\xi_k^\mu \sigma_k
+
\sqrt{\beta}\left(z_1 x_{\hat\mu}^{\mu}
+z_2 x_{\hat\mu}^{\hat\mu}
\right)
\sum_{\ell\in \partial\hat\mu}\xi_\ell^{\hat\mu} \sigma_\ell
\right]\right\rangle_{z_1,z_2} \ ,
\nonumber\\ \eea
where $\hat\mu=\mu+P/2$ and $\bra \cdot \ket_{z_1,z_2}$ denotes the average over the distribution 
$f(z_1)f(z_2)$ where $f$ is a Gaussian distribution with zero mean and unit variance.
For all $\mu=1,\ldots,P/2$ we have
\bea
z_1 x_\mu^\mu+z_2 x_\mu^{\mu+P/2}=\insm z_1-\insp z_2\ , \\
z_1 x_{\hat\mu}^{\mu}
+z_2 x_{\hat\mu}^{\hat\mu}
=\insm z_1+\insp z_2\ .
\eea
Next we apply to variables $z_1, z_2$ the transformation
$\by=\bX \bz$
with 
\bea
\bX=\left(
\begin{array}{ll}
\insm & -\insp\\
\insm & \insp
\end{array}
\right)
\quad\quad 
\by=\left(
\begin{array}{l}
y_1 \\
y_2
\end{array}
\right)
\quad\quad
\bz=\left(
\begin{array}{l}
z_1 \\
z_2
\end{array}
\right)
\eea
and rewrite (\ref{eq:combined_av}) in terms of the variables $(y_1,y_2)$.
Using the Jacobian of the transformation $(z_1,z_2)\to (y_1,y_2)$
\be
J=\left(
\begin{array}{ll}
\insm & -\insp\\
\insm & \insp
\end{array}
\right)\ ,
\ee
we have
\bea
p(\bsigma)=\prod_{\mu=1}^{P/2} \left\langle
\exp\left[\sqrt{\beta}
y_1\sum_{k\in \partial\mu}\xi_k^\mu \sigma_k
+
\sqrt{\beta}y_2 
\sum_{\ell\in \partial\hat\mu}\xi_\ell^{\hat\mu} \sigma_\ell
\right]
\right\rangle_\by \ ,
\eea 
where 
\bea
\bra (\cdots) \ket_\by=\detx\int \frac{dy_1 dy_2}{2\pi} (\cdots) e^{-\frac{1}{2}
\by^T \bC^{-1} \by
}
\label{eq:integral}
\eea
and 
\bea
\bC^{-1}=(\bX^{-1})^T\bX^{-1}=\left(
\begin{array}{cc}
1 & -k \\
-k & 1
\end{array}
\right) \ .
\label{eq:Cmatrix}
\eea
We can finally write $P(\bsigma)$ as a product of $P/2$ factors
\be
p(\bsigma)=\prod_{\mu=1}^{P/2} f_{\mu\hat{\mu }}(\{\sigma_k, k\in\partial \mu\},\{\sigma_\ell, \ell\in\partial \hat\mu\})\ , 
\label{eq:psigmaf1}
\ee
each involving a pair of (complementary) clones $\mu, \hat \mu$
\be
f_{\mu\hat{\mu}}(\{\sigma_k, k\in\partial \mu\},\{\sigma_\ell, \ell\in\partial \hat\mu\})= 
\left\langle \exp\left[\sqrt{\beta}\left(y_1
\sum_{k\in \partial\mu}\xi_k^\mu \sigma_k
+y_2
\sum_{\ell\in \partial\hat\mu}\xi_\ell^{\hat\mu} \sigma_\ell
\right)\right]
\right\rangle_{\by}\ .
\label{eq:fmu1}
\ee

\section{Derivation of the cavity equations on a factor tree}
\label{app:BB2}
\begin{figure}[htb!]
\centering
\includegraphics[width=0.48\textwidth]{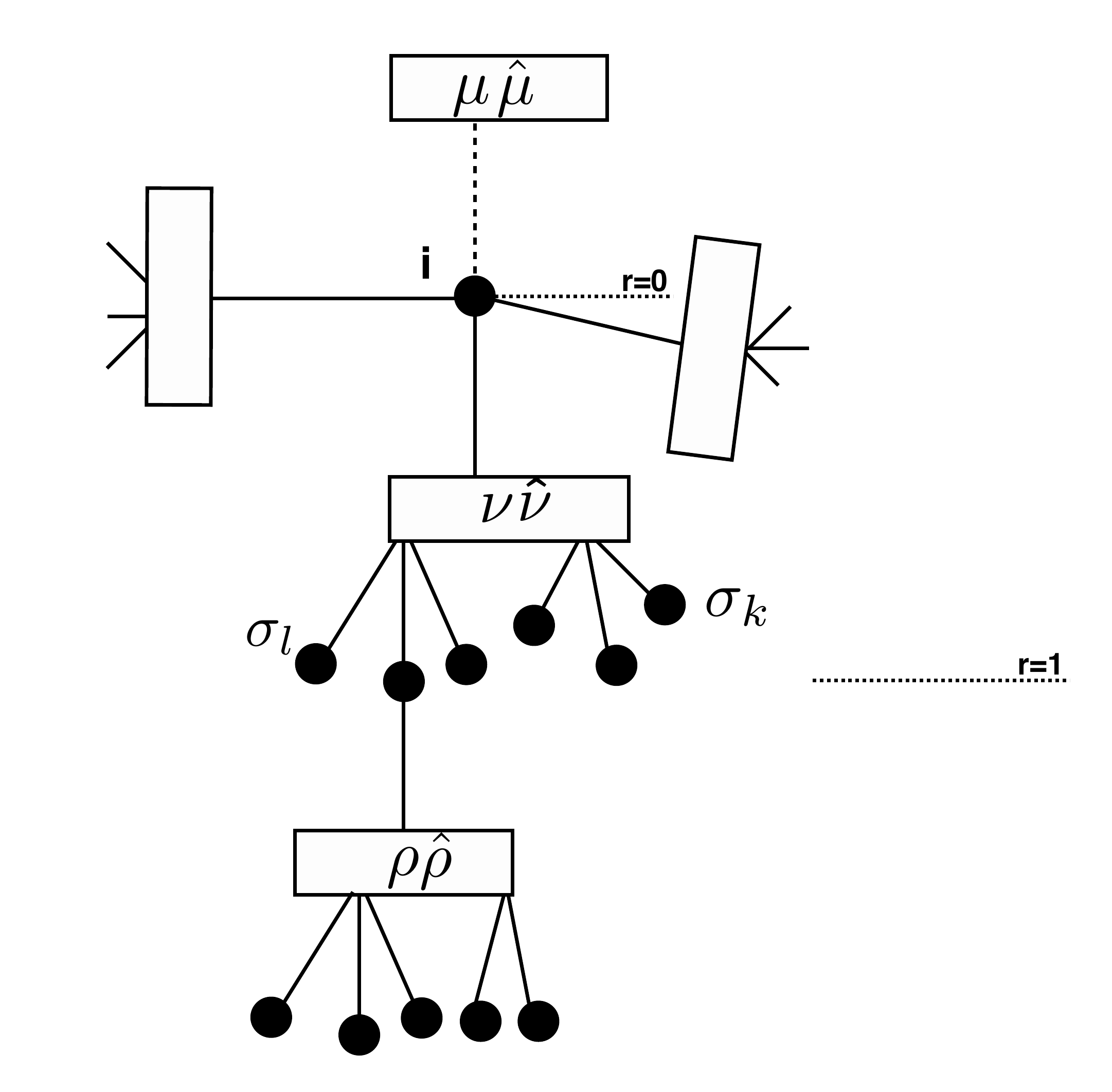}
\caption{Schematic representation of the factor tree $T_i(r)$.}
\label{fig:graph2}
\end{figure}

We consider the tree schematically represented in fig. \ref{fig:graph2}. 
We start by removing the factor $\mu$ and its complementary factor $\hat\mu$ and 
we calculate the partition function of the tree $T_i(r)$ rooted in $i$ 
with depth $r$, in the absence of the pair $\mu,\hat\mu$
\bea
Z_{i\mu\hat\mu }^{(0)}=\sum_{\{\sigma_{n\in T_i(r)}\}} \prod_{\nu\in T_i(r)\setminus \mu,\hat\mu} f_{\nu\hat{\nu}}(\{\sigma_{k\in\partial \nu}\},\{\sigma_{\ell\in \partial\hat{\nu}}\})\ ,
\eea
where the $0$-index highlights that the root is the $0$-th layer of the tree. 
We can rewrite the above more economically as 
\bea
Z_{i\mu\hat\mu }^{(0)}
&=&\sum_{\{\sigma_{n\in T_i(r)}\}} \prod_{\nu\in T_i(r)\setminus \mu,\hat\mu} 
f_{\nu\hat{\nu}}(\{\sigma_{k\in\partial \nu\hat\nu}\})\ ,
\eea
where we introduced the notation $\partial \nu\hat\nu=\partial\nu \cup \partial \hat\nu$ and $\sigma_k$ denotes any spin interacting with either $\nu$ or $\hat\nu$.
Next, we compute the same quantity as above but fixing the spin $\sigma_i$.
This gives the constrained partition function $Z_{i\mu\hat\mu}^{(0)}(\sigma_i)$ 
which represents the un-normalised marginal 
distribution of $\sigma_i$ 
in the absence of factors $\mu,\hat\mu$. Its normalised version gives the 
cavity distribution $P_{\setminus \mu\hat\mu}(\sigma_i)$, 
also defined as the ``message'' that site $i$ sends to factors 
$\mu,\hat\mu$ 
$$
Z_{i\mu\hat\mu}^{(0)}(\sigma_i)=\sum_{\{\sigma_{n\in T_i(r)\setminus i}\}} 
\prod_{\nu\in \partial i\setminus \mu,\hat\mu} \left[
f_{\nu\hat{\nu}}(\sigma_i,\{\sigma_{k\in\partial \nu\hat\nu\setminus i}\})
\prod_{k \in \partial \nu\hat\nu \setminus i}
\prod_{\rho\in T_k(r-1)} f_{\rho\hat{\rho}}(\{\sigma_{s\in\partial \rho\hat\rho\setminus k}\})
\right]\ ,
$$
where we separated the contributions of the factors directly attached to 
$i$ from those in the rest of the tree, which are independent of 
$\sigma_i$. 
Rearranging the sums in the expression above we obtain
\bea
&&\hspace*{-2cm}Z_{i\mu \hat{\mu}}^{(0)}(\sigma_i)
= \prod_{\nu\in\partial i\setminus\mu\hat{\mu}} 
\left[\sum_{\{\sigma_{k\in\partial \nu\hat\nu\setminus i}\}} 
f_{\nu\hat{\nu}}(\sigma_i, \{\sigma_{k\in \partial \nu\hat{\nu}\setminus i}\}) 
\prod_{k\in \partial \nu\hat{\nu} \setminus i }
\sum_{\{\sigma_{n\in T_k(r-1)\setminus k}\}} 
\prod_{\rho\in T_k(r-1)\setminus\nu,\hat\nu}
f_{\rho\hat{\rho}}(\{\sigma_{s\in \partial \rho\hat{\rho}}\})\right]\ .
\eea
The last sum in the expression,
is nothing but the partition function 
$Z_{k\nu\hat{\nu}}^{(1)}(\sigma_k) $ of the sub-tree rooted in $k$ of depth $r-1$, 
where the spin $\sigma_k$ has been fixed. This leads 
to the following recursive relation 
\bea
Z_{i\mu \hat{\mu}}^{(0)}(\sigma_i)
= \prod_{\nu\in \partial i\setminus\mu\hm}
\left[ \sum_{\{\sigma_{k\in\partial \nu\hat\nu\setminus i}\}}
f_{\nu\hat{\nu}}(\sigma_i, \{\sigma_{k\in \partial \nu\hat\nu \setminus i}\}) 
\prod_{k\in \partial \nu\hat{\nu}\setminus i }
Z_{k\nu\hat{\nu}}^{(1)}(\sigma_k)\right] \ ,
\eea
and more in general, one can write a recursion for the (un-normalised) cavity 
distribution at layer $r$ in terms of those at layer $r+1$
\bea
Z_{i\mu\hat{\mu}}^{(r)}(\sigma_i)&=&
\prod_{\nu\in \partial i\setminus\mu\hat{\mu}}\left[ 
\sum_{\{\sigma_{k\in\partial \nu\hat\nu\setminus i}\}}
f_{\nu\hat{\nu}}(\sigma_i, \{\sigma_{k\in \partial \nu\hat\nu \setminus i}\}) 
\prod_{k\in \partial \nu\hat{\nu}\setminus i }
Z_{k\nu\hat{\nu}}^{(r+1)}(\sigma_k)\right]\ .
\nonumber\\
\eea
The stationary solutions of the recursion equations then satisfy
\bea
Z_{i\mu\hat{\mu}}(\sigma_i)&=&
\prod_{\nu\in \partial i\setminus\mu\hat{\mu}}\left[ 
\sum_{\{\sigma_{k\in\partial \nu\hat\nu\setminus i}\}}
f_{\nu\hat{\nu}}(\sigma_i, \{\sigma_{k\in \partial \nu\hat\nu \setminus i}\}) 
\prod_{k\in \partial \nu\hat{\nu}\setminus i }
Z_{k\nu\hat{\nu}}(\sigma_k)\right]\ .
\nonumber\\
\eea
Recalling the meaning of $Z_{i\mu\hm}(\sigma_i)$ as the un-normalised marginal 
distribution of $\sigma_i$ where factors $\mu,\hat{\mu}$ have been removed, 
one finally gets 
\bea
P_{\setminus \mu\hat{\mu}}(\sigma_i)&=&
\prod_{\nu\in \partial i\setminus\mu\hat{\mu}} P_{\nu\hat{\nu}}(\sigma_i)\ ,
\label{eq:rec1}\\
P_{\nu\hat{\nu}}(\sigma_i)&=&
\sum_{\{\sigma_{k\in\partial \nu\hat\nu\setminus i}\}}
f_{\nu\hat{\nu}}(\sigma_i, \{\sigma_{k\in \partial \nu\hat\nu \setminus i}\}) 
\prod_{k\in \partial \nu\hat{\nu}\setminus i }
P_{\setminus\nu\hat{\nu}}(\sigma_k)\ .
\label{eq:rec2}
\eea

\section{Population dynamics algorithm} 
\label{appendix:algorithm}
In this section we provide fomulae to compute the cavity fields in our population dynamics algorithm. 
Starting form the parametrisation of $P_{\setminus \mu\hat{\mu}}(\sigma_i)$ introduced in Sec. \ref{sec:disorder}
\begin{eqnarray}
P_{\setminus \mu\hat{\mu}}(\sigma_i) \propto{\rm e}^{\beta \phi_{i\to\mu\hat{\mu}}\sigma_i}\ ,
\label{eq:par}
\end{eqnarray}
we can express the cavity field $\phi_{i\to\mu\hat\mu}$ in terms of the cavity marginals
\begin{eqnarray}
\phi_{i\to\mu\hat{\mu}}=\frac{1}{2\beta}\sum_{\sigma}\sigma_i\log( P_{\setminus \mu\hat{\mu}}(\sigma_i))=\\
=\frac{1}{2\beta}\log\left[\frac{P_{\setminus \mu\hat{\mu}}(+1)}{P_{\setminus \mu\hat{\mu}}(-1)}\right].
\label{eq:phiexpr}
\end{eqnarray}
which satisfies the recursion 
\bea
P_{\setminus \mu\hat{\mu}}(\sigma_i)=\prod_{\nu\in \partial i\setminus\mu\hat{\mu}}\sum_{\{\sigma_{k\in\partial \nu\setminus i}\},\{\sigma_{\ell\in\partial \hat{\nu}}\}}f_{\nu\hat\nu}(\sigma_i, \{\sigma_{k\in \partial \nu \setminus i}\}, \{\sigma_{\ell\in \partial \hat{\nu} }\}) 
\prod_{k\in \partial \nu\hat{\nu}\setminus i }P_{\setminus \nu\hat{\nu}}(\sigma_k)\ .
\eea
obtained by combining (\ref{eq:rec1}) and (\ref{eq:rec2}). 
Inserting the explicit expression for the factors 
\be
f_{\nu\hat{\nu}}(\sigma_i,\{\sigma_k, k\in\partial \nu\},\{\sigma_\ell, \ell\in\partial \hat\nu\})= 
\left\langle \exp\left[\sqrt{\beta}\left(y_1\xi^\nu_i\sigma_i+y_1
\sum_{k\in \partial\nu\setminus i}\xi_k^\nu \sigma_k
+y_2
\sum_{\ell\in \partial\hat\nu}\xi_\ell^{\hat\nu} \sigma_\ell
\right)\right]
\right\rangle_{\by}\ , \label{eq:fmu1}
\ee
we obtain
\bea
P_{\setminus \mu\hat{\mu}}(\sigma_i)&=&\prod_{\nu\in \partial i\setminus\mu\hat{\mu}}\sum_{\{\sigma_{k\in\partial\nu\setminus i}\},\{\sigma_{\ell\in\hat\nu}\}} \int 
\mathcal{D}{\bf y}\exp\left[\sqrt{\beta}\left(y_1\xi^\nu_i\sigma_i+y_1
\sum_{k\in \partial\nu\setminus i}\xi_k^\nu \sigma_k
+y_2\sum_{\ell\in \partial\hat\nu}\xi_\ell^{\hat\nu} \sigma_\ell\right)\right]
\nonumber\\
&&\times
\prod_{k\in \partial \nu\hat{\nu}\setminus i }P_{\setminus \nu\hat{\nu}}(\sigma_k)\ ,
\label{eq:ps1}
\eea
with $\mathcal{D}{\bf y}=\detx \frac{\rmd y_1 \rmd y_2}{2\pi} \rme^{-\frac{1}{2}
\by^T \bC^{-1} \by
}$.
Expressing everything in terms of the fields $\phi_{k\to \nu\hat\nu}$ and introducing the variables $X_\nu=\sum_{k\in\partial\nu\setminus i}\xi_k^\nu \sigma_k$ and $\hat{X}_\nu=\sum_{\ell\in\partial\hat{\nu}}\xi_\ell^{\hat{\nu}} \sigma_\ell$, yields
\begin{flalign}
P_{\setminus \mu\hat{\mu}}(\sigma_i)=\prod_{\nu\hat{\nu}\in \partial i\setminus\mu\hat{\mu}}\sum_{X_\nu,\hat{X}_\nu} \int \mathcal{D}{\bf y} {\rm e}^{\sqrt{\beta}(y_1\xi^\nu_i\sigma_i+y_1X_\nu+y_2\hat{X}_\nu)}\sum_{\{\sigma_{k}\}}\prod_{k\in\partial\nu\setminus i} {\rm e}^{\beta\phi_{k\to\nu\hat\nu}\sigma_k}\delta_{X_\nu,\sum_{k\in\partial\nu\setminus i}\xi_k^\nu \sigma_k}
\nonumber\\ \times\sum_{\{\sigma_{\ell}\}}\prod_{\ell\in\partial\hat\nu}{\rm e}^{\beta\phi_{\ell\to\nu\hat\nu}\sigma_\ell}\delta_{\hat{X}_\nu,\sum_{\ell\in\partial\hat{\nu}}\xi_\ell^{\hat{\nu}} \sigma_\ell} \ .
\label{eq:ps2}
\end{flalign}
Finally, we can integrate over ${\bf y}$
\bea
 \int \mathcal{D}{\bf y}(\cdots)=\detx\int \frac{\rmd y_1 \rmd y_2}{2\pi}( \cdots )\rme^{-\frac{1}{2}
\by^T \bC^{-1} \by}\ ,
\eea
using
\bea
\bC^{-1}=\left(
\begin{array}{cc}
1 & -k \\
-k & 1
\end{array}
\right) \ .
\label{eq:Cmatrix}
\eea
The integral in \eqref{eq:ps2} can be written in the form 
\bea 
\detx\int \frac{\rmd y_1 \rmd y_2}{2\pi} {\rm e}^{-\frac{1}{2}
\by^T \bC^{-1} \by + {\bf J}^T\by}\ ,
\eea 
with $ {\bf J}= (\sqrt{\beta}
(\xi^\nu_i\sigma_i+X_\nu),\sqrt{\beta}
\hat{X}_\nu)$ and ${\bf y}=(y_1,y_2) $.
Solving it by Gaussian integration gives
\bea 
\detx\int \frac{\rmd y_1 \rmd y_2}{2\pi} {\rm e}^{-\frac{1}{2}
\by^T \bC^{-1} \by + {\bf J}^T\by} = {\rm e}^{\frac{1}{2} {\bf J}^T \bC {\bf J}}\ ,
\eea 
with 
\bea
\bC=\left(
\begin{array}{ll}
\frac{1}{1-k^2} & \frac{k}{1-k^2} \\
\frac{k}{1-k^2} & \frac{1}{1-k^2} 
\end{array}
\right)\ .
\label{eq:Cmatrixinverse}
\eea
Hence, we get
\bea 
{\rm e}^{\frac{1}{2} {\bf J}^T \bC {\bf J}} = \exp\bigg({\frac{\beta}{2(1-k^2)}\left((\xi^\nu_i\sigma_i+X_\nu)^2 + 2k \hat{X}_\nu(\xi^\nu_i\sigma_i+X_\nu) +\hat{X}_\nu^2\right)\bigg)}\ .
\eea
We insert this result in \eqref{eq:ps2}, obtaining
\begin{flalign}
P_{\setminus \mu\hat{\mu}}(\sigma_i)=\prod_{\nu\hat{\nu}\in \partial i\setminus\mu\hat{\mu}}\sum_{X_\nu,\hat{X}_\nu} \exp\bigg({\frac{\beta}{2(1-k^2)}\left((\xi^\nu_i\sigma_i+X_\nu)^2 + 2k \hat{X}_\nu(\xi^\nu_i\sigma_i+X_\nu) +\hat{X}_\nu^2\right)\bigg)}\times\nonumber\\ \times
\sum_{\{\sigma_{k}\}}\prod_{k\in\partial\nu\setminus i} {\rm e}^{\beta\phi_{k\to\nu\hat\nu}\sigma_k}\delta_{X_\nu,\sum_{k\in\partial\nu\setminus i}\xi_k^\nu \sigma_k}\sum_{\{\sigma_{\ell}\}}\prod_{\ell\in\partial\hat\nu}{\rm e}^{\beta\phi_{\ell\to\nu\hat\nu}\sigma_\ell}\delta_{\hat{X}_\nu,\sum_{\ell\in\partial\hat{\nu}}\xi_\ell^{\hat{\nu}} \sigma_\ell}\ .
\label{eq:ps3}
\end{flalign}
\begin{figure}[htb!]
\centering
\includegraphics[width=0.8\textwidth]{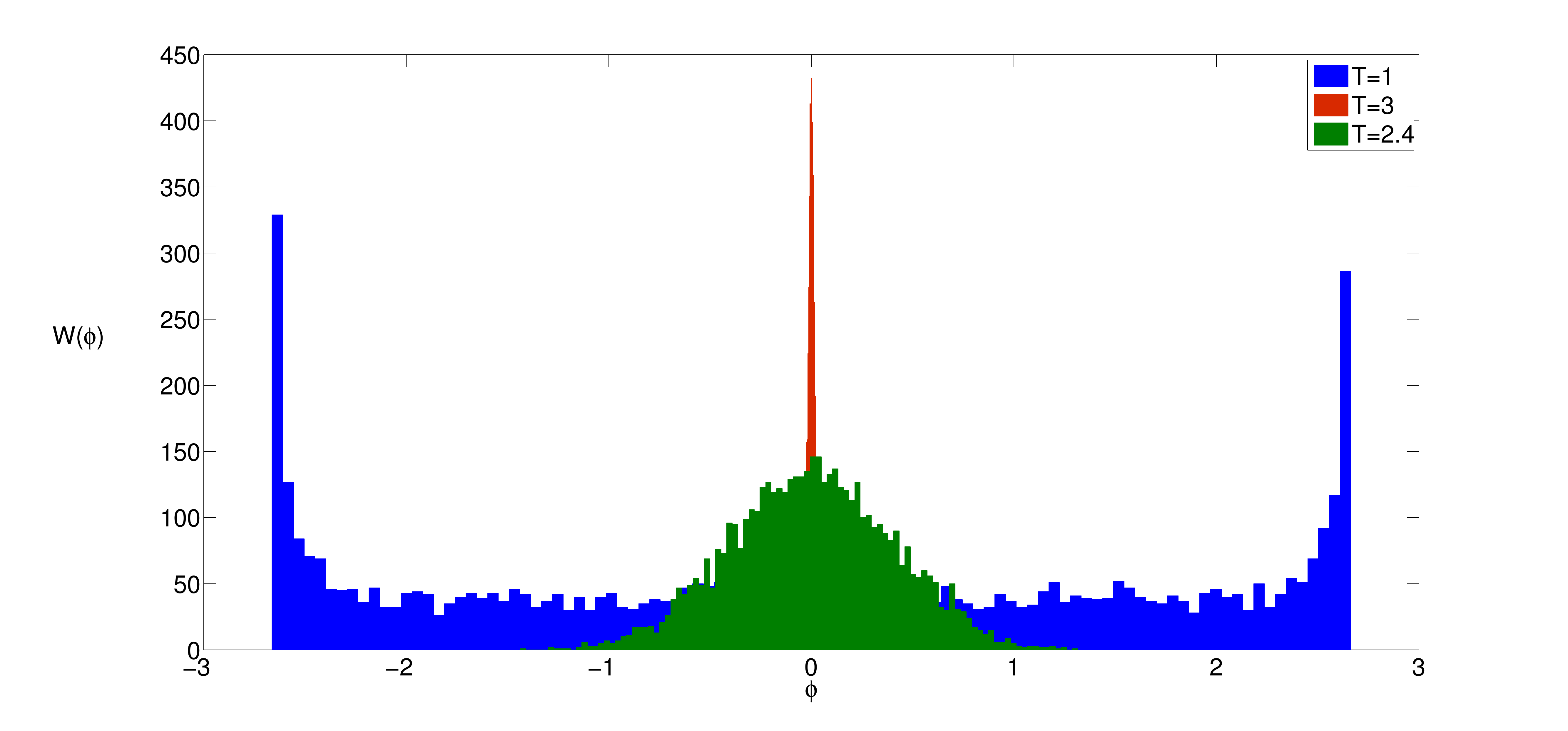}
\caption{Plot of  $W(\phi)$ fields distribution for a regular random graph with $L=2=K$, $k=0.5$ and bias $a=0$ for $\beta=0.05,1$.}
\label{fig:fieldscav}
\end{figure}
Lastly, we plug the expression in \eqref{eq:phiexpr} in order to get the expression for the update of the cavity field
\begin{flalign}
\phi_{i\to\mu\hat{\mu}}= \frac{1}{2\beta}\sum_{\nu\hat{\nu}\in \partial i\setminus\mu\hat{\mu}}\log\bigg\{ \bigg[  \sum_{X_\nu,\hat{X}_\nu} \exp\bigg({\frac{\beta}{2(1-k^2)}\left((\xi^\nu_i+X_\nu)^2 + 2k \hat{X}_\nu(\xi^\nu_i\sigma_i+X_\nu) +\hat{X}_\nu^2\right)\bigg)}\times\nonumber\\ \times
\sum_{\{\sigma_{k}\}}\prod_{k\in\partial\nu\setminus i} {\rm e}^{\beta\phi_{k\to\nu\hat\nu}\sigma_k}\delta_{X_\nu,\sum_{k\in\partial\nu\setminus i}\xi_k^\nu \sigma_k}\sum_{\{\sigma_{\ell}\}}\prod_{\ell\in\partial\hat\nu}{\rm e}^{\beta\phi_{\ell\to\nu\hat\nu}\sigma_\ell}\delta_{\hat{X}_\nu,\sum_{\ell\in\partial\hat{\nu}}\xi_\ell^{\hat{\nu}} \sigma_\ell}\bigg]\bigg / \nonumber\\ \bigg[  \sum_{X_\nu,\hat{X}_\nu} \exp\bigg({\frac{\beta}{2(1-k^2)}\left((-\xi^\nu_i+X_\nu)^2 + 2k \hat{X}_\nu(\xi^\nu_i\sigma_i+X_\nu) +\hat{X}_\nu^2\right)\bigg)}\times\nonumber\\ \times
\sum_{\{\sigma_{k}\}}\prod_{k\in\partial\nu\setminus i} {\rm e}^{\beta\phi_{k\to\nu\hat{\nu}}\sigma_k}\delta_{X_\nu,\sum_{k\in\partial\nu\setminus i}\xi_k^\nu \sigma_k}\sum_{\{\sigma_{\ell}\}}\prod_{\ell\in\partial\hat\nu}{\rm e}^{\beta\phi_{\ell\to\nu\hat\nu}\sigma_\ell}\delta_{\hat{X}_\nu,\sum_{\ell\in\partial\hat{\nu}}\xi_\ell^{\hat{\nu}} \sigma_\ell}\bigg]\bigg\} \ .
\label{eq:updatecode}
\end{flalign}
This expression for the fields update will be used in the population dynamics algorithm.
\subsection*{Algorithm}
We summarise here the main steps of the algorithm:
\begin{enumerate}
\item Define the degree distributions $Q(e)$ and $P_d(d)$ - in our algorithm we used a regular graph with vertex degree $L$ and factor degree $K$.
\item Generate the links $\bxi=(\xi_1,\dots,\xi_e)$ with $e$ i.i.d. random variables with probability distribution $P(\xi)= \frac{1+a}{2}\delta_{\xi,1}+\frac{1-a}{2}\delta_{\xi,-1}$, dependent on the parameter $a$. 
\item Extract a population composed of $M$ fields $\phi_i$, $i=1,\ \dots M$ uniformly in the interval $[-f_{max},f_{max}]$: their histogram defines the zero-step approximation of the  field distribution $W_{\psi}^0(\Psi)$.
\item Start the iteration: choose $e,d$ and generate $\xi$'s.
\item Choose $e-1$ fields randomly: compute the updated field $\phi_{{\rm new}}$using \eqref{eq:updatecode}.
\item Choose randomly one field $\phi_i$ and replace it with the field just computed.
\end{enumerate}
\begin{figure}[htb!]
\centering\includegraphics[trim= 0cm 0cm 2cm 1.19cm, clip, width=0.8\textwidth]{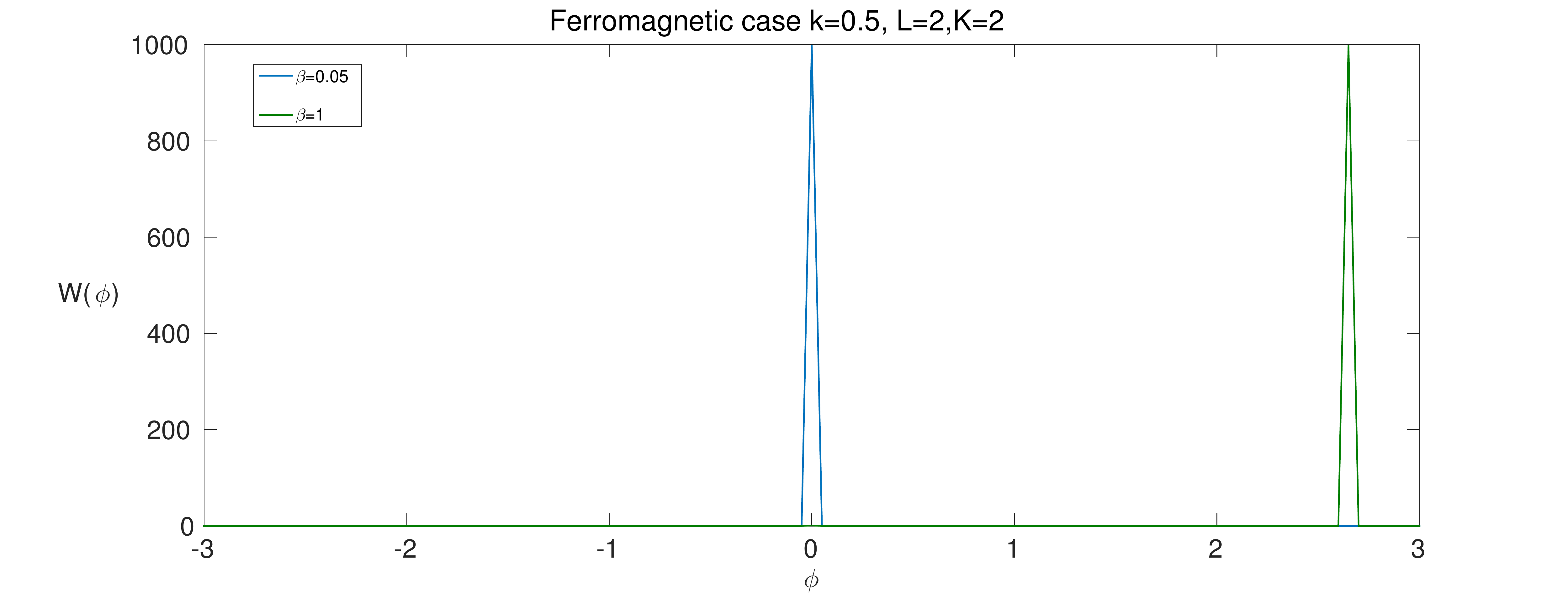}
\caption{Ferromagnetic case $\xi=1$ $(a=1)$ on a regular graph $L=2,K=2$ with $k=0.5$ for $\beta=1$ (blue), $0.05$ (green). 
When $\beta=1$ the distribution is peaked around $\phi=2.66$ as predicted analytically from \eqref{eq:phic}.}
\label{fig:fieldferro}
\end{figure}
In figure \ref{fig:fieldscav} we plot the field distribution $W(\phi)$ in different regions of the phase diagram fixing $k=0.5$ and varying the temperature for 
disordered patterns drawn from a symmetric distribution i.e. $a=0$. Starting from the high temperature regime, where the distribution is delta-peaked in $\phi=0$, 
the variance increases as the temperature is decreased, when crossing the critical line.
For ordered interactions, i.e. $a=1$, the distribution, plotted in figure \ref{fig:fieldferro}, shows a transition from a delta peak in 
$\phi=0$ at high temperature, 
to a peak at non-zero values, when the critical line is crossed.

\end{document}